\documentclass[fleqn,11pt]{article}
\usepackage[numbers,sort&compress]{natbib}
\usepackage[footnotesep=0.4in]{geometry}
\usepackage{graphicx}
\usepackage{amsmath}
\usepackage{amsfonts}
\usepackage{caption}
\usepackage{float}
\usepackage{graphics}
\usepackage{subfigure}
\usepackage{float}
\usepackage{booktabs}
\usepackage{epstopdf}
\usepackage[utf8]{inputenc}
\usepackage{multirow}
\usepackage{array}
\usepackage{hyperref}
\usepackage{verbatim}
\usepackage{diagbox}
\usepackage{makecell}
\usepackage{amsmath,amssymb}
\usepackage{amsthm}
\theoremstyle{plain}

\theoremstyle{definition}
\newtheorem{remark}{Remark}

\hypersetup{hypertex=true,
colorlinks=true,
linkcolor=black,
anchorcolor=black,
citecolor=black}
\makeatletter
\newcommand\figcaption{\def\@captype{\textbf{figure}}\caption}
\newcommand\tabcaption{\def\@captype{table}\caption}
\begin{document}

\captionsetup[figure]{labelfont={bf},labelformat={default},labelsep=period,name={Fig.}}
\date{}
\title{Novel Transition Mechanisms of Vector Localized Waves Induced by the Fourth-Order Effect}

\author{Rong Han$^{1}$, Muchen Dong$^{2}$ and Lei Wang$^{2\,,}$\thanks{Corresponding author: 50901924@ncepu.edu.cn}
\\{\em \small  1. \,School of Control and Computer Engineering, North China Electric Power University, Beijing 102206, China}
\\{\em \small  2.\,School of Mathematics and Physics, North China Electric Power University, Beijing 102206, China}
}
\maketitle

\vspace{-5mm}
\begin{abstract}
We investigate novel vector localized wave solutions in the coupled Lakshmanan-Porsezian-Daniel equations, which describe the dynamics of the Heisenberg ferromagnetic spin chain. We present Tajiri-Watanabe breathers, rogue waves and resonant modes in both the degenerate and non-degenerate regions, together with the degenerate beating solitons. The fourth-order effect induces state transitions in both regions. In particular, the degenerate breathers can be transformed into solitons, whereas such transitions are absent in the coupled Hirota equations. Moreover, beating solitons can be converted into stable solitons only in the degenerate region, the phenomena not found in the Manakov system. We further uncover the state transitions of the resonant modes and derive the corresponding transition conditions for each branch. We derive the physical spectra and subsequently identify the state transition conditions in the spectral domain for both the degenerate and non-degenerate cases. These spectra provide an additional characterization of the transition dynamics. Finally, direct numerical simulations are performed to verify the validity of the exact solutions.

\vspace{2mm}
\end{abstract}

\newpage

\vspace{3mm}
\section{Introduction}
Breathers are nonlinear localized structures characterized by periodic oscillations and have been widely observed in various physical fields, such as optics~\cite{I1,I2}, water wave tanks~\cite{I3}, Bose-Einstein condensates~\cite{I4,I5}. The finite-period breathers can generally be classified into three types. The Akhmediev breather~(AB), also known as a temporal homoclinic-orbit solution, is periodic in space and localized in time~\cite{I6,I7}. In contrast, the Kuznetsov-Ma soliton (KMS) is periodic in time and localized in space, with vanishing group velocity~\cite{breather5,KMS6}. A more general breather with nonzero group velocity is referred to as the Tajiri-Watanabe~(TW) breather~\cite{TW}. In the infinite period limit, these breather solutions degenerate into rogue waves~(RWs), which are localized in both space and time~\cite{RW}. As the classic scalar model, the nonlinear Schr\"odinger equation~(NLSE)~\cite{NLSE11} has provided a fundamental basis for the study of these breathers and extensive results have been reported in~\cite{NLSE1,NLSE2,NLSE3,NLSE4,NLSE5}. It should be noted that the NLSE describes only the nonlinear dynamics of a scalar wave field. In many physical systems, however, multiple wave components are nonlinearly coupled. Vector integrable models therefore provide a natural framework for describing nonlinear interactions in diverse physical scenarios~\cite{breather1,breather2,breather3,Tbreather4}.

The Manakov system~\cite{Manakov1}, as a two-component generalization of the NLSE, is given by
\begin{equation}\label{eq:Manakov}
\begin{aligned}
i\psi_{t}^{(j)}+\frac{1}{2}\psi_{xx}^{(j)}+\psi^{(j)}(|\psi^{(1)}|^{2}+|\psi^{(2)}|^{2})=0,\quad j=1,2,
\end{aligned}
\end{equation}
where $\psi^{(j)}(x,t)$ describe the orthogonally polarised complex waves while the variables $x$ and $t$ are the propagation
distance and the retarded time in a moving frame, respectively~\cite{Manakov2}. The Manakov system can describe the evolution of localized waves in two-mode nonlinear fiber~\cite{Manakov3}, two-component Bose-Einstein condensate~\cite{Manakov4}, and two-directional ocean waves~\cite{Manakov5}. Extensive studies of this system have uncovered the vector breather and rogue wave dynamics that are generally unattainable in the NLSE~\cite{Manakov6,Manakov7,Manakov8,Manakov9}. Beyond these localized waves, a distinctive class of the vector breathers known as beating solitons~(BSs), can be derived from dark-bright soliton solutions via the SU(2) rotation symmetry of the Manakov system~\cite{dfBS}. Subsequent studies have investigated the classification and resonant modes of vector BSs~\cite{beating1,Glesh1}. Recently, the non-degenerate localized waves characterized by distinct spectral parameters have attracted considerable attention. Their existence conditions, spectral characteristics, and excitation mechanisms have been systematically elucidated~\cite{non-AB1,non-AB2,non-AB3,non-AB4,non-KMS1,non-RW1,non-RW2}. Building on these developments, vector super-regular breathers~(SRBs) and their connection to modulation instability (MI) have also been explored
~\cite{SRB1}. Moreover, the non-degenerate localized waves have been extended to other vector systems, such as the coupled Hirota equations~\cite{pan1,pan2}, vector derivative NLSEs~\cite{DNLSE}, three-wave resonant interaction system~\cite{TWRI,TWRI1}, long wave-short wave resonance systems~\cite{LS}, and three-component Manakov system~\cite{TNLSE}.

The Manakov system neglects higher-order effect and is therefore inadequate for high-intensity ultrashort pulses in two-mode optical fibers. Higher-order vector models provide a more accurate description of localized wave dynamics in nonlinear systems such as microstructured optical fibers~\cite{CH11} and fiber lasers~\cite{CH12}. A representative example incorporating third-order effect is the coupled Hirota~(CH) equations, which can be written as
\begin{equation}\label{eq:CH}
\begin{aligned}
i\psi_{t}^{(j)}+\frac{1}{2}\psi_{xx}^{(j)}+M\psi^{(j)}
+i\varepsilon[\psi_{xxx}^{(j)}+3M\psi_{x}^{(j)}+3P\psi^{(j)}]=0,\quad j=1,2.
\end{aligned}
\end{equation}
Here, for concision we introduce $M=|\psi^{(1)}|^{2}+|\psi^{(2)}|^{2}$, $P=\psi_{x}^{(1)}\psi^{(1)*}+\psi_{x}^{(2)}\psi^{(2)*}$. The envelope of wave fields $\psi^{(j)}(x,t)$ and the variables $x$ and $t$ represent the similar physical meaning as those in Eqs.~(\ref{eq:Manakov}). The CH equations support both the degenerate and non-degenerate vector localized waves, including TWs, ABs, KMSs, RWs, and SRBs~\cite{pan1,pan2}. Notably, breather-to-soliton transitions are restricted to the non-degenerate regions, where various solitons can be observed. No corresponding transitions are found in the degenerate regions. Although Eqs.~(\ref{eq:Manakov})--(\ref{eq:CLPDE}) share the same spatial spectral problem in their Lax pairs, it remains unclear whether this transition mechanism persists induced by the fourth-order effect, particularly in the degenerate regions.

We further consider the simultaneous propagation of optical fields in a two-mode optical fibers with fourth-order effect, the wave dynamics of the system can be governed by the coupled Lakshmanan-Porsezian-Daniel~(CLPD) equations~\cite{1}. This model in dimensionless form is given by:
 \begin{equation}\label{eq:CLPDE}
 \begin{aligned}
 i\psi_{t}^{(j)}+\frac{1}{2}\psi_{xx}^{(j)}+M\psi^{(j)}
 +\varepsilon[\psi_{xxxx}^{(j)}+4M\psi_{xx}^{(j)}+2\psi^{(j)}_{x}(M_{x}+2P)+2\psi^{(j)}(3M^{2}+P^{*}_{x}+2Q)]=0,\quad j=1,2.
 \end{aligned}
 \end{equation}
Where $Q=\psi_{xx}^{(1)}\psi^{(1)*}+\psi_{xx}^{(2)}\psi^{(2)*}$, $M$ and $P$ are defined in the same way as those in Eqs.~(\ref{eq:CH}). Here, the envelope of wave fields $\psi^{(j)}(x,t)$ and the variables $x$ and $t$ represent the similar physical meaning as those in Eqs.~(\ref{eq:Manakov}). As a vector generalization of the scalar LPD equation, Eqs.~(\ref{eq:CLPDE}) describe complex coupled dynamics in Heisenberg ferromagnetic chains~\cite{Hei1,Hei2}. The parameter $\varepsilon$ characterizes fourth-order effect beyond the Manakov limit ($\varepsilon=0$), enabling a more accurate description of fewcycle ultrashort pulse propagation in optical fibers~\cite{4,5}. Vector breathers and RWs in Eqs.~(\ref{eq:CLPDE}) have been extensively investigated~\cite{CLPD1,CLPD2,CLPD3,CLPD4}. We note that the reported state transitions are restricted to the equal-wavenumber backgrounds, under which the solutions of Eqs.~(\ref{eq:CLPDE}) in fact reduce to those in the scalar LPD equation~\cite{CLPD3}. Recently, Ref.~\cite{CLPD5} has been devoted to the vector ABs and SRBs, with non-degeneracy identified through unequal background wavenumbers. However, we should point out that the degenerate and non-degenerate regions are distinguished by a non-zero critical relative wavenumber.

In this paper, we investigate vector degenerate and non-degenerate localized waves in the CLPD equations, including TWs, RWs, BSs, and resonant modes. We show that both the degenerate and non-degenerate localized waves can be transformed into solitons. In contrast to the CH equations, where such transitions are absent in the degenerate regions, highlighting a new mechanism induced by the fourth-order effect. Most notably, we report for the first time the state transitions between BSs and stable solitons, a phenomenon not supported by the Manakov system. We also uncover the state transitions in resonant modes and derive the corresponding transition conditions for the individual branch. These findings demonstrate previously unexplored dynamics induced by the fourth-order effect. Furthermore, we derive the physical spectra of the localized waves. This enables the transition conditions to be identified in the spectral domain for both the degenerate and non-degenerate cases. Finally, direct numerical simulations confirm the validity of the exact solutions.

The remainder of this paper is organized as follows. Sec.~\ref{eq:sec2} reformulates the Lax pair and develops a generalized Darboux transformation for the CLPD equations. In Sec.~\ref{eq:sec3}, the exact analytical expressions of the fundamental solutions are derived, together with their corresponding state transition conditions. Sec.~\ref{eq:sec4} focuses on the non-degenerate TW and RW resonant modes, with particular emphasis on the state transitions of their individual branches. In Sec.~\ref{eq:sec5}, we present the fundamental BSs, the resonant modes and interactions between two type BSs, as well as the associated state transition solitons. The analytical physical spectra of the localized waves under the fourth-order effect are discussed in Sec.~\ref{eq:sec6}. Sec.~\ref{eq:sec7} provides representative numerical simulations to verify the validity of the analytical results. Finally, the main conclusions are summarized in Sec.~\ref{eq:sec8}.


\vspace{3mm}
\section{DT} \label{eq:sec2}
\vspace{2mm}
In this section, we reformulate Lax pair of the CLPD equations and derive vector degenerate and non-degenerate breathers and RWs, as well as the degenerate BSs via the Darboux transformation~(DT) method.

\subsection{Lax pair}
Similar to the procedure of DT in \cite{1}, we obtain the exact analytical expression for vector degenerate and non-degenerate localised waves corresponding to three individual eigenvalues. We represent Eqs.~$(\ref{eq:CLPDE})$ as the condition of compatibility of two linear equations with $3\times3$ matrix operators:
\begin{equation}\label{eq:lax1}
\begin{aligned}
\Psi_{x}=U\Psi,\quad \Psi_{t}=V\Psi,
\end{aligned}
\end{equation}
where $\Psi=(R,S,W)^{\top}$ ($\top$ means a matrix transpose), and

\begin{equation}
U =i \begin{pmatrix}
\lambda & \psi^{(1)*} & \psi^{(2)*} \\
\psi^{(1)} & 0 & 0 \\
\psi^{(2)} & 0 & 0
\end{pmatrix},\label{eq:U}
\end{equation}
\begin{equation}
V=\lambda^{4}V_{4}+\lambda^{3}V_{3}+\lambda^{2}V_{2}+\lambda V_{1}+V_{0},\label{eq:V}
\end{equation}
with
\begin{equation}
\begin{aligned}
&V_{4}=-\frac{1}{2}i\varepsilon(I+\Lambda),\quad V_{3}=-i\varepsilon Q,\\
&V_{2}=\frac{1}{4}i\Lambda+i\varepsilon\Lambda Q^{2}-\varepsilon \Lambda Q_{x}, \quad V_{1}=\varepsilon[Q_{x},Q]+\frac{1}{2}iQ+2i\varepsilon Q^{3}+i\varepsilon Q_{xx},\\
&V_{0}=-\frac{1}{2}i\Lambda Q^{2}-3i\varepsilon\Lambda Q^{4}-i\varepsilon\Lambda(QQ_{xx}+Q_{xx}Q-Q_{x}^{2})+\frac{1}{2}\Lambda Q_{x}+3\varepsilon\Lambda (Q^{2}Q_{x}+Q_{x}Q^{2})+\varepsilon\Lambda Q_{xxx},
\end{aligned}
\end{equation}
and
\[
I = \begin{pmatrix}
1 & 0 & 0\\
0 & 1 & 0\\
0 & 0 & 1\\
\end{pmatrix},\qquad
\Lambda = \begin{pmatrix}
1 & 0 & 0\\
0 & -1 & 0\\
0 & 0 & -1\\
\end{pmatrix},\qquad
Q= \begin{pmatrix}
0 & \psi^{(1)*} & \psi^{(2)*}\\
\psi^{(1)} & 0 & 0\\
\psi^{(2)} & 0 & 0\\
\end{pmatrix}.
\]
In the above equations, the asterisk $*$ denotes the complex conjugation, $\lambda$ is the spectral parameter. It is easy to show that Eqs.~(\ref{eq:CLPDE}) can be exactly reproduced from the compatibility condition
\begin{equation}
\begin{aligned}
U_{t}-V_{x}+[U,V]=0.
\end{aligned}
\end{equation} \label{eq:condition1}

\subsection{Localized waves solutions}
In the subsequent discussion, we derive the fundamental and higher-order vector localized wave solutions for Eqs.~(\ref{eq:CLPDE}), which includes TWs, RWs and BSs by means of DT method. The seed solutions of Eqs.~(\ref{eq:CLPDE}) are vector background plane waves given by
\begin{equation}\label{eq:seed solution}
\begin{aligned}
\psi_{0}^{(j)}=a_{j}\exp[i(\beta_{j}x-\omega_{j}t)], \quad j=1,2,
\end{aligned}
\end{equation}
where
\begin{equation}
\begin{aligned}
\omega_{j}=\frac{\beta_{j}^{2}}{2}-A+4\varepsilon(A-a_{j}^{2})(\beta_{1}^{2}+\beta_{1}\beta_{2}+\beta_{2}^{2})+\varepsilon[\beta_{j}^{2}(12a_{j}^{2}-\beta_{j}^{2})-6A^{2}],
\end{aligned}
\end{equation}
with $A=a_{1}^{2}+a_{2}^{2}$. Here, $a_{j}$ are two real amplitudes and $\beta_{j}$ are the wavenumbers of background waves. The vector eigenfunctions $\Psi[1]=(R[1],S[1],W[1])^{\top}$ of the linear system (\ref{eq:lax1}) can be written as
\begin{equation}\label{PSI}
\begin{aligned}
&R[1]=\varphi_{1a}+\varphi_{1b}+\varphi_{1c},\\
&S[1]=\psi_{0}^{(1)}(\frac{\varphi_{1a}}{\beta_{1}+\chi_{1a}}+\frac{\varphi_{1b}}{\beta_{1}+\chi_{1b}}+\frac{\varphi_{1c}}{\beta_{1}+\chi_{1c}}),\\
&W[1]=\psi_{0}^{(2)}(\frac{\varphi_{1a}}{\beta_{2}+\chi_{1a}}+\frac{\varphi_{1b}}{\beta_{2}+\chi_{1b}}+\frac{\varphi_{1c}}{\beta_{2}+\chi_{1c}}).
\end{aligned}
\end{equation}
Meanwhile, three fundamental solutions are given as follows:
\begin{equation}\label{phi}
\begin{aligned}
\varphi_{1a}&=c_{11}\exp\{i[\chi_{1a}x+(\Theta_{1a}+\Theta_{0})t]\},\\
\varphi_{1b}&=c_{12}\exp\{i[\chi_{1b}x+(\Theta_{1b}+\Theta_{0})t]\},\\
\varphi_{1c}&=c_{13}\exp\{i[\chi_{1c}x+(\Theta_{1c}+\Theta_{0})t]\},\\
\end{aligned}
\end{equation}
with
\begin{equation}
\begin{aligned}
&\Theta_{0}=-A-\frac{\lambda^{2}}{4}-6\varepsilon A^{2}+4\varepsilon(a_{1}^{2}\beta_{1}^{2}+a_{2}^{2}\beta_{2}^{2}),\\
&\Theta_{1l}=-\varepsilon\chi_{1l}^{4}+(\frac{1}{2}+4\varepsilon A)\chi_{1l}^{2}-4\varepsilon(a_{1}^{2}\beta_{1}+a_{2}^{2}\beta_{2})\chi_{1l},\quad l=a,b,c.
\end{aligned}
\end{equation}
Then fundamental breather solutions $\psi^{(1)}[1]$ and $\psi^{(2)}[1]$ are given by Lax pair (\ref{eq:lax1}) as the following forms
\begin{equation}\label{DT1}
\begin{aligned}
\psi^{(1)}[1]=\psi_{0}^{(1)}+\frac{(\lambda^{*}[1]-\lambda[1])R^{*}[1]S[1]}{|R[1]|^{2}+|S[1]|^{2}+|W[1]|^{2}},\\
\psi^{(2)}[1]=\psi_{0}^{(2)}+\frac{(\lambda^{*}[1]-\lambda[1])R^{*}[1]W[1]}{|R[1]|^{2}+|S[1]|^{2}+|W[1]|^{2}}.\\
\end{aligned}
\end{equation}

\begin{remark}
Especially, the fundamental vector RWs solutions are in forms of Eqs.~(\ref{DT1}) when $\lambda_{0}^{*}\neq\lambda_{0}$ ($\lambda_{0}$ is the spectral parameter corresponding to the eigenvalue $\chi_{a}=\chi_{b}$ at the branching point) and $\Psi_{RW}=\lim\limits_{\epsilon\rightarrow0} \frac{\Psi(\chi_{0}+\epsilon)-\Psi(\chi_{0})}{\epsilon}$ ($\epsilon\ll0$ is a small real parameter).
\end{remark}

Higher-order breather solutions can be obtained via the iteration of DT from the fundamental breather solution Eqs.~(\ref{DT1}). After
performing the transformation, we obtain the $n$th-order breather solutions:
\begin{equation}\label{DT2}
\begin{aligned}
\psi^{(1)}[n]
&=
\psi^{(1)}[n-1]
+
\left(\lambda[n]^{*}-\lambda[n]\right)
\frac{
R^{*}[n]S[n]
}{
\left|R[n]\right|^2
+
\left|S[n]\right|^2
+
\left|W[n]\right|^2
},
\qquad \\
\psi^{(2)}[n]
&=
\psi^{(2)}[n-1]
+
\left(\lambda[n]^{*}-\lambda[n]\right)
\frac{
R^{*}[n]W[n]
}{
\left|R[n]\right|^2
+
\left|S[n]\right|^2
+
\left|W[n]\right|^2
},
\end{aligned}
\end{equation}
where
\begin{equation}
\begin{aligned}
\Psi_{n}[n-1]
=
T[n]
T[n-1]
\cdots
T[2]
T[1]
\Psi[n]|_{\lambda=\lambda_{n}},
\end{aligned}
\end{equation}
with
\begin{equation}
\begin{aligned}
T[n]
=
I+
\frac{
\lambda[n]^{*}-\lambda[n]
}{
\lambda-\lambda[n]^{*}
}
\frac{
\Psi_{n}[n-1]
\Psi_{n}^{\dagger}[n-1]}
{
\Psi_{n}^{\dagger}[n-1]
\Psi_{n}[n-1]
},\quad \Psi[n]=T[n]\Psi,
\end{aligned}
\end{equation}
and
\begin{equation}
\begin{aligned}
&R[n]
=
c_{n1}\varphi_{na}
+
c_{n2}\varphi_{nb}
+
c_{n3}\varphi_{nc},\\
&S[n]
=
\psi^{(1)}[n-1]
\left(
c_{n1}\frac{\varphi_{na}}
{\chi_{na}+\beta_1}
+
c_{n2}\frac{\varphi_{nb}}
{\chi_{nb}+\beta_1}
+
c_{n3}\frac{\varphi_{nc}}
{\chi_{nc}+\beta_1}
\right),\\
&W[n]
=
\psi^{(2)}[n-1]
\left(
c_{n1}\frac{\varphi_{na}}
{\chi_{na}+\beta_2}
+
c_{n2}\frac{\varphi_{nb}}
{\chi_{nb}+\beta_2}
+
c_{n3}\frac{\varphi_{nc}}
{\chi_{nc}+\beta_2}
\right).
\end{aligned}
\end{equation}

\begin{remark}
It is known that the vector BSs represent special case of the degenerate vector TWs, when the wavenumber difference between the two plane wave backgrounds is zero, i.e., $\beta_{1}-\beta_{2}=0$, Eqs.~(\ref{DT1}) reduce to BS solutions. Here, we emphasize the vector eigenfunctions $\Psi[k]=(R[k],S[k],W[k])^{\top}$ of BSs are different from those for the breathers (\ref{PSI}), we provide additional expressions as follows:

\begin{equation}
\begin{aligned}
&R[k]'=\varphi'_{ka}+\varphi'_{kb},\\
&S[k]'=\psi_{0}^{(1)}(\frac{\varphi'_{ka}}{\chi_{a}}+\frac{\varphi'_{kb}}{\chi_{b}}+\varphi'_{kc}),\\
&W[k]'=\psi_{0}^{(2)}(\frac{\varphi'_{ka}}{\chi_{a}}+\frac{\varphi'_{kb}}{\chi_{b}}-\varphi'_{kc}),\\
\end{aligned}
\end{equation}
where
\begin{equation}
\begin{aligned}
\varphi'_{ka}&=c_{k1}\exp\{i[\chi_{ka}x+(\Theta'_{0}+\Theta'_{ka})t]\},\\
\varphi'_{kb}&=c_{k2}\exp\{i[\chi_{kb}x+(\Theta'_{0}+\Theta'_{kb})t]\},\\
\varphi'_{kc}&=c_{k3}\exp\{i(\chi_{kc}x+\Theta'_{0}t)\},
\end{aligned}
\end{equation}
with
\begin{equation}
\begin{aligned}
\Theta'_{0}=-2a^{2}-\frac{\lambda^{2}}{4}-24\varepsilon a^{4},\quad
\Theta'_{kl}=-24\varepsilon a^{4}-\varepsilon \chi_{kl}^{4}+(\frac{1}{2}+8\varepsilon a^{2})\chi_{kl}^{2},\quad l=a,b.
\end{aligned}
\end{equation}
Then the higher-order DT solutions of vector BSs are still given by Eqs.~(\ref{DT2}).
\end{remark}

\vspace{3mm}
\section{Vector localized waves and state transitions} \label{eq:sec3}
\vspace{2mm}
In this section, we investigate two representative localized waves, TWs and RWs, focusing on their dynamical evolution. The degenerate and non-degenerate regions are distinguished by the critical relative wavenumber, following the criterion previously established in the Manakov system~\cite{non-AB1,non-AB2,non-AB3,non-AB4,non-KMS1}. We further construct second-order exact solutions through the nonlinear superposition of two TW breathers. Under the fourth-order effect, the state transitions of both fundamental and second-order solutions are analyzed. In particular, the transformed soliton solutions identified in the degenerate regions are not supported in the CH equations~\cite{pan1}.

\subsection{Vector TWs and RWs}
In this subsection, we discuss the dynamical behaviours of the vector degenerate and non-degenerate localized waves including TWs and RWs of Eqs.~(\ref{eq:CLPDE}). Generally, the exact analytical expressions for the fundamental vector TWs, obtained in compact form using a DT scheme, are given by
\begin{equation}\label{eq:GB}
\begin{aligned}
\psi_{TW}^{(j)}[1]=\rho_{j}\psi_{0}^{(j)}\psi_{tw}^{(j)}, \quad j=1,2,
\end{aligned}
\end{equation}
where
\begin{equation}\label{eq:GB1}
\begin{aligned}
\rho_{j}=\sqrt{\frac{(\chi_{b}^{*}+\beta_{j})(\chi_{a}^{*}+\beta_{j})}{(\chi_{b}+\beta_{j})(\chi_{a}+\beta_{j})}},\quad
\psi_{tw}^{(j)}=\frac{\kappa \cosh(\Gamma+i\vartheta_{j})+\varpi\cos(\Omega+i\gamma_{j})}{\kappa \cosh\Gamma+\varpi\cos\Omega},
\end{aligned}
\end{equation}
with
\begin{equation}
\begin{aligned}
\Gamma=\alpha x+\operatorname{Im}(\Delta\delta)t+\Gamma_{0}, \quad
\Omega=\omega x+\operatorname{Re}(\Delta\delta)t+\Omega_{0},
\end{aligned}
\end{equation}
and
\begin{equation}\label{eq:GB3}
\begin{aligned}
&\Delta\delta=\delta_b-\delta_a,\quad \delta_m=-\varepsilon\chi_m^4
+\left(\frac12+4\varepsilon A\right)\chi_m^2
-4\varepsilon\mu_1\chi_m,
\qquad m=a,b,\\
&\Gamma_{0}=\frac{1}{2}\left|\frac{\chi_{bi}}{\chi_{ai}}\right|, \quad
\Omega_{0}=\arg(\frac{1}{\chi_{a}^{*}-\chi_{b}}),\quad
\mu_1=a_1^2\beta_1+a_2^2\beta_2.
\end{aligned}
\end{equation}
Without loss of generality, we set real parameters $\alpha\geq0, \omega\geq0$. The other parameters in Eq.~(\ref{eq:GB}) are
\begin{equation}\label{GB5}
\begin{aligned}
&\kappa=
\frac{1}{\sqrt{\chi_{ai}\chi_{bi}}},
\quad
\varpi=
\frac{2}{\left|\chi_a^*-\chi_b\right|},\\
&\vartheta_{j}=
\frac{1}{2i}
\ln
\left[
\frac{
(\chi_a^*+\beta_j)(\chi_b+\beta_j)
}{
(\chi_a+\beta_j)(\chi_b^*+\beta_j)
}
\right],\quad
\gamma_j=
-\ln
\left|
\frac{\chi_a^*+\beta_j}
{\chi_b+\beta_j}
\right|.
\end{aligned}
\end{equation}
Subscripts $r$ and $i$ denote the real and imaginary parts of the complex parameter $\chi_{a}$, respectively. $\chi_{a}$ and $\chi_{b}$ are eigenvalues associated with the same spectral parameter $\lambda$ and will be discussed in detail below.

It is well known that the spectral parameters $\lambda$ and eigenvalues $\chi$ play crucial roles in DT. We investigate the relationship between them to gain deeper insight into the characteristics of the non-degenerate solutions and the resonant modes. The relation between $\lambda$ and $\chi$ can be derived in following form:
\begin{equation}\label{lamda}
\begin{aligned}
\lambda=\chi-\frac{a_{1}^{2}}{\beta_{1}+\chi}-\frac{a_{2}^{2}}{\beta_{2}+\chi}.
\end{aligned}
\end{equation}
From the mathematical perspective, Eq.~(\ref{lamda}) has three roots $\chi_{a},~\chi_{b},~\chi_{c}$.  There is a relationship between two roots, which is
\begin{equation}
\begin{aligned}
\chi_{b}=\chi_{a}+\omega+i\alpha.
\end{aligned}
\end{equation}
By use of trace of matrix $U$, the third eigenvalue can be expressed as
\begin{equation}
\begin{aligned}
\chi_{c}=\chi_{a}-\omega-i\alpha-\frac{a_{1}^{2}}{\chi_{a}+\beta_{1}}-\frac{a_{2}^{2}}{\chi_{a}+\beta_{2}}.
\end{aligned}
\end{equation}
Then the eigenvalues of the Eqs.~(\ref{eq:CLPDE}) obey the relations as follows
\begin{equation}
\begin{aligned}
1+\sum_{j=1}^{2} \frac{a_{j}^{2}}{(\chi_{a}+\beta_{j})(\chi_{b}+\beta_{j})}=0,
\end{aligned}
\end{equation}
which admits four roots $\chi_{n}~(n=1,2,3,4)$ corresponding to $\chi_{a}$. When $a_{1}=a_{2}=a$ and $\beta_{1}=-\beta_{2}=\beta$~($a$ and $\beta$ are all nonzero real numbers), these roots reduce to
\begin{equation}\label{kai}
\begin{aligned}
&\chi_{1}=-\frac{\omega+i\alpha}{2}+\sqrt{\mu-\sqrt{\nu}}, \quad
\chi_{2}=-\frac{\omega+i\alpha}{2}-\sqrt{\mu-\sqrt{\nu}},\\
&\chi_{3}=-\frac{\omega+i\alpha}{2}+\sqrt{\mu+\sqrt{\nu}}, \quad
\chi_{4}=-\frac{\omega+i\alpha}{2}-\sqrt{\mu+\sqrt{\nu}},\\
\end{aligned}
\end{equation}
with
\begin{equation}
\begin{aligned}
\mu=\beta^{2}-a^{2}+\frac{(\omega+i\alpha)^{2}}{4},\quad
\nu=a^{4}-4a^{2}\beta^{2}+(\omega+i\alpha)^{2}\beta^{2}.
\end{aligned}
\end{equation}

Eqs.~(\ref{kai}) give the eigenvalues associated with the vector TW solutions. In the limit $\beta_{1}-\beta_{2}=0$, they reduce to those of vector BSs, which will be discussed in Sec.~\ref{eq:sec5}. For the Manakov system, the degenerate and non-degenerate regions are separated by a critical relative wavenumber $\beta_{c}^{2}$~\cite{non-AB1}. Since the matrix $U$ in the Lax pairs of Eqs.~(\ref{eq:Manakov}) and (\ref{eq:CLPDE}) are identical, the same criterion $\beta_{c}^{2}=\frac{4a^{4}}{16a^{2}-(\omega+i\alpha)^{2}}$ is applied in Eqs.~(\ref{eq:CLPDE}). Accordingly, the TW solutions are the degenerate for $\beta^{2}\leq\beta_{c}^{2}$ and the non-degenerate for $\beta^{2}>\beta_{c}^{2}$. Their existence regions can therefore be determined from the eigenvalues. According to the sign of $\nu$, the eigenvalues are divided into two groups, $\chi_{1},~\chi_{2}$ and $\chi_{3},~\chi_{4}$. The solutions are classified as the degenerate if the corresponding eigenvalues generate identical wave structures, otherwise, they are the non-degenerate. This criterion differs from that in Ref.~\cite{CLPD5}, where degeneracy is determined by unequal wavenumbers $\beta_{1}\neq\beta_{2}$ and instead emphasizes the underlying eigenvalues. The resulting classification is summarized in Table~\ref{tab:region}.

\begin{table*}[htbp]
    \centering
    \renewcommand{\arraystretch}{1.25}

    \captionsetup{font=small}
    \begin{scriptsize}
    \caption{Determination of the degenerate and non-degenerate regions.}
    \label{tab:region}
    \begin{tabular*}{\textwidth}
        {@{\extracolsep{\fill}} c c c c c}
        \hline
        \hline

        \makecell{Regions}
        &
        \makecell{Critical relative wavenumber}
        &
        \makecell{Eigenvalues}
        &
        \makecell{Wave modes}
        &
        \makecell{Degree of degeneracy}
        \\

        \hline

        Degenerate      & $\beta^{2}\leq\beta_{c}^{2}$ & $\nu~(\eta)\geq0$ & $\psi^{(j)}[1](\chi_{1})=\psi^{(j)}[1](\chi_{2})$ & 1 \\
        Non-degenerate    & $\beta^{2}>\beta_{c}^{2}$   & $\nu~(\eta)<0$ & $\psi^{(j)}[1](\chi_{1})\neq\psi^{(j)}[1](\chi_{2})$ & 2 \\

        \hline
        \hline
    \end{tabular*}
    \end{scriptsize}
\end{table*}

\begin{figure}[H]
	\centering
{\includegraphics[width=400 bp,height=4 cm]{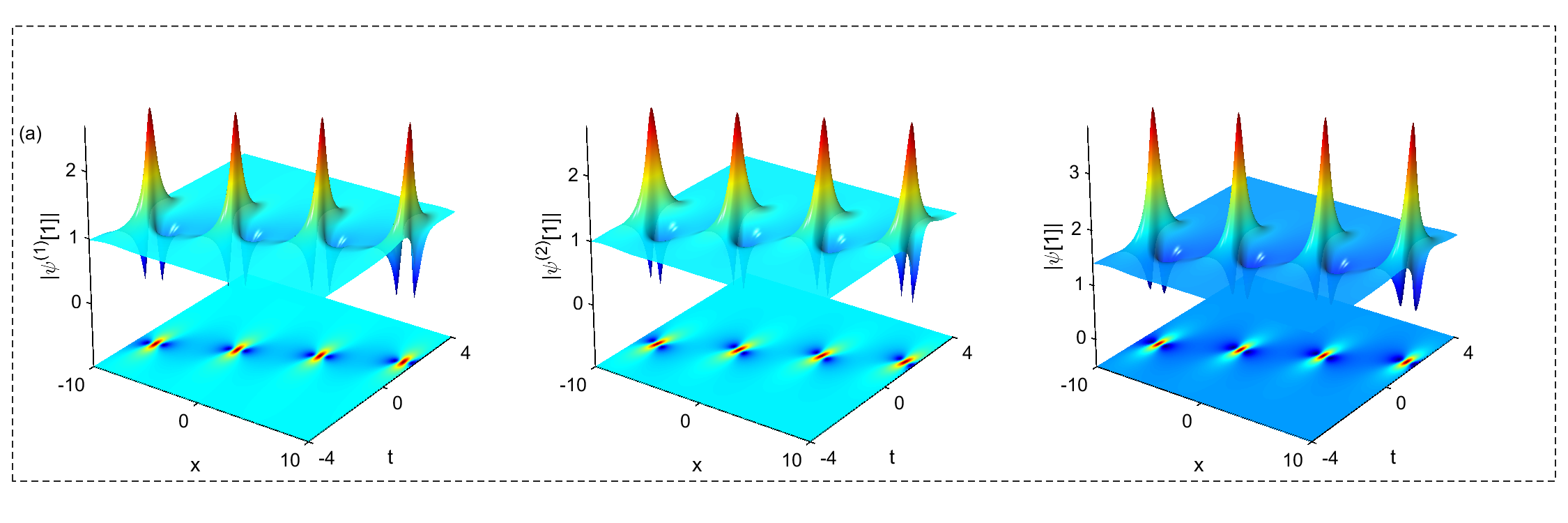}}
{\includegraphics[width=400 bp,height=8 cm]{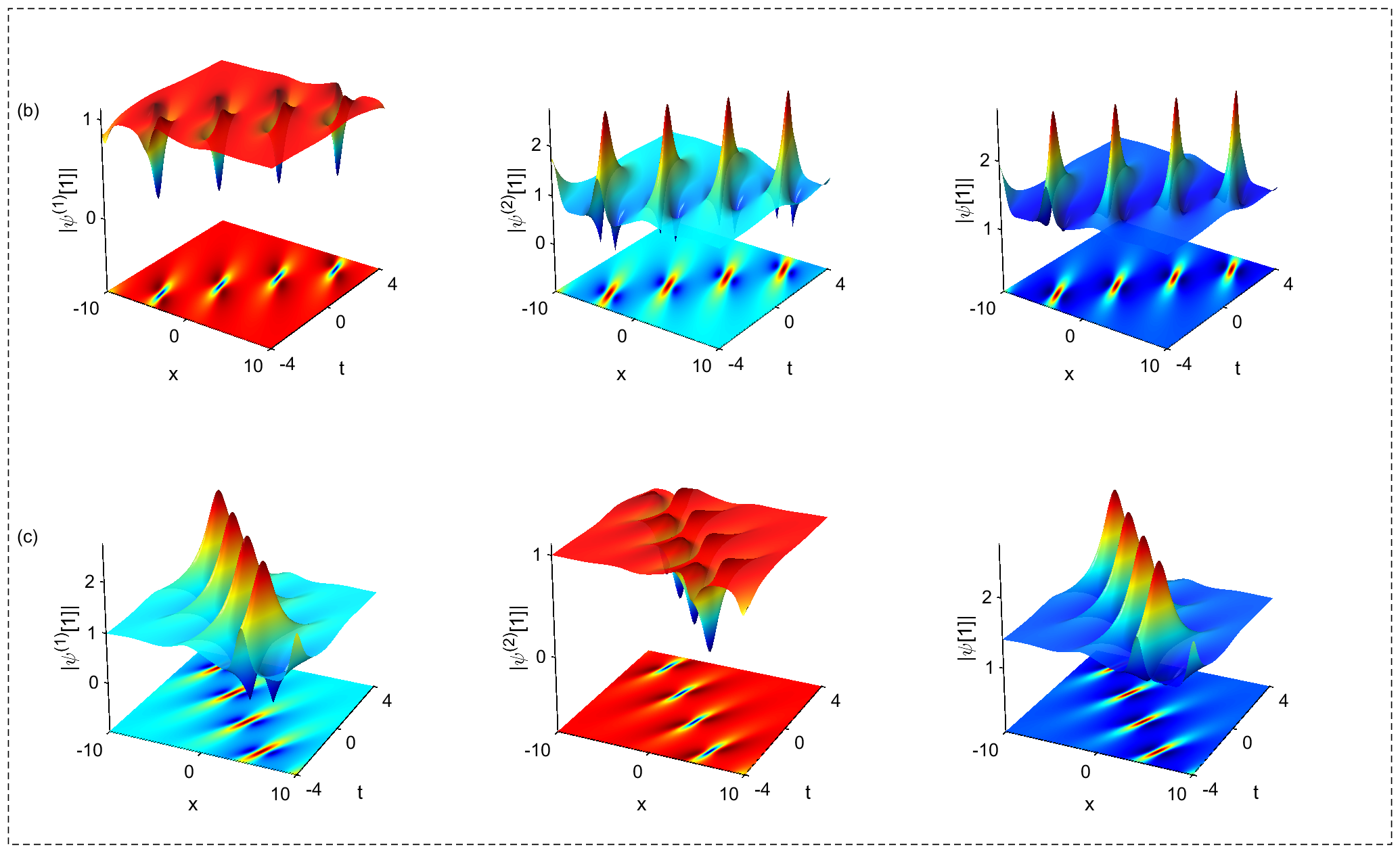}}
\caption{
\footnotesize
\linespread{1.2}\selectfont
The amplitude profiles $|\psi^{(1)}[1]|,~|\psi^{(2)}[1]|$ and $|\psi[1]|$ of vector TW solutions in the degenerate case: (a) $\chi_{1}~(\chi_{2}), \beta=0.3$ and the non-degenerate case: (b) $\chi_{1}$, (c) $\chi_{2}$ with $\beta=1$. Here, the total amplitude is $|\psi[1]|=\sqrt{|\psi^{(1)}[1]|^{2}+|\psi^{(2)}[1]|^{2}}$. Other parameters are $a=1$, $\omega=1$, $\alpha=0.3$, $\varepsilon=0.01$ and $c_{11}=c_{12}=1$.}
	\label{GB}
\end{figure}

Physically, the solutions (\ref{eq:GB}) describe the propagation of two TW breathers on their respective plane wave backgrounds (\ref{eq:seed solution}). From Eqs.~(\ref{eq:GB1})-(\ref{GB5}), we find vector TW solutions (\ref{eq:GB}) are controlled by the amplitudes $a_{j}$, wavenumbers $\beta_{j}$, real parameters $\alpha,~\omega$ and fourth-order effect coefficient $\varepsilon$. Among all the parameters, the relative wavenumber $\beta_{1}-\beta_{2}$ is significant since it cannot be eliminated through Galilean transformation~\cite{non-AB2,non-KMS1}. Indeed, when $\beta_{1}=\beta_{2}$, Eqs.~(\ref{eq:CLPDE}) become decoupled, while Eq.~(\ref{eq:GB}) reduce to the TW solutions of the scalar LPD equation. For simplicity, we can set $\beta_{1}=-\beta_{2}=\beta,~a_{1}=a_{2}=a$. When we take into account only two eigenvalues $\chi_{a}$ and $\chi_{b}$, there are the degenerate and non-degenerate TW solutions in Figs.~\ref{GB}, which depend on four parameters $a,~\beta,~\omega,~\alpha$.

As shown in Figs.~\ref{GB}, distinct eigenvalues give rise to rich dynamical behaviors. From Eqs.~(\ref{kai}), when $\beta^{2}\leq\beta_{c}^{2}$,  equivalently $\nu\geq0$, the eigenvalues satisfy $\chi_{1}=\chi_{2}^{*}$ and $\chi_{3}=\chi_{4}^{*}$. Considering the pair $\chi_{1}$ and $\chi_{2}$, the corresponding solutions exhibit identical bright-bright profiles, differing only by a $90^{\circ}$ shift in propagation direction, such that $\psi^{(j)}[1](\chi_{1})=\psi^{(j)}[1](\chi_{2})$. These solutions are therefore classified as the degenerate TWs, as shown in Figs.~\ref{GB}~(a). A distinctive feature of the degenerate TWs is the absence of dark components. When $\beta^{2}>\beta_{c}^{2}~(\nu<0)$, the eigenvalues satisfy $\chi_{1}=\chi_{3}^{*}$ and $\chi_{2}=\chi_{4}^{*}$, forming the pairs $(\chi_{1},\chi_{3})$ and $(\chi_{2},\chi_{4})$, the corresponding wave profiles are reversed in both structure and propagation direction. Figs.~\ref{GB}~(b) show the dark-bright structure, whereas Figs.~\ref{GB}~(c) exhibit the complementary bright-dark structure. These solutions are therefore classified as the non-degenerate TWs with a degeneracy degree 2.

Under the fourth-order effect, the TW solutions exhibit distinctive dynamical behaviors. In the following, we analyze the underlying mechanism and derive the corresponding conditions. The propagation speeds of the hyperbolic function $\Gamma$ and the trigonometric function $\Omega$ are $v_{\Gamma}=-\frac{\!Im(\Delta\delta)}{\alpha}$ and $v_{\Omega}=-\frac{Re(\Delta\delta)}{\omega}$, respectively. When $v_{\Gamma}=v_{\Omega}$, the breathers can be transformed into solitons. We derive the state transition condition as follows:
\begin{equation}\label{GB_ST1}
\begin{aligned}
\operatorname{Im}\!\Lambda=0,
\end{aligned}
\end{equation}
with $\Lambda=\frac{\Delta\delta}{\omega+i\alpha}$. Then the fourth-order effect parameter in Eq.~(\ref{GB_ST1}) are explicitly given by
\begin{equation}\label{GB_ST2}
\begin{aligned}
\varepsilon_{tr}=-\frac{\operatorname{Im}\!(\chi_{a}+\chi_{b})}
{2\operatorname{Im}\![4A(\chi_{a}+\chi_{b})-\chi_{a}^{3}\chi_{a}^{2}\chi_{b}+\chi_{a}\chi_{b}^{2}\chi_{b}^{3}]}.
\end{aligned}
\end{equation}
And velocity of the transformed soliton are
\begin{equation}\label{GB_ST3}
\begin{aligned}
V_{tr}=-\Lambda.
\end{aligned}
\end{equation}

\begin{figure}[H]
	\centering
{\includegraphics[width=400 bp,height=4 cm]{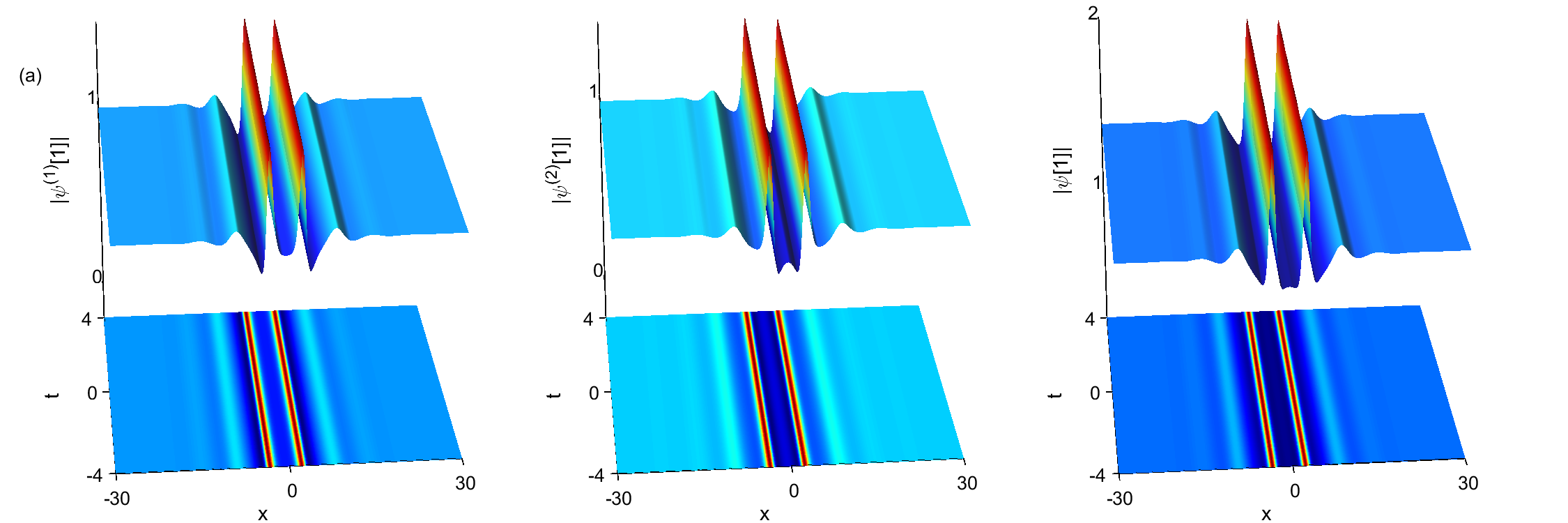}}
{\includegraphics[width=400 bp,height=4 cm]{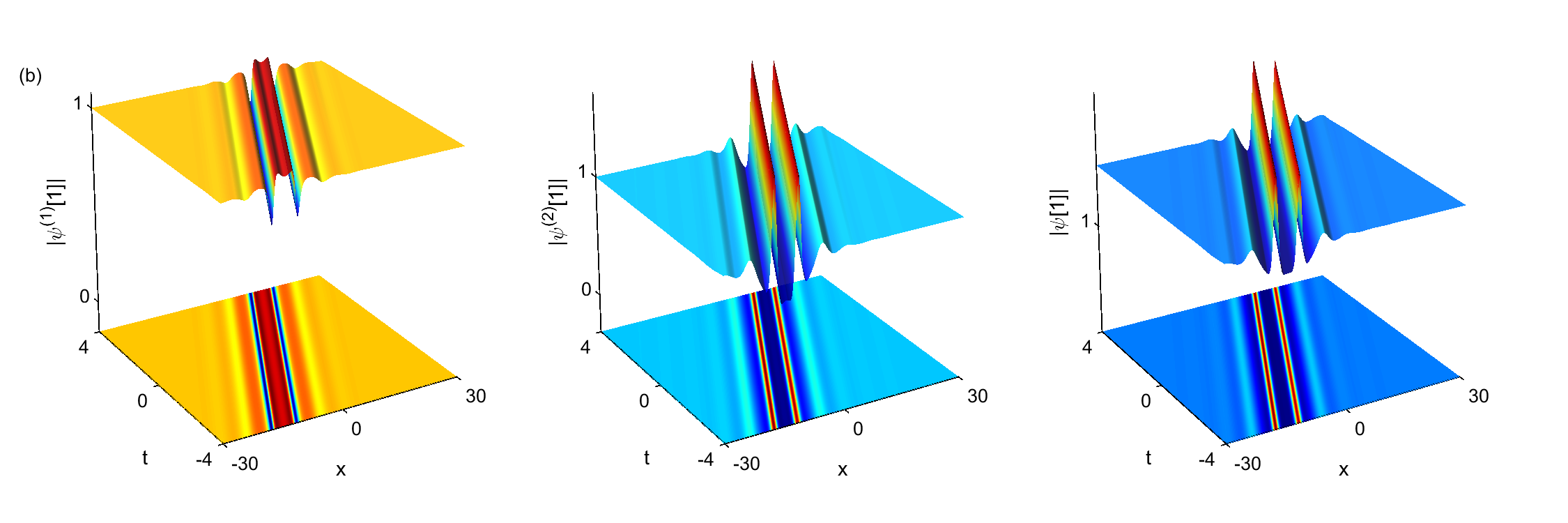}}
\caption{\footnotesize The amplitude profiles $|\psi^{(1)}[1]|,~|\psi^{(2)}[1]|$ and $|\psi[1]|$ of the multipeak solitons from vector TWs breathers in (a) the degenerate case: $\beta=0.3, \varepsilon_{tr}=-0.0472$ and (b) the non-degenerate case: $\beta=1, \varepsilon_{tr}=-0.2396$. Other parameters are the same as those in Figs.~\ref{GB}. }
	\label{abtsGB}
\end{figure}

Such state transition dynamics are absent in the Manakov system due to its intrinsic model constraints. By contrast, Figs.~\ref{abtsGB} show that Eqs.~\ref{eq:CLPDE} support transformations in both the degenerate and non-degenerate regions, whereas the corresponding degenerate solitons are not found in the CH equations~\cite{pan1}. Specifically, the degenerate TWs evolve into the bright-bright solitons, while the non-degenerate TWs are transformed into the dark-bright solitons. This difference highlights the richer conversion behaviors induced by the fourth-order effect compared with the third-order case. Since TWs generally possess nonzero group velocities, the resulting solitons are typically moving. Under the transformed condition, however, the velocity in the degenerate region approaches zero $V_{tr}\approx0$, yielding almost static solitons, whereas the non-degenerate case retain the finite propagation velocity $V_{tr}\neq0$. The velocities of the transformed solitons from the degenerate and non-degenerate TWs are governed by the parameter $\varepsilon_{tr}$, resulting in distinct dynamical behaviors.

We consider the limits $\alpha\rightarrow0$ and $\omega\rightarrow0$, under which the vector TW solutions reduce to the vector RW solutions that can be expressed in rational form
\begin{equation}\label{eq:RW}
\begin{aligned}
\psi_{RW}^{(j)}[1]=\psi_{0}^{(j)}[1+(\lambda^{*}-\lambda)\frac{\xi_{j}(M_{j}+N_{j})}{\zeta_{j}(M_{j}+N_{j})+H_{j}}], \quad j=1,2,
\end{aligned}
\end{equation}
where
\begin{equation}
\begin{aligned}
&M_j=
\left[
\chi_{0i} x+\mathcal{G}_j t+1
\right]^2,\quad
N_j=
\left[
(\chi_{0r}+\beta_j)x+\Lambda_j t
\right]^2,\\
&H_j=
\varsigma_i^2t^2+(x+\varsigma_rt)^2
+
\sum_{\substack{\ell=1\\ \ell\neq j}}^{2}
\frac{a_\ell^2(M_\ell+N_\ell)}
{\left[(\chi_r+\beta_\ell)^2+\chi_i^2\right]^2},
\end{aligned}
\end{equation}
and
\begin{equation}
\begin{aligned}
&\mathcal{G}_j=(\chi_r+\beta_j)\varsigma_i+\chi_i\varsigma_r, \quad
\quad
\Lambda_j=(\chi_r+\beta_j)\varsigma_r-\chi_i\varsigma_i,\\
&\zeta_j=
\frac{a_j^2}
{\left[(\chi_r+\beta_j)^2+\chi_i^2\right]^2},\quad
\xi_j=
\frac{
\varsigma_i t+i(x+\varsigma_rt)
}{
\left(\chi_r+\beta_j+i\chi_i\right)^2
\left[
\chi_i x+\mathcal{G}_jt+1
+i\left((\chi_r+\beta_j)x+\Lambda_jt\right)
\right]
}.
\end{aligned}
\end{equation}

\begin{figure}[H]
	\centering
{\includegraphics[width=400 bp,height=4 cm]{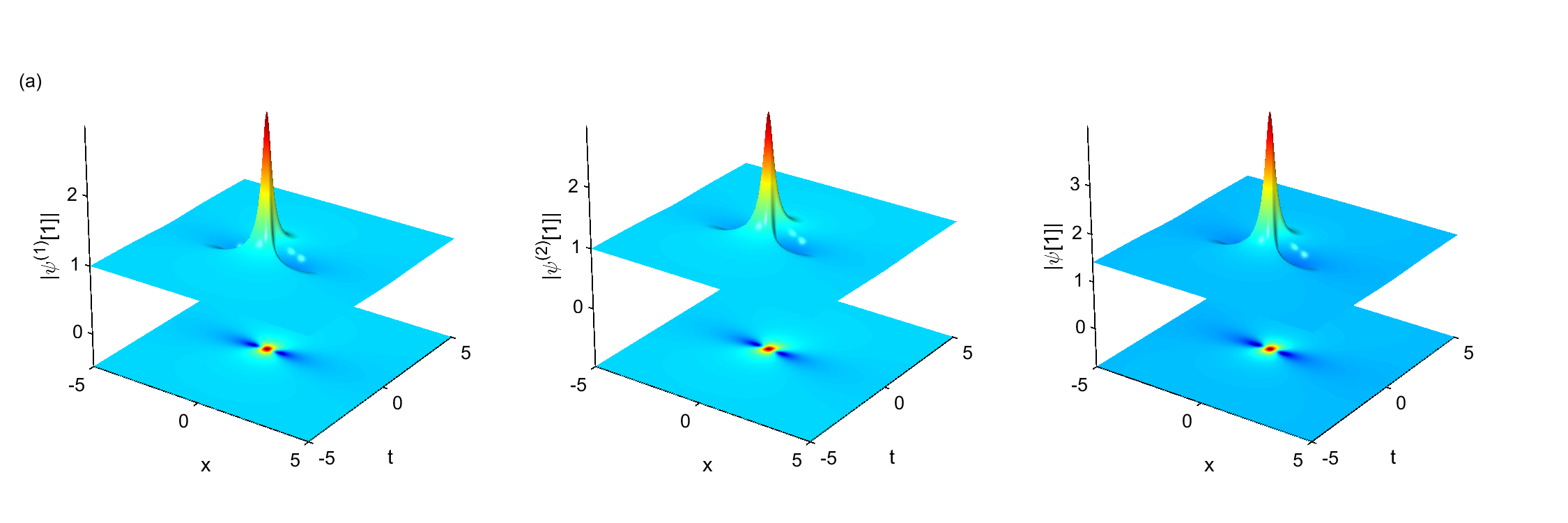}}
{\includegraphics[width=400 bp,height=4 cm]{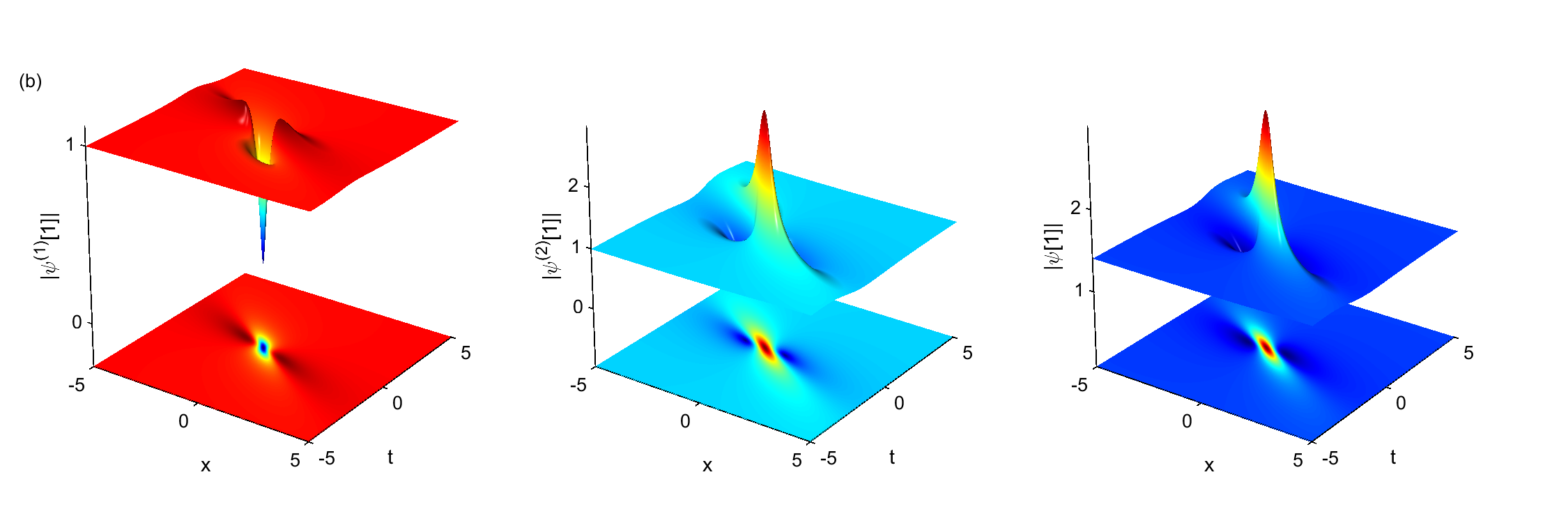}}
\caption{%
\footnotesize
\linespread{1.2}\selectfont
The amplitude profiles $|\psi^{(1)}[1]|,~|\psi^{(2)}[1]|$ and $|\psi[1]|$ of vector RWs in (a) the degenerate case: $\beta=0.1$ and (b) the non-degenerate case: $\beta=1$. Other parameters are $a=1, \varepsilon=0.01, c_{11}= c_{12}=1$.
}
	\label{RW}
\end{figure}

Vector RWs are obtained as the breathers limit at a spectral branch point, where $\chi_{0}$ follows from Eqs.~(\ref{kai}) by setting $\omega=\alpha=0$. The criterion of the degenerate and non-degenerate RWs follows Table~\ref{tab:region}. The degenerate and non-degenerate RWs can be classified according to the eigenvalues and are readily distinguished by the wave profiles, as shown in Figs.~\ref{RW}. Defining $\zeta=\beta^{2}-a^{2}$ and $\eta=a^{2}-4\beta^{2}$, the branch points are given by $\chi_{0}=\pm\sqrt{\zeta\pm a\sqrt{\eta}}$. The corresponding critical relative wavenumber is $\beta_{c0}^{2}=\frac{a^{2}}{4}$. For $\eta\geq0$, equivalently $\beta^{2}\leq\beta_{c0}^{2}$, the solutions are the degenerate RWs (bright-bright), otherwise, they are the non-degenerate RWs (dark-bright). The similar degenerate RWs are reported in~\cite{CLPD2}, but their definition is fundamentally different from that adopted in the present work.

In particular, we focus on the state transition of vector RWs. We have
\begin{equation}
\begin{aligned}
\mathcal{D}(\chi_{0})=F_{2}\chi_{0}^{2}+F_{1}\chi_{0},
\end{aligned}
\end{equation}
with
\begin{equation}
\begin{aligned}
F_{2}=\frac{1}{2}+\varepsilon(6a^{2}-\beta^{2}-\lambda_{0}^{2}),\quad
F_{1}=-2\varepsilon a^{2}\lambda_{0},
\end{aligned}
\end{equation}
where $\lambda_{0}$ is the spectral parameter associated with the corresponding eigenvalue $\chi_{0}$. The characteristic line $\mathcal{D}$ coefficient is given by
\begin{equation}
\begin{aligned}
\rho_{0}=\chi_{0}+2\varepsilon[(6a^{2}-\beta^{2}-\lambda_{0}^{2})\chi_{0}-a^{2}\lambda_{0}].
\end{aligned}
\end{equation}
Thus, the state transition condition is
\begin{equation} \label{RWTR}
\begin{aligned}
\operatorname{Im}\!\rho_{0}=0.
\end{aligned}
\end{equation}
The fourth-order parameter governing the RW state transition is given by
\begin{equation} \label{RW-ST}
\begin{aligned}
\varepsilon_{s}=-\frac{\operatorname{Im}\!(\chi_{0})}{2\operatorname{Im}\![(6a^{2}-\beta^{2}-\lambda^{2})\chi_{0}-2a^{2}\lambda_{0}]}.
\end{aligned}
\end{equation}
The velocity of the transformed soliton can be derived exactly,
\begin{equation}
\begin{aligned}
V_{s}=-Re(\rho_{s}), \quad \rho_{s}=\rho_{0}|_{\varepsilon=\varepsilon_{s}}.
\end{aligned}
\end{equation}

\begin{figure}[H]
	\centering
{\includegraphics[width=400 bp,height=4 cm]{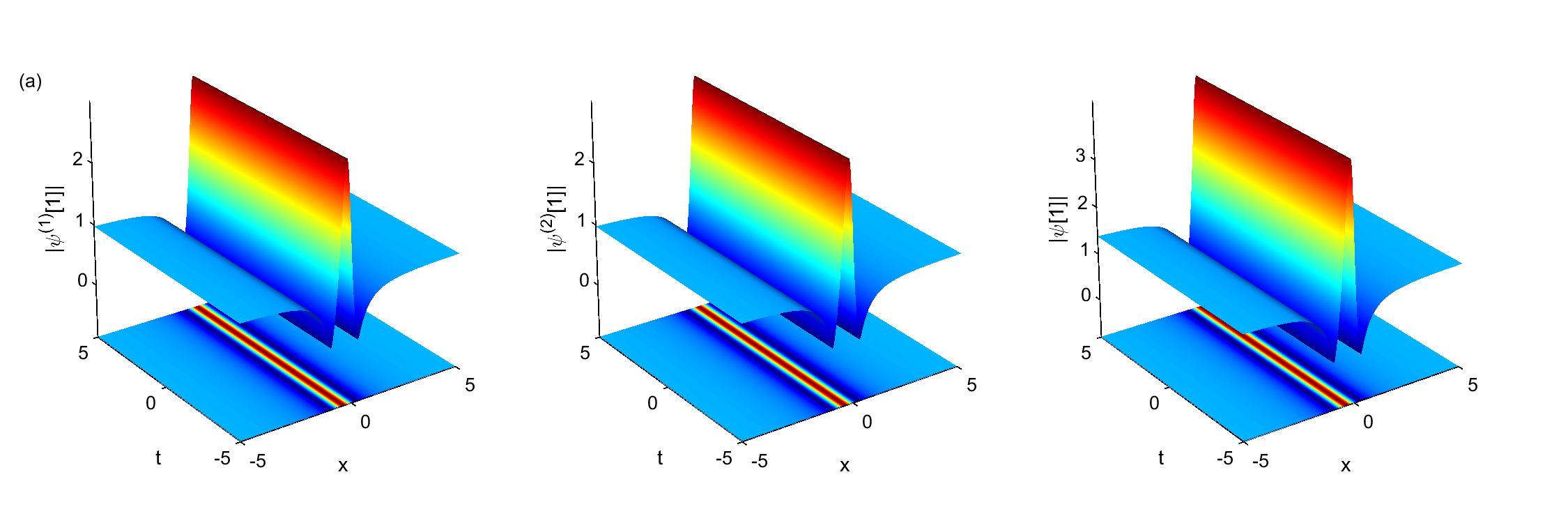}}
{\includegraphics[width=400 bp,height=4 cm]{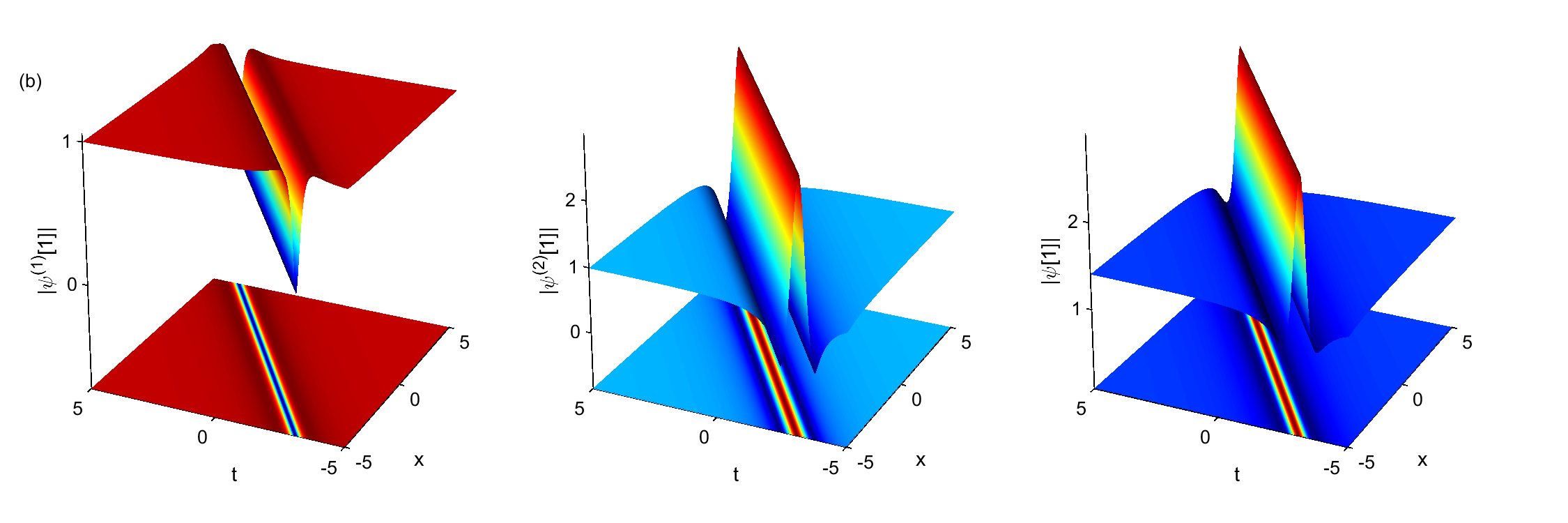}}
\caption{\footnotesize The amplitude profiles $|\psi^{(1)}[1]|,~|\psi^{(2)}[1]|$ and $|\psi[1]|$ of the rational solitons from the vector RWs in (a) the degenerate case: $\beta=0.1, \varepsilon_{c}=-0.0419$ and (b) the non-degenerate case: $ \beta=1, \varepsilon_{s}=-0.1102$. Here, $\varepsilon_{c}$ corresponding to the static transformed soliton described by Eq.~(\ref{RW-ST}) when $\beta=0.1$. Other parameters are the same as those in Figs.~\ref{RW}.}
	\label{abts1}
\end{figure}

As illustrated in Figs.~\ref{abts1}, the degenerate and non-degenerate transformed solitons exhibit distinct structures and propagation velocities. The degenerate case yields static bright-bright solitons, whereas the non-degenerate case produces moving dark-bright solitons. For the degenerate RWs, the relevant eigenvalue is purely imaginary, causing the real part of the characteristic line parameter to vanish and hence the propagation velocity to be zero. In comparison, the nonzero real part of the spectral branch point in the non-degenerate case leads to the finite velocity. This difference highlights the distinct state transition dynamics from the degenerate and non-degenerate RWs induced by the fourth-order effect. As depicted in Figs.~\ref{abts1}, however, state transitions occur in both the degenerate and non-degenerate regions for the CLPD equations, whereas the degenerate transformed solutions are absent in the CH equation~\cite{pan1}.

\subsection{Interaction}
Similar to the scalar LPD equation, higher-order vector breather solutions for Eqs.~\ref{eq:CLPDE} can be constructed through iterative DT~\cite{Lou2024}. The second-order solutions arise from the nonlinear superposition of two fundamental breathers.

\begin{figure}[H]
	\centering
{\includegraphics[width=400 bp,height=3 cm]{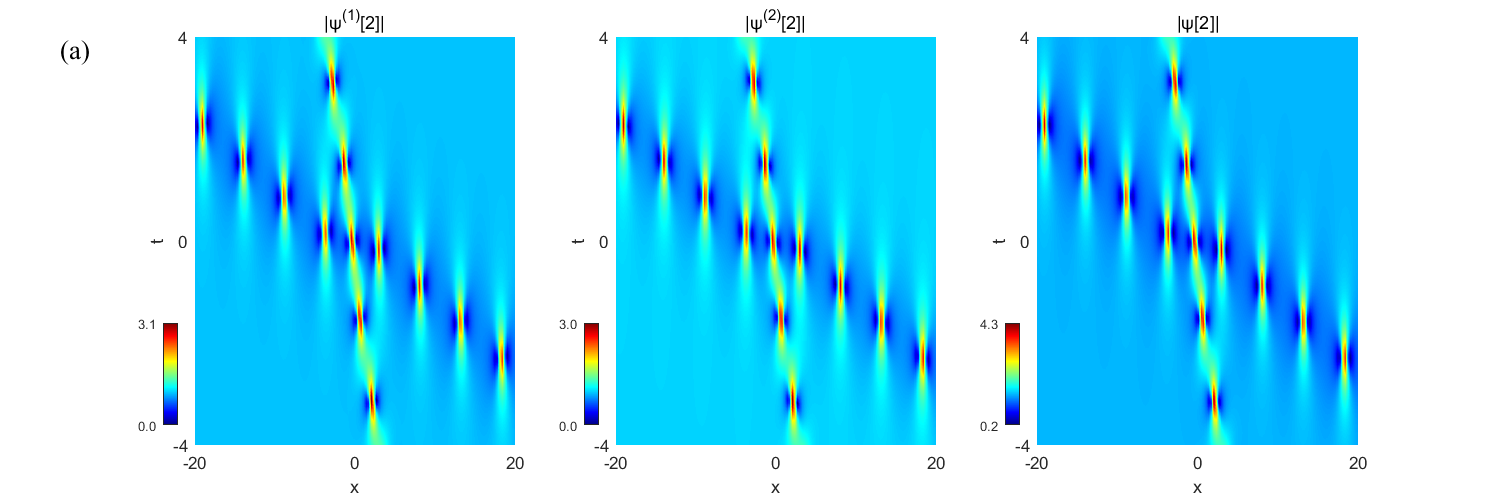}}
{\includegraphics[width=400 bp,height=3 cm]{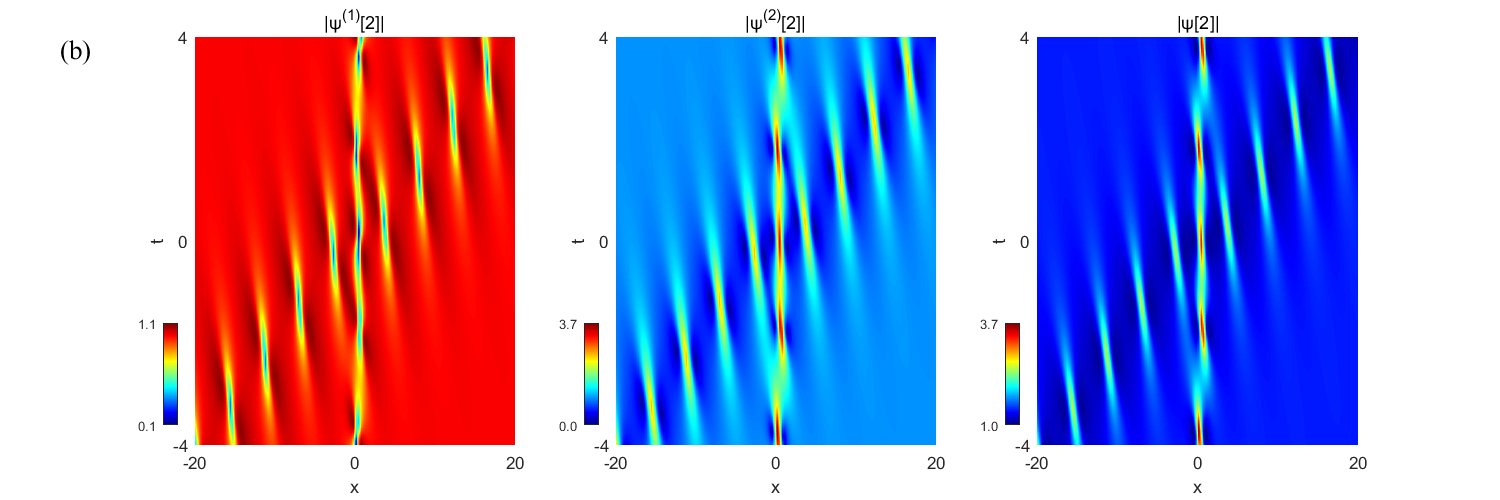}}
\caption{\footnotesize Amplitude profiles $|\psi^{(1)}[2]|,~|\psi^{(2)}[2]|$ and $|\psi[2]|$ of the second-order solutions formed by the nonlinear superposition of two TWs in (a) the degenerate case: $\beta=0.1$ and (b) the non-degenerate case: $\beta=1$. Here, the total amplitude is $|\psi[2]|=\sqrt{|\psi^{(1)}[2]|^{2}+|\psi^{(2)}[2]|^{2}}$. Other parameters are $a=1, \varepsilon=0.01,\omega_{1}=1, \omega_{2}=1.2, \alpha_{1}=1, \alpha_{2}=0.1, c_{11}=c_{12}=c_{21}=c_{22}=1$.}
	\label{erjiejie}
\end{figure}

As shown in Figs.~\ref{erjiejie}, the two components of the degenerate second-order solutions exhibit identical X-shaped interaction patterns formed by bright-bright TW breathers, with no dark breather structures present. In comparison, the non-degenerate second-order solutions are formed by dark-bright breathers, leading to distinct interaction patterns in two components and consequently richer dynamics. These differences provide a clear signature for distinguishing the non-degenerate second-order solutions from their degenerate counterparts.

\begin{figure}[H]
	\centering
{\includegraphics[width=400 bp,height=3 cm]{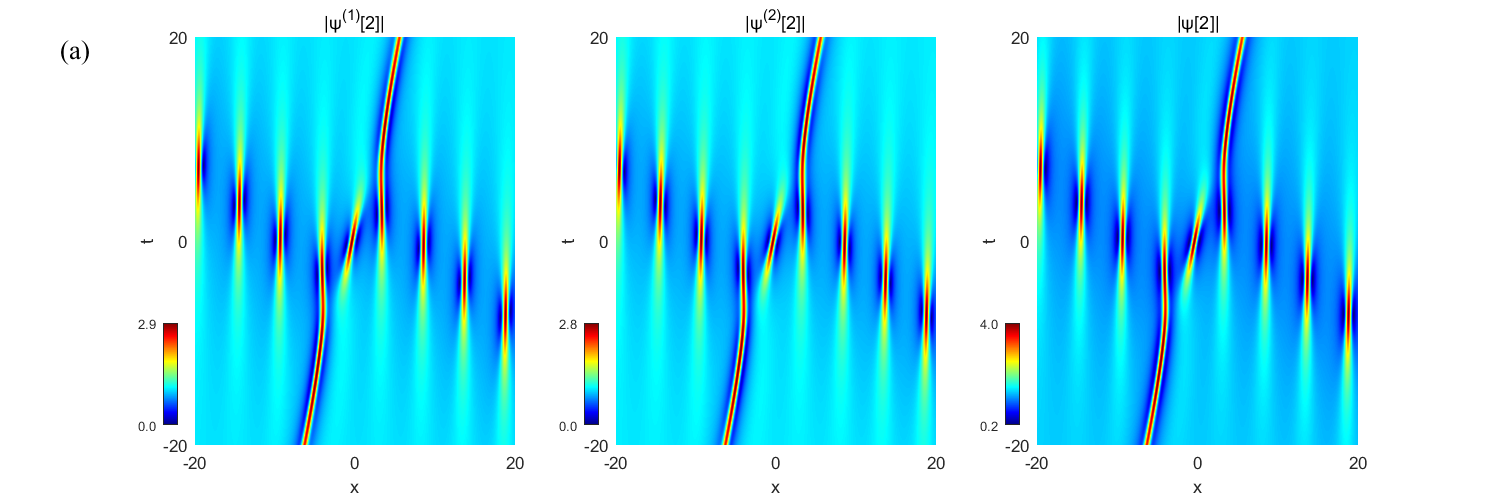}}
{\includegraphics[width=400 bp,height=3 cm]{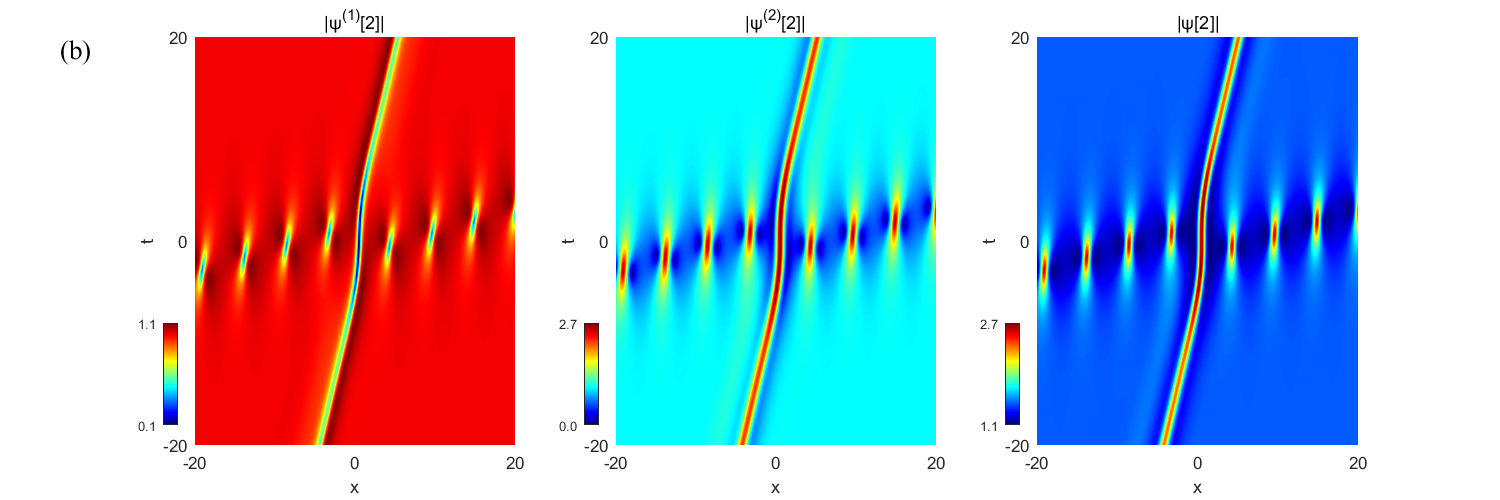}}
\caption{\footnotesize Amplitude profiles $|\psi^{(1)}[2]|,~|\psi^{(2)}[2]|$ and $|\psi[2]|$ of the second-order solutions formed by the nonlinear superposition of a transformed solitons and a TW breather in (a) the degenerate case: $\varepsilon_{tr}=-0.0427$ and (b) the non-degenerate case: $\varepsilon_{tr}=-0.0624$. Other parameters are the same as those in Figs.~\ref{erjiejie}.}
	\label{erjiejie1}
\end{figure}

The state transition condition for each fundamental TW breather is given by Eq.~(\ref{GB_ST2}). When two TW breathers constitute a second-order solution with different parameters, their transition conditions generally differ. Accordingly, two TW breathers cannot be simultaneously transformed into solitons under the same fourth-order parameter. The transition conditions for two TWs are given by Eq.~(\ref{GB_ST1}), from which the corresponding fourth-order parameters can be calculated by Eq.~(\ref{GB_ST2}) directly. Representative cases for both the degenerate and non-degenerate second-order solutions are displayed in Figs.~\ref{erjiejie1}, where one of two breathers is transformed into multipeak solitons.

\section{Resonance modes} \label{eq:sec4}
In this section, we discuss the dynamic features of the resonant modes and their state transition. We particularly emphasize that the number of fundamental solutions (\ref{phi}) influences the structure of the localized waves. The standard localized wave solutions only require two fundamental solutions corresponding to eigenvalues $\chi_{a}$ and $\chi_{b}$. When the third eigenvalue $\chi_{c}$ is introduced, the resulting solutions exhibit the resonant modes, which will be studied in detail below.

\subsection{Resonance and transition patterns}
We known that the solutions (\ref{DT1}) depend on the background amplitudes $a_{1}, a_{2}$, wavenumbers $\beta_{1}, \beta_{2}$, the two real constants $(\omega,\alpha)$, the fourth-order effect coefficient $\varepsilon$ and the mode coefficients $(c_{11},c_{12},c_{13})$. We set $a_{1}=a_{2}=a=1$, since solutions with arbitrary $a$ can be obtained through the scaling transformation. However, parameters $(\omega,\alpha)$, $(c_{11},c_{12},c_{13})$ and $\chi_{a}$ play key roles in the formation of the wave modes. Typically, single breather or BS only exist when one of the mode coefficients $(c_{11},c_{12},c_{13})$ is zero. In particular, when three coefficients $(c_{11},c_{12},c_{13})$ are non-zero, the breather solutions exist the resonance patterns.

\begin{table*}[htbp]
    \centering
    \renewcommand{\arraystretch}{1.25}

    \captionsetup{font=small}
    \begin{scriptsize}
    \caption{Branch Analysis of the Resonant Solutions.}
    \label{tab:III}
    \begin{tabular*}{\textwidth}
        {@{\extracolsep{\fill}} c c c c c}
        \hline
        \hline

        \makecell{Branches}
        &
        \makecell{Analytical expression}
        &
        \makecell{Coefficient}
        &
        \makecell{Eigenvalues}
        &
        \makecell{State transition condition}
        \\

        \hline

        type-I     & $\psi_{TW}^{(j)}$ & $(1,1,0)$ & $\chi_{a}$ and $\chi_{b}$ & $\operatorname{Im}\!\Lambda=0$ \\
        type-II    & $\psi_{II}^{(j)}$   & $(1,0,1)$ & $\chi_{a}$ and $\chi_{c}$ & $\operatorname{Im}\!\Xi=0$   \\
        type-III    & $\psi_{III}^{(j)}$   & $(0,1,1)$ & $\chi_{b}$ and $\chi_{c}$ & $\operatorname{Im}\!\Upsilon=0$   \\

        \hline
        \hline
    \end{tabular*}
    \end{scriptsize}
\end{table*}

Generally, we consider $\omega\neq0, \alpha\neq0$, the resonance modes can be divided into three branches using the coefficients $(c_{11},c_{12},c_{13})$, i.e., branch-I with $(c_{11},c_{12},c_{13})=(1,1,0)$, branch-II with $(c_{11},c_{12},c_{13})=(1,0,1)$ and branch-III with $(c_{11},c_{12},c_{13})=(0,1,1)$. Obviously, the expression of the first branch and the corresponding state transition condition are given by Eqs.~(\ref{eq:GB}) and (\ref{GB_ST2}). We summarize the formation mechanisms of three types of breathers in Table \ref{tab:III}. Next, we will sequentially present the expressions for the exact solutions of other two branchs in the resonance modes. Here, we give out the expression of the second branch in the resonant modes.
\begin{equation}\label{GBreII}
\begin{aligned}
\psi_{II}^{(j)} = \rho'_j\psi_0^{(j)} \frac{\mathcal{H}\cosh\left(\Gamma_{1}+i\sigma_j\right) + \mathcal{P}\cos\left(\Omega_{1}+i\gamma'_j\right) }{\mathcal{H}\cosh\Gamma_{1} + \mathcal{P}\cos\Omega_{1}}, \qquad j=1,2,
\end{aligned}
\end{equation}
with
\begin{equation}
\begin{aligned}
&\Gamma_{1} = -\alpha x - \operatorname{Im}\left(\Delta\delta\right)t + \frac{1}{2}\ln\left(\frac{G_c}{G_a}\right),\quad
\Omega_{1}  = \omega x + \operatorname{Re}\left(\Delta\delta\right)t + \arg(\mathcal{H}),\\
&\mathcal{H} = 1+ \frac{a_1^2} {\left(\chi_a^*+\beta_1\right) \left(\chi_c+\beta_1\right)} + \frac{a_2^2} {\left(\chi_a^*+\beta_2\right) \left(\chi_c+\beta_2\right)},\quad \mathcal{P}=|\mathcal{H}|.
\end{aligned}
\end{equation}
and
\begin{equation}
\begin{aligned}
&G_m = 1+ \frac{a_1^2}{\left|\chi_m+\beta_1\right|^2} + \frac{a_2^2}{\left|\chi_m+\beta_2\right|^2}, \quad \Delta\delta=\delta_{c}-\delta_{a},\\
&\delta_n = -\varepsilon\chi_n^4 + \left( \frac{1}{2}+4\varepsilon A \right)\chi_n^2 - 4\varepsilon\mu_1\chi_n - \frac{\lambda^2}{4} - A - 6\varepsilon A^2 + 4\varepsilon\mu_2, \quad n=a,c,\\
&\sigma_j = \frac{1}{2i} \ln \left[ \frac{ \left(\chi^{*}_{c}+\beta_j\right) \left(\chi_a+\beta_j\right) }{ \left(\chi_c+\beta_j\right) \left(\chi_a^*+\beta_j\right) } \right], \quad
\rho'_{j}=\sqrt{\frac{(\chi_{c}^{*}+\beta_{j})(\chi_{a}^{*}+\beta_{j})}{(\chi_{c}+\beta_{j})(\chi_{a}+\beta_{j})}},\quad
\gamma'_j = \ln \left| \frac{\chi_c+\beta_j} {\chi_a+\beta_j} \right|.
\end{aligned}
\end{equation}

The temporal oscillation of breathers occurs because $\Gamma_{1}$ and $\Omega_{1}$ generally propagate with different velocities. The breather solutions to be transformed into solitons, the velocities of the two characteristic lines $\Gamma_{1}$ and $\Omega_{1}$ must be equal, which yields the following transition condition:
\begin{equation}\label{TWII_ST}
\begin{aligned}
\operatorname{Im}\!\Xi=0,
\end{aligned}
\end{equation}
with $\Xi=\frac{\Delta \delta}{\omega+i\alpha}$.
Then the fourth-order effect parameter in Eq.~(\ref{TWII_ST}) are explicitly given by
\begin{equation}\label{TW_II}
\begin{aligned}
\varepsilon_{tr1}=-\frac{\operatorname{Im}\!(\chi_{a}+\chi_{c})}
{2\operatorname{Im}\![4A(\chi_{a}+\chi_{c})-\chi_{a}^{3}\chi_{a}^{2}\chi_{c}+\chi_{a}\chi_{c}^{2}\chi_{c}^{3}]}.
\end{aligned}
\end{equation}

Next, the third branch in the resonant modes is expressed by
\begin{equation}\label{GBreIII}
\begin{aligned}
\psi_{III}^{(j)} = \widehat{\rho}_j\psi_0^{(j)} \frac{ \mathcal{G}\cosh\left(\Gamma_{2}+i\varrho_j\right) + \mathcal{O}\cos\left(\Omega_{2}+i\tau_j\right) }{ \mathcal{G}\cosh\Gamma_{2} + \mathcal{O}\cos\Omega_{2}}, \qquad j=1,2,
\end{aligned}
\end{equation}
with
\begin{equation}
\begin{aligned}
&\Gamma_{2}
=\alpha x+
\operatorname{Im}\!\left(\Delta\delta\right)t
+
\frac{1}{2}\ln\!\left(\frac{M_b}{M_c}\right),\quad
\Omega_{2}=-\omega x-
\operatorname{Re}\!\left(\Delta\delta\right)t
+\arg N,\\
&\mathcal{G}=\sqrt{M_bM_c},
\quad
\mathcal{O}=|N|.
\end{aligned}
\end{equation}
and
\begin{equation}
\begin{aligned}
&\widehat{\rho}_j = \frac{1}{2i} \ln \left[ \frac{ \left(\chi_c^*+\beta_j\right) \left(\chi_b+\beta_j\right) }{ \left(\chi_c+\beta_j\right) \left(\chi_a^*+\beta_j\right) } \right], \quad
\tau_j = \ln \left| \frac{\chi_c+\beta_j} {\chi_b+\beta_j} \right|,\quad
\varrho_j = \frac{1}{2i} \ln \left[ \frac{ \left(\chi^{*}_{c}+\beta_j\right) \left(\chi_a+\beta_j\right) }{ \left(\chi_c+\beta_j\right) \left(\chi_a^*+\beta_j\right) } \right], \\
&\delta_r=
-\varepsilon\chi_r^4+
\left(\frac{1}{2}+4\varepsilon A\right)\chi_r^2
-4\varepsilon\mu_1\chi_r
-\frac{\lambda^2}{4}-A-
6\varepsilon A^2+4\varepsilon\mu_2,\quad \Delta\delta=\delta_{c}-\delta_{b},\\
&M_r=
1+\sum_{\ell=1}^{2}
\frac{a_\ell^2}
{\left|\chi_r+\beta_\ell\right|^2},
\quad
N=1+
\sum_{\ell=1}^{2}
\frac{a_\ell^2}
{\left(\chi_b+\beta_\ell\right)
 \left(\chi_c^*+\beta_\ell\right)},\quad r=b,c.
\end{aligned}
\end{equation}
We find that the spatial and temporal oscillations of the third branch are controlled by $\Gamma_{2}$ and $\Omega_{2}$. Thus, the state transition condition is given by
\begin{equation}\label{TWIII_ST}
\begin{aligned}
\operatorname{Im}\!\Upsilon=0,
\end{aligned}
\end{equation}
with $\Upsilon=\frac{\Delta \delta}{\omega+i\alpha}$. Then the fourth-order effect parameter in Eq.~(\ref{TWIII_ST}) are explicitly given by
\begin{equation}\label{TW_III}
\begin{aligned}
\varepsilon_{tr2}=-\frac{\operatorname{Im}\!(\chi_{b}+\chi_{c})}
{2\operatorname{Im}\![4A(\chi_{b}+\chi_{c})-\chi_{b}^{3}\chi_{b}^{2}\chi_{c}+\chi_{b}\chi_{c}^{2}\chi_{c}^{3}]}.
\end{aligned}
\end{equation}

\begin{figure}[H]
	\centering
{\includegraphics[width=400 bp,height=3 cm]{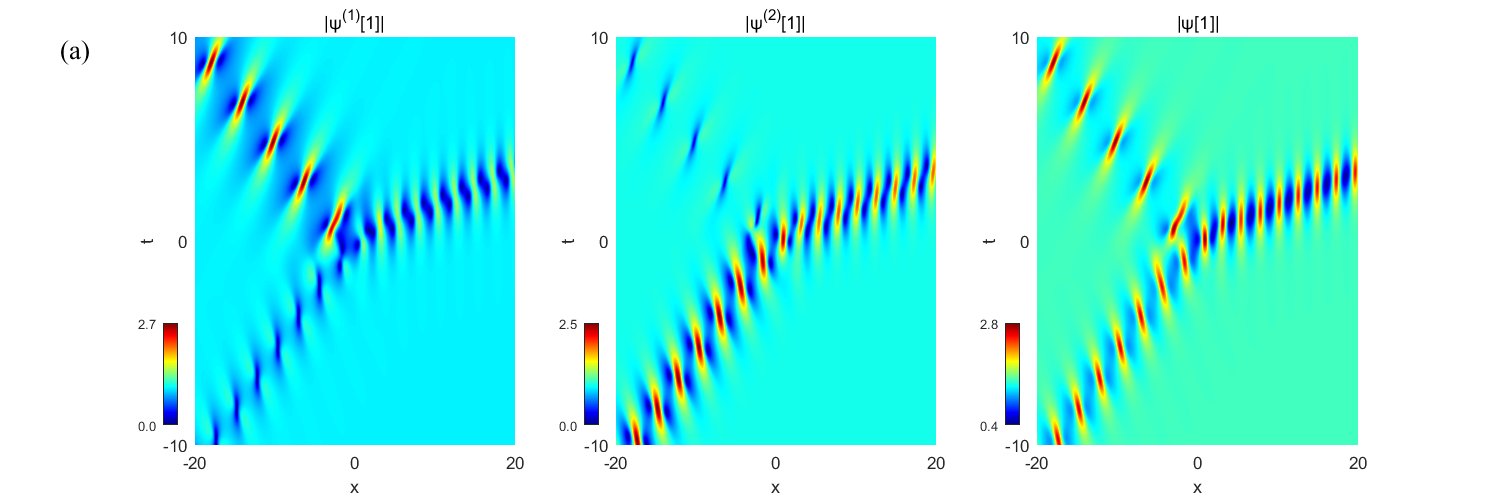}}
{\includegraphics[width=400 bp,height=3 cm]{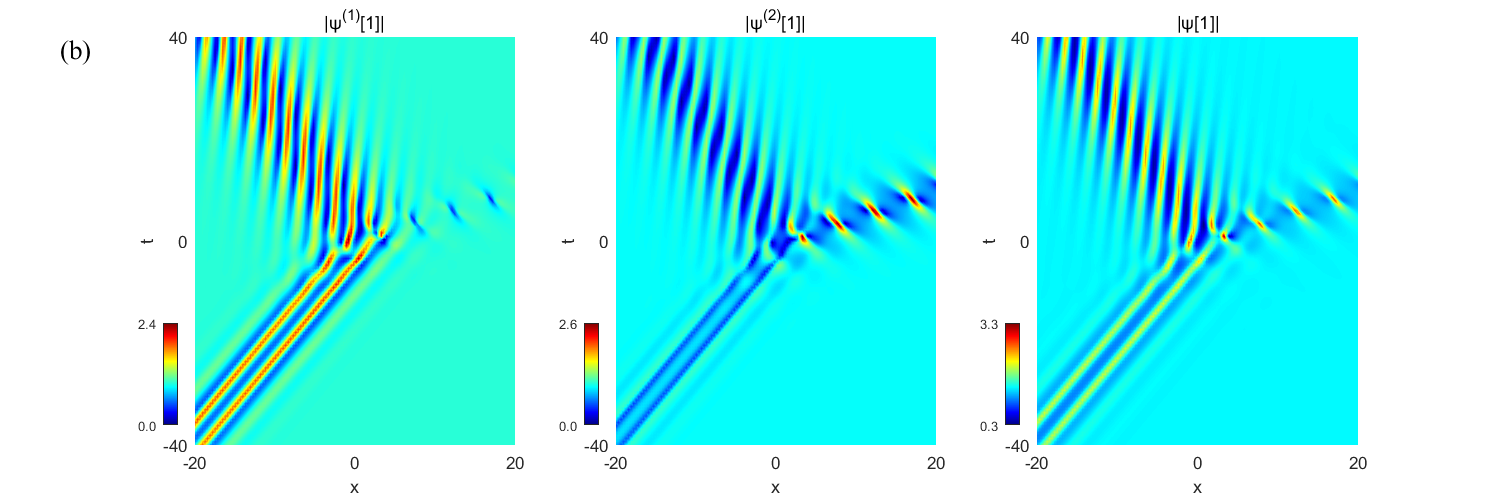}}
\caption{\footnotesize Amplitude profiles $|\psi^{(1)}[1]|, |\psi^{(2)}[1]|$ and $|\psi[1]|$ of the TW resonance modes. Panels (a) show the fusion between two TWs when $\varepsilon=0.01$. Panels (b) show the fusion between a TW and the multipeak soliton from TW branch in the resonance modes when $\varepsilon=-0.4384$. Other parameters are $a=1, \beta=1, \alpha=0.1, \omega=1, c_{11}=c_{12}=c_{13}=1$.}
    \label{GB_resonant}
\end{figure}

Based on the above analysis, we conclude that three branches in the resonant modes cannot be simultaneously transformed into solitons because each branch requires different transition parameters. The state transitions of the degenerate resonant modes are similar to that of the corresponding individual solutions discussed above and are therefore not repeated here. We instead focus on the state transitions of the non-degenerate resonant modes. As shown in Figs.~\ref{GB_resonant}~(a), we observe that TW resonant modes have three breather branches expressed by Eqs.~(\ref{eq:GB}), (\ref{GBreII}) and (\ref{GBreIII}). The resonant modes belong to the non-degenerate solutions, accordingly, the structures in the two components are visually distinct. In the case of Eqs.~(\ref{GB_ST2}), we can see one branch in the resonant modes turns to be solitons in Figs.~\ref{GB_resonant}~(b).

\begin{figure}[H]
	\centering
{\includegraphics[width=400 bp,height=3 cm]{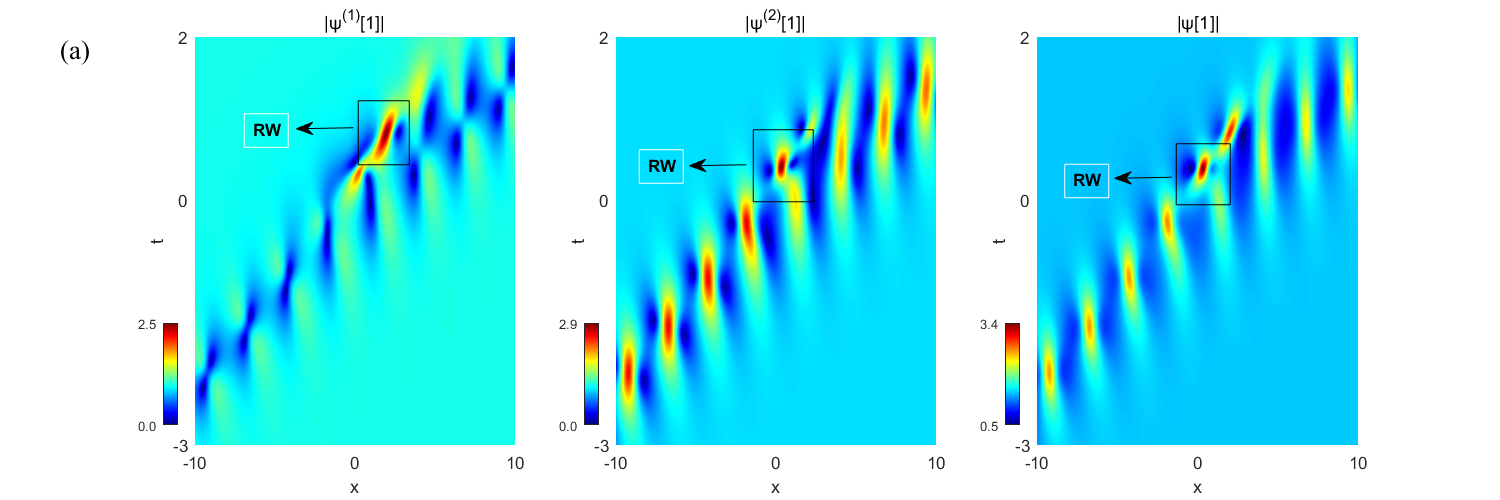}}
{\includegraphics[width=400 bp,height=3 cm]{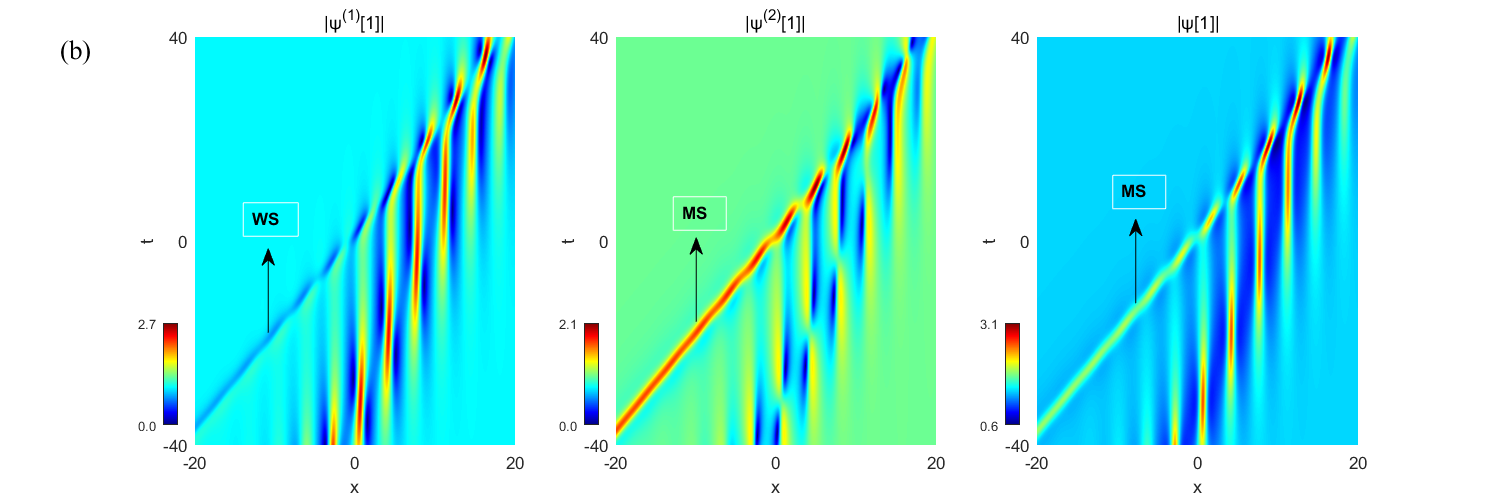}}
\caption{\footnotesize Amplitude profiles $|\psi^{(1)}[1]|, |\psi^{(2)}[1]|$ and $|\psi[1]|$ of the RW resonance modes. Panels (a) show the fusion between a TW and a RW when $\varepsilon=0.01$. Panels (b) show the fusion between a TW and the rational solitons from RW branch in the resonance modes when $\varepsilon=-0.0731$. The short forms `WS' and `MS' represent W-type and M-type solitons, respectively. Other parameters are $a=1, \beta=1, c_{11}=c_{12}=c_{13}=1$.}
    \label{RW_resonant}
\end{figure}

We then explore the resonant modes of vector RWs, which consist of two breather branches and a RW branch. The analytical expressions of all three branches can be derived in the same manner. The RW branch is given by Eq.~(\ref{eq:RW}), while two breather branches follow Eqs.~(\ref{GBreII}) and (\ref{GBreIII}) in the limit $\omega=\alpha=0$ and are therefore omitted for brevity. We primarily focus on RW branch in the resonant modes and corresponding state transition, as plotted in Figs.~\ref{RW_resonant}.

\subsection{Resonance interaction}
\begin{figure}[H]
	\centering
{\includegraphics[width=400 bp,height=3 cm]{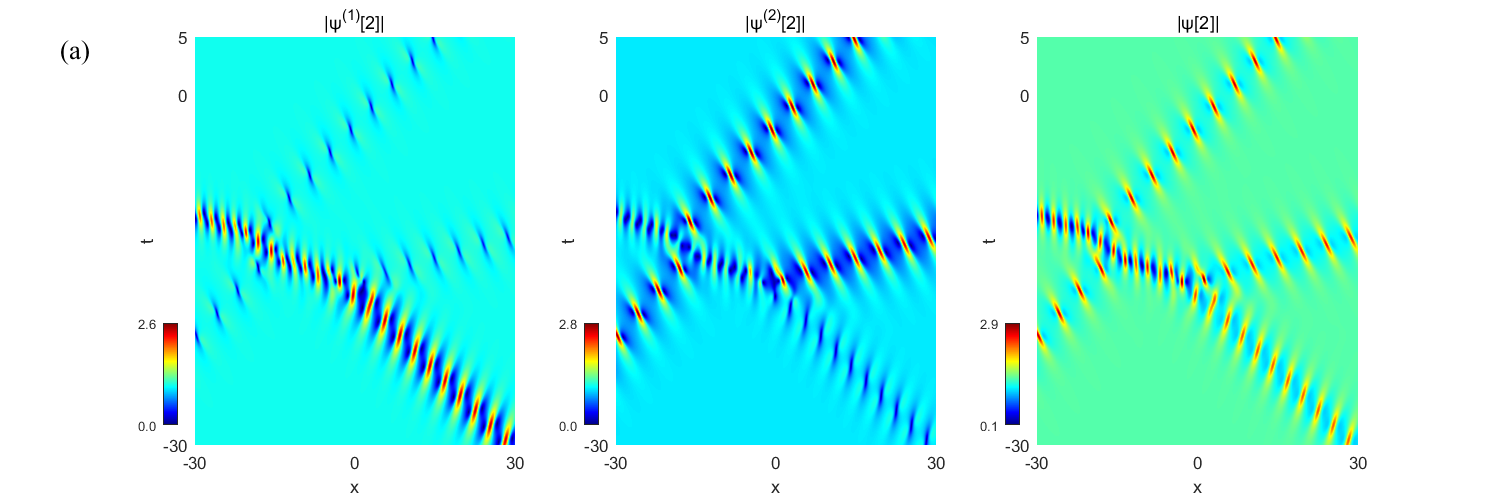}}
{\includegraphics[width=400 bp,height=3 cm]{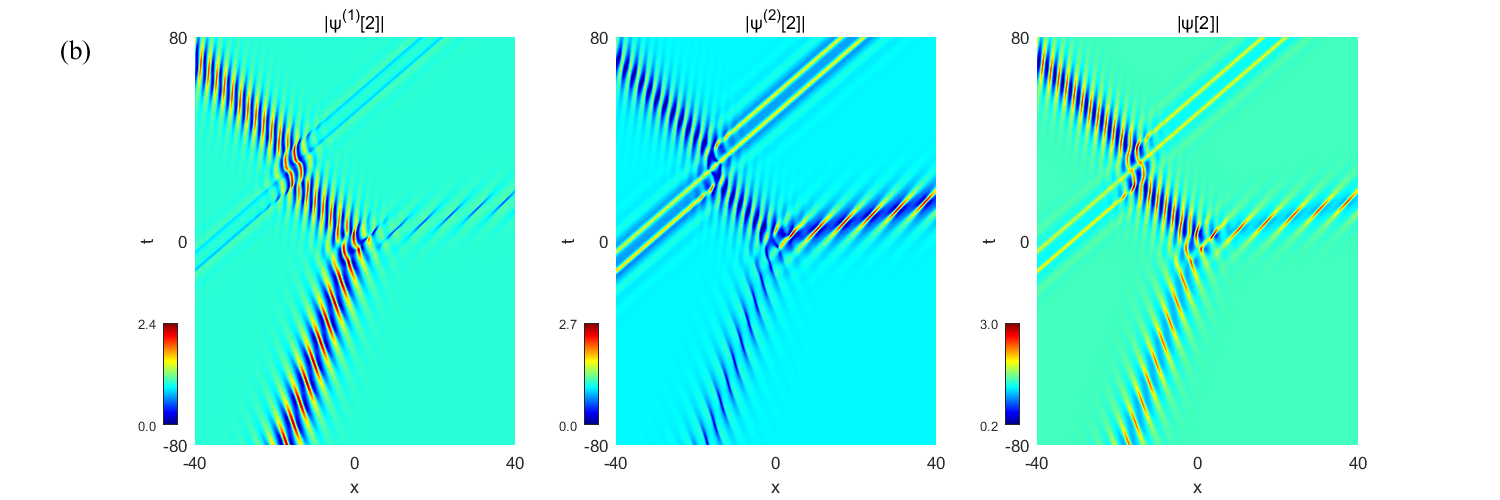}}
\caption{\footnotesize Amplitude profiles $|\psi^{(1)}[2]|, |\psi^{(2)}[2]|$ and $|\psi[2]|$ of the second-order solutions. Panels (a) show the interaction between one TW breather and the resonance modes when $\varepsilon=0.01$. Panels (b) show the interaction between one transformed soliton from TW breather and the resonance modes when $\varepsilon_{tr}=-0.0973$. Other parameters are $\beta=1, \alpha_{1}=\alpha_{2}=0.3, \omega_{1}=\omega_{2}=1, c_{11}=c_{12}=c_{21}=c_{22}=c_{23}=1, c_{13}=0$.}
	\label{rb1}
\end{figure}

In this subsection, we discuss the interaction between the resonant modes and a general TW breather, together with their state transitions. Their nonlinear superposition yields the second-order solution shown in Figs.~\ref{rb1}. When TW breather collides with one branch in the resonant modes, the collision induces positional shifts among them while preserving their individual wave profiles.

\section{BS solutions and their interaction} \label{eq:sec5}
BSs are a special class of vector degenerate TW solutions ($\beta_{1}-\beta_{2}=0$) and can be derived from the dark-bright soliton solutions via the SU(2) rotation symmetry of the Manakov system~\cite{beating1}. Vector BSs are essentially breathers with complementary internal beating structures between two components, while its total intensity exhibits soliton-like profile. Such solutions are confined to the degenerate regions, because the formation requires the wavenumber difference between the two wave components to vanish. Beating dark-dark solitons have been experimentally observed in two-component Bose-Einstein condensates~\cite{BS1}, while analogous oscillatory energy transfer has been predicted in coupled two-color optical fields~\cite{BS2}. In this section, we present vector BSs and their resonant modes under the fourth-order effect. In addition, we study the nonlinear superpositions of BSs, exhibiting conventional X-shaped collision and short-lived structure through parameter variation. Because the fourth-order effect can induce state transitions, we further investigate the transition dynamics of the above solutions in detail.

\subsection{Fundamental BSs and state transitions}
In this subsection, vector BSs are classified into Type-I and Type-II according to eigenvalues and wave component modes. The two types of BSs differ in localization width, propagation velocity, internal beating frequency and peak-to-valley contrast. Nevertheless, they share the same underlying formation mechanism, i.e., Type-I is excited by $\chi_a$, while Type-II are associated with $\chi_b$. The classification of vector BSs is summarized in Table \ref{tab:BS}.

\begin{table*}[htbp]
    \centering
    \renewcommand{\arraystretch}{1.25}
    \captionsetup{font=small}
    \begin{scriptsize}
    \caption{Classification of BSs.}
    \label{tab:BS}
    \begin{tabular*}{\textwidth}
        {@{\extracolsep{\fill}} c c c c}
        \hline
        \hline

        \makecell{Classification}
        &
        \makecell{Coefficient}
        &
        \makecell{Eigenvalue}
        &
        \makecell{Wave component modes}
        \\

        \hline

        Type-I
        & $(1,0,1)$
        & $\chi_{k}=\chi_{a1}~(\chi_{b2})$
        & $\chi_{a}$ and zero modes
        \\

        Type-II
        & $(0,1,1)$
        & $\chi_{k}=\chi_{a2}~(\chi_{b1})$
        & $\chi_{b}$ and zero modes
        \\

        \hline
        \hline
    \end{tabular*}
    \end{scriptsize}
\end{table*}

Two types of BSs can be presented in a unified form:
\begin{equation}\label{BS}
\begin{aligned}
\psi^{(j)}_{BS}=\psi_{0}^{(j)}(\psi_{DS}\pm\psi_{BS}), \quad j=1,2,
\end{aligned}
\end{equation}
where $\psi_{DS},\psi_{BS}$ are the dark and bright solitons of Eqs.~(\ref{eq:CLPDE}), respectively. They are given by
\begin{equation}
\begin{aligned}
\psi_{DS}&=1-\frac{2ia\lambda_{i}e^{-\chi_{ki}(x+\chi_{kr}t)}}{N_{k}\cosh{[\chi_{ki}(x+\chi_{kr}t)+\theta_{k}]}},\quad
\psi_{BS}=\frac{-2ia\lambda_{i}e^{-i\mathcal{C}}}{N_{k}\cosh{[\chi_{ki}(x+\chi_{kr}t)+\theta_{k}]}}.
\end{aligned}
\end{equation}
Subscripts $r$ and $i$ denote the real and imaginary parts of the corresponding parameters. The coefficients $N_{k}$ and $\theta_{k}$ in these expressions are
\begin{equation}
\begin{aligned}
&\mathcal{C}=\chi_{kr}x+\delta_{kr}t+(a^{2}+24\varepsilon a^{4})t+\frac{\lambda^{2}}{4},\\
&N_{k}=2a\sqrt{2(|\chi_{k}|^{2}+2a^{2})}, \quad
\theta_{k}=\frac{1}{2}ln\frac{2a^{2}}{|\chi_{k}|^{2}+2a^{2}},\quad k=a,b.
\end{aligned}
\end{equation}
To provide a clearer analysis of the vector BS solutions, we reformulate their eigenvalues as follow:
\begin{equation}\label{BSvalue1}
\begin{aligned}
&\chi_{a1}=\frac{1}{2}(-\omega-i\alpha-\sqrt{2i\omega\alpha-\alpha^{2}+\omega^{2}-8a^{2}}),\\
&\chi_{a2}=\frac{1}{2}(-\omega-i\alpha+\sqrt{2i\omega\alpha-\alpha^{2}+\omega^{2}-8a^{2}}).
\end{aligned}
\end{equation}

\begin{figure}[H]
	\centering
{\includegraphics[width=400 bp,height=3 cm]{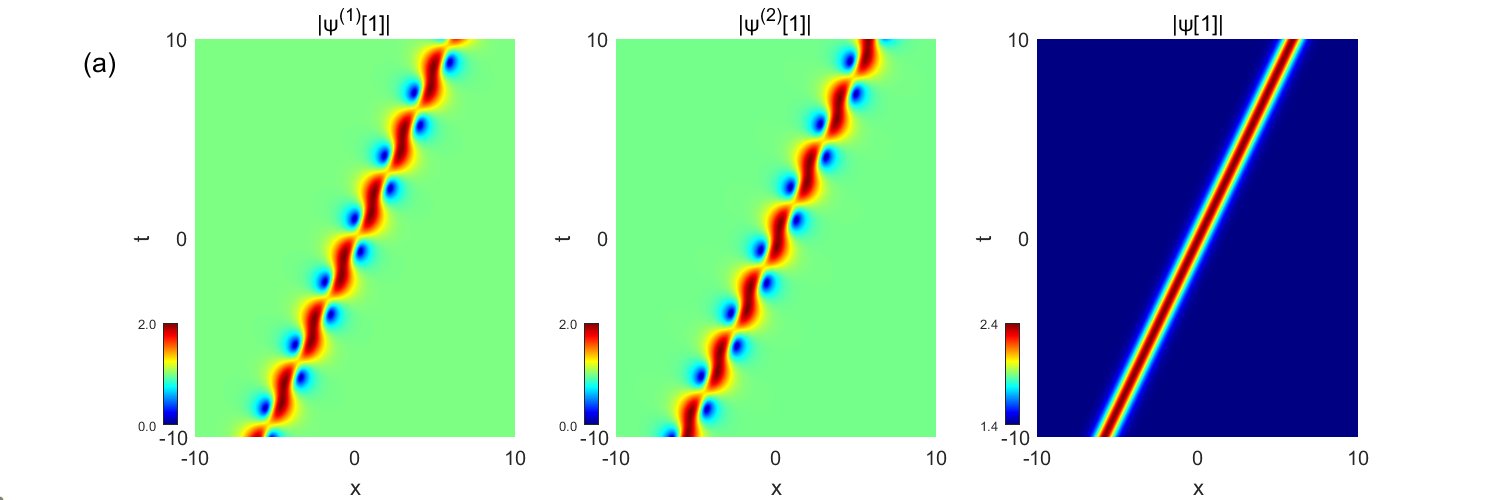}}
{\includegraphics[width=400 bp,height=3 cm]{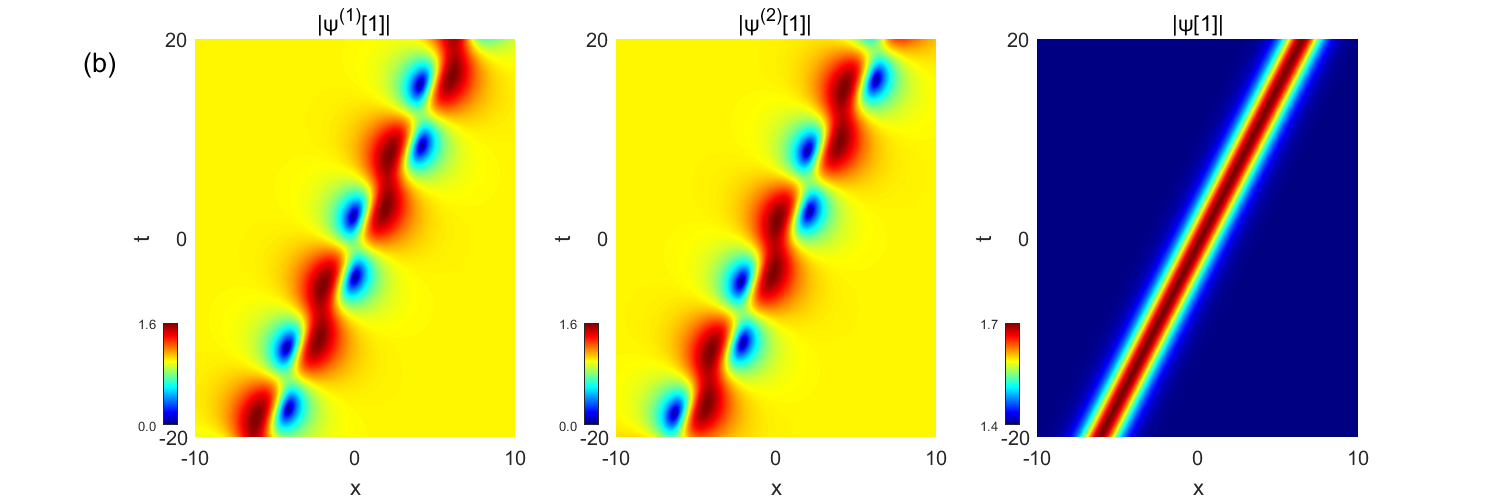}}
\caption{\footnotesize Amplitude profiles of (a) Type-I and (b) Type-II moving BSs. The coefficients in (a)-(b) are (1,0,1) and (0,1,1), respectively. Parameters are: $a=1, \varepsilon=0.01, \alpha=1$ and $\omega=1$.  The three panels from left to right show $|\psi^{(1)}[1]|$, $|\psi^{(2)}[1]|$ and the total amplitude $|\psi[1]|=\sqrt{|\psi^{(1)}[1]|^{2}+|\psi^{(2)}[1]|^{2}}$, respectively.}
	\label{beating1}
\end{figure}

We have $\chi_{a1}=-\chi_{b2}$ and $\chi_{a2}=-\chi_{b1}$. The parameter $\chi_k$ corresponds to the same spectral parameter $\lambda(\chi_{k})=\chi_{k}-\frac{2a^{2}}{\chi_{k}}$. Direct calculation further yields the third eigenvalue $\chi_c=0$. Under equal amplitude and wavenumber conditions, vector BSs arise from the interplay between a nonzero symmetric mode $(1,\frac{a}{\chi_{k}},\frac{b}{\chi_{k}})^{\top}$ and a zero antisymmetric mode $(0,1,-1)^{\top}$. The latter introduces opposite signs in the fundamental solutions $(R,S,W)^{\top}$, producing complementary peak-valley beating structures in two components. Vector BSs associated with $\chi_{a1},~\chi_{b2}$ and $\chi_{a2},~\chi_{b1}$ have identical structures but propagate with opposite velocities and are therefore classified as the same type. Specifically, $\chi_{a1}$ and $\chi_{b2}$ correspond to Type-I BSs, whereas $\chi_{a2}$ and $\chi_{b1}$ belong to Type-II BSs, as shown in Figs.~\ref{beating1}. We find the parameter $\omega$ controls the group velocity of the vector BSs. Since $\omega\neq0$, the vector BSs exhibit finite propagation velocities.

\begin{figure}[H]
	\centering
{\includegraphics[width=350 bp,height=3 cm]{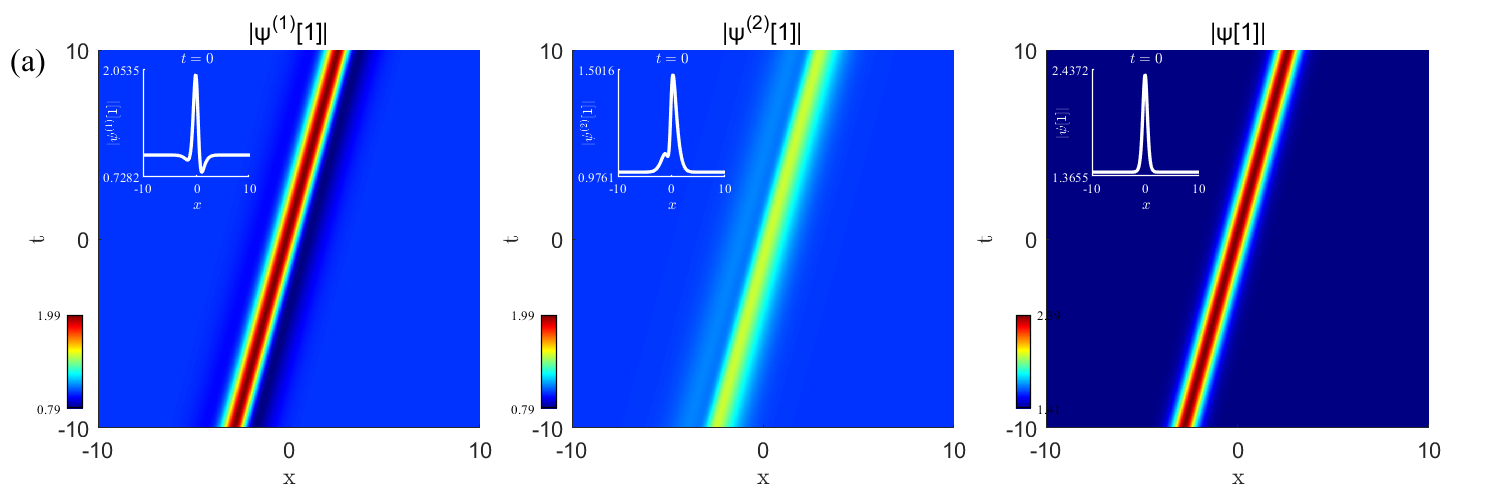}}
{\includegraphics[width=350 bp,height=3 cm]{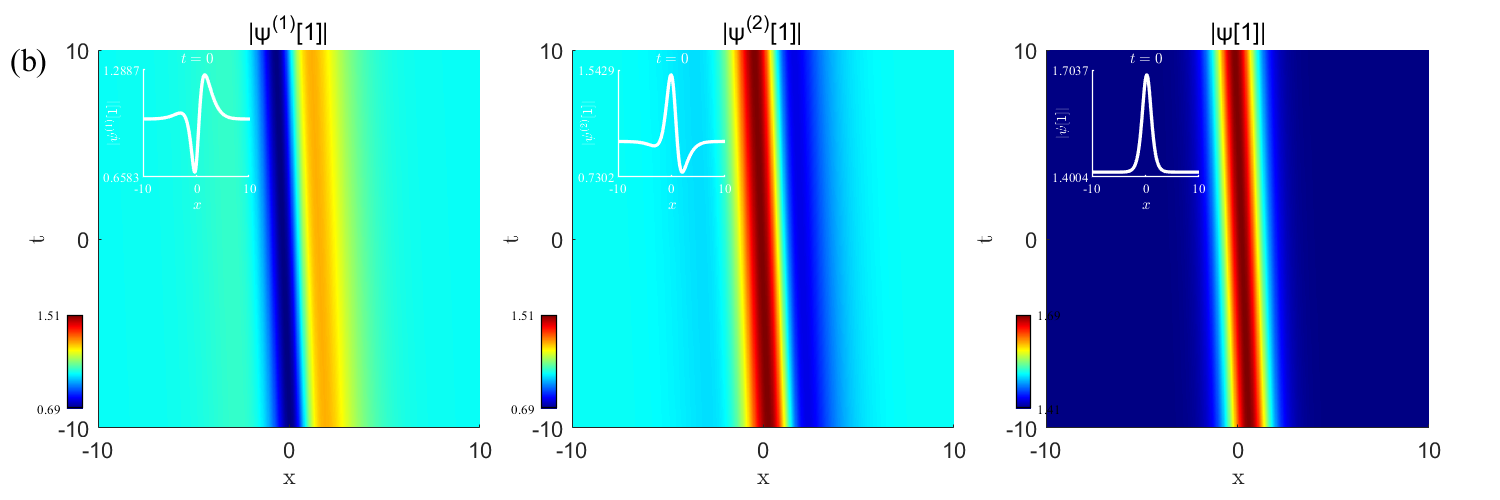}}
\caption{\footnotesize Amplitude profiles $|\psi^{(1)}[1]|,~|\psi^{(2)}[1]|$ and $|\psi[1]|$ of the kink solitons from the moving BSs. The fourth effect coefficient is given in (a): $\varepsilon_{m}=-0.0484$ and in (b): $\varepsilon_{m}=-0.0586$. Other parameters are the same as those in Figs.~\ref{beating1}. The insets in the two columns show the profiles of the kink solitons at $t=0$, revealing clear differences from the bright solitons shown in the last column.}
	\label{beating2}
\end{figure}

Vector BSs-to-solitons transition originates from velocity locking $V_{g}=V_{P}$ between the group velocity $V_{g}=-\frac{\Theta_{li}}{\chi_{li}}$ and the phase velocity $V_{P}=-\frac{\Theta_{lr}}{\chi_{lr}}$. Then the transition condition is given by
\begin{equation} \label{BS_ST1}
\begin{aligned}
\operatorname{Im}\!\frac{\Theta(\chi_{l})}{\chi_{l}}=0, \quad l=a,b,
\end{aligned}
\end{equation}
where
\begin{equation}
\begin{aligned}
\Theta(\chi_{l})=-\varepsilon\chi_{l}^{4}+(\frac{1}{2}+8\varepsilon a^{2})\chi_{l}^{2}.
\end{aligned}
\end{equation}
And the fourth-order term in Eq.~(\ref{BS_ST1}) is expressed by
\begin{equation}\label{BS_ST3}
\begin{aligned}
\varepsilon_{m}=\frac{1}{2(3[\chi_{l,r}]^{2}-[\chi_{l,i}]^{2}-8a^{2})}.
\end{aligned}
\end{equation}
The moving BSs can be transformed into kink solitons by means of Eq.~(\ref{BS_ST3}), as illustrated in Figs.~\ref{beating2}. Here, the velocity of the transformed soliton can be calculated exactly and is given by $V_{m}=-2\varepsilon_{m}\chi_{lr}|\chi_{l}|^{2}$.

\begin{figure}[H]
	\centering
{\includegraphics[width=400 bp,height=3 cm]{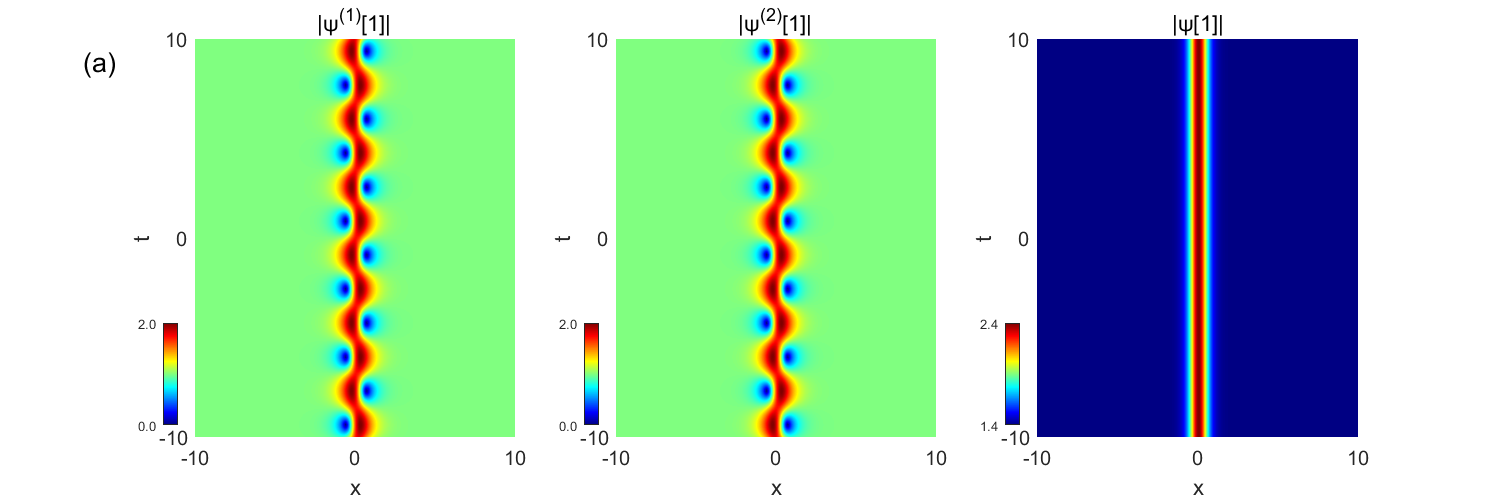}}
{\includegraphics[width=400 bp,height=3 cm]{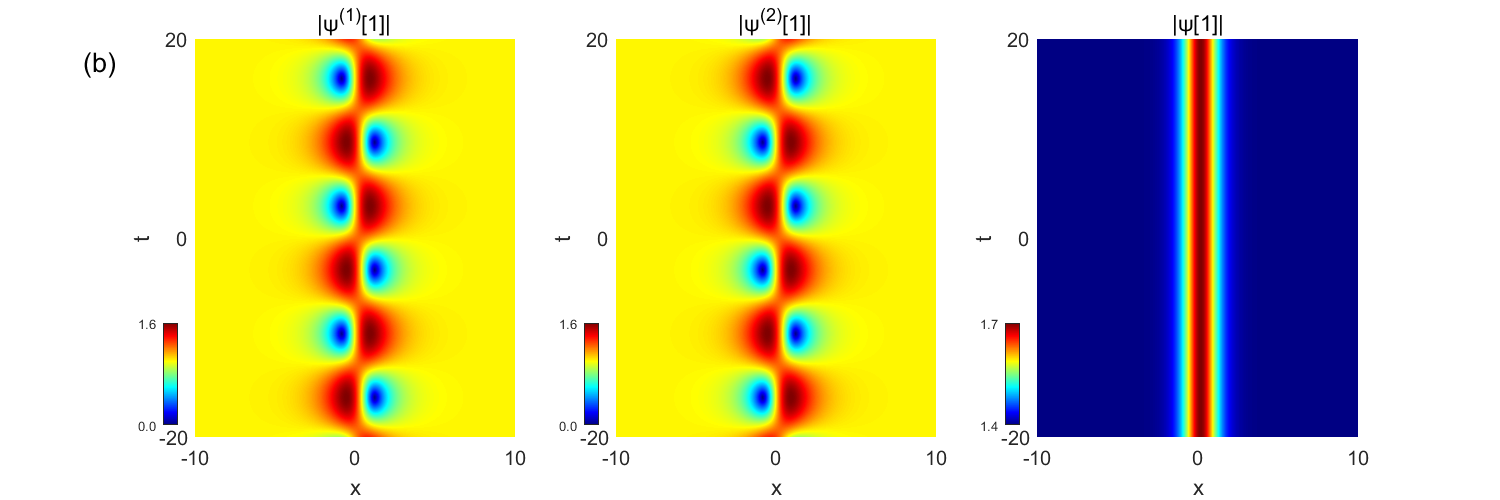}}
\caption{\footnotesize Amplitude profiles $|\psi^{(1)}[1]|, |\psi^{(2)}[1]|$ and $|\psi[1]|$ of (a) Type-I and (b) Type-II static BSs. The coefficients in (a)-(b) are (1,0,1) and (0,1,1), respectively. Parameters are: $a=1, \varepsilon=0.01, \alpha=1$ and $\omega=0$.}
	\label{beating3}
\end{figure}

The moving BSs satisfy $V_{g}\neq0,~\omega\neq0$. On the other hand, when $\omega=0$, the eigenvalues (\ref{BSvalue1}) are purely imaginary $\chi_{a}=\frac{1}{2}(-i\alpha\mp\sqrt{-\alpha^{2}-8a^{2}})$. The resulting BSs have zero velocity $V_{g}=0$, as shown in Figs.~\ref{beating3}. Clearly, the static BSs can be regarded as special cases of the moving BSs. To further elucidate the properties of BSs, we next consider the state transition solitons arising from the static BSs.

\begin{figure}[H]
	\centering
{\includegraphics[width=350 bp,height=3 cm]{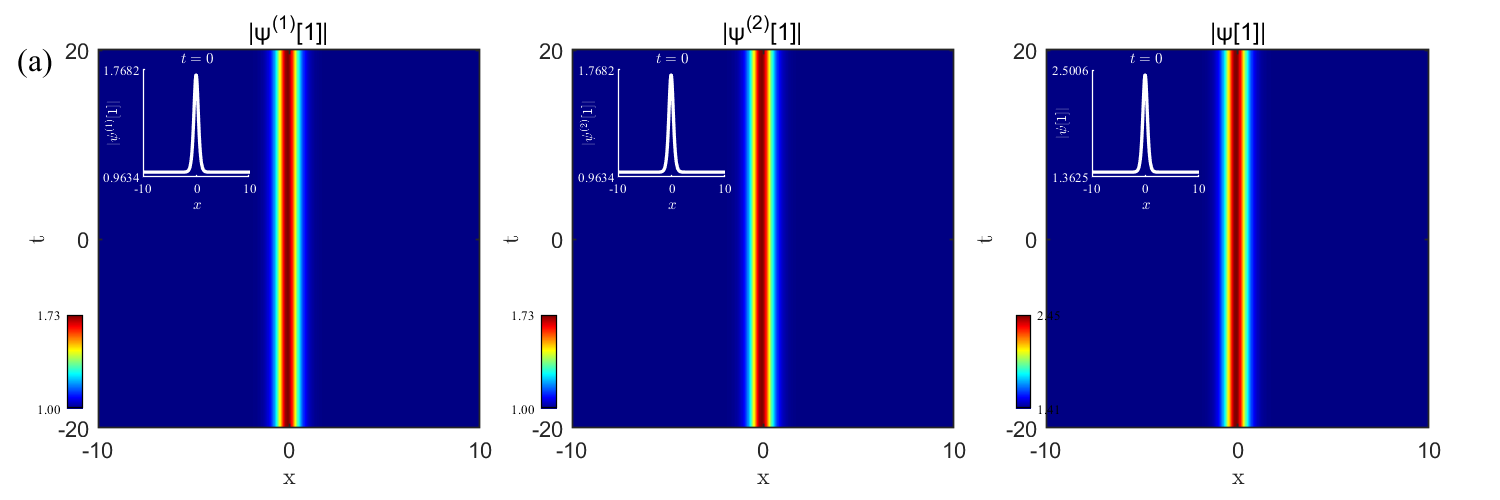}}
{\includegraphics[width=350 bp,height=3 cm]{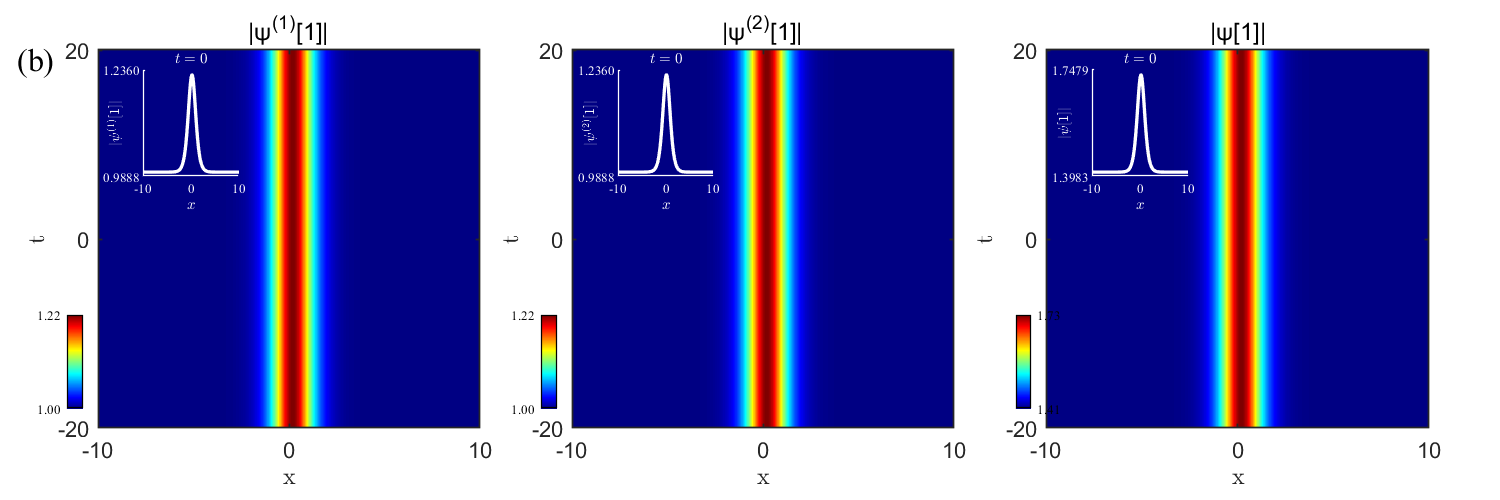}}
\caption{\footnotesize Amplitude profiles $|\psi^{(1)}[1]|,~|\psi^{(2)}[1]|$ and $|\psi[1]|$ of the static kink solitons from the static BSs. the fourth-effect coefficient in (a): $\varepsilon_{t}=-0.0417$ and in (b): $\varepsilon_{t}=-0.0556$. Other parameters are the same as those in Figs.~\ref{beating3}. The insets in the three columns show the profiles of the static kink solitons at $t=0$.}
	\label{beating4}
\end{figure}

We find that the state transition condition for the static BSs is still given by Eq.~(\ref{BS_ST1}) when $\omega=0$ and the fourth-effect coefficient is $\varepsilon_{t}=-\frac{1}{2([\chi_{li}]^{2}+8a^{2})}$. Both the static BSs and their transformed solitons have zero velocity. As depicted in Figs.~\ref{beating4}, the two types of transformed solitons differ in both width $L=\frac{1}{|\chi_{l,i}|}$ and the maximal soliton amplitudes $|\psi^{(j)}|=\sqrt{a^{2}+\frac{\chi_{l,i}}{2}}$. Although the static kink solitons may resemble bright solitons in profile, they are fundamentally different. Namely, the former exists on a nonzero background with an asymptotic phase difference $\Delta\phi=\pi$, whereas the latter exists on a zero background and has no well-defined asymptotic phase. Although the moving and static BSs share the same internal oscillatory structure, the solitons transformed from the former are asymmetric with respect to $x$, whereas that from the latter are completely symmetric.

\subsection{BS resonance and interaction}
In this subsection, we study BS resonant modes and interaction. We then analyze their state transitions induced by the fourth-order effect. We first discuss the BS resonant modes with the coefficient $(1,1,1)$ and parameters $(\alpha,~\omega)$. When the mode coefficients are chosen as $(1,1,1)$, the resulting solution is not described by Type-I or Type-II BS. Instead, the solution forms a three mode composite patterns consisting of two BSs and a breather. The control parameter $\alpha$ alters the sign symmetry of the spectral roots, thereby giving rise to two distinct resonance modes.

\begin{figure}[H]
	\centering
{\includegraphics[width=400 bp,height=3 cm]{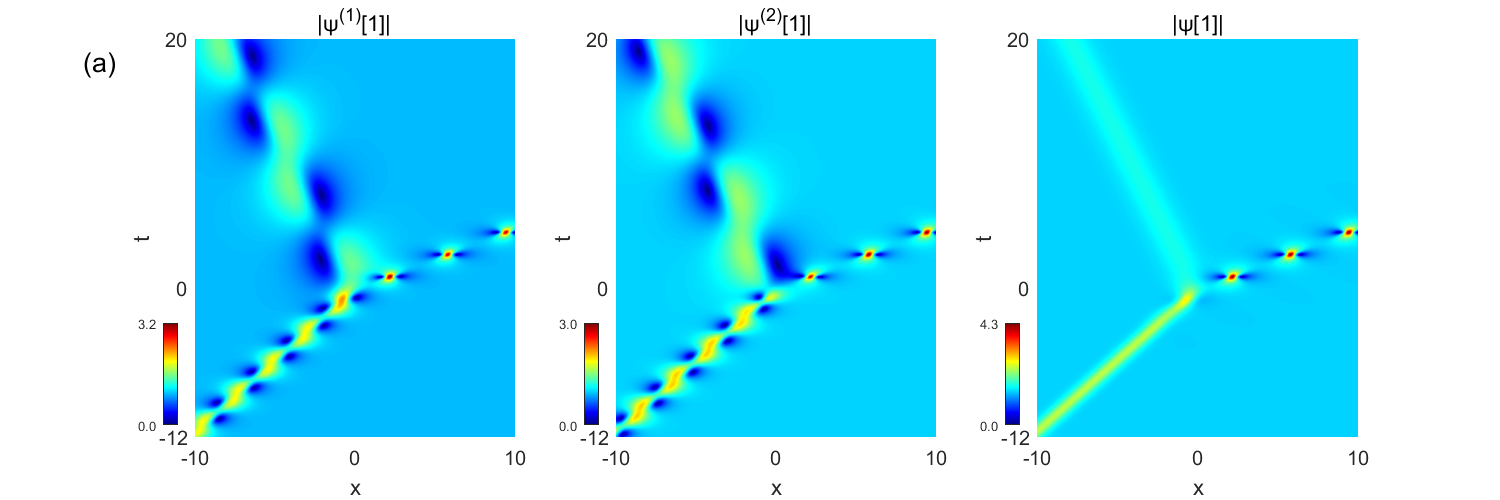}}
{\includegraphics[width=400 bp,height=3 cm]{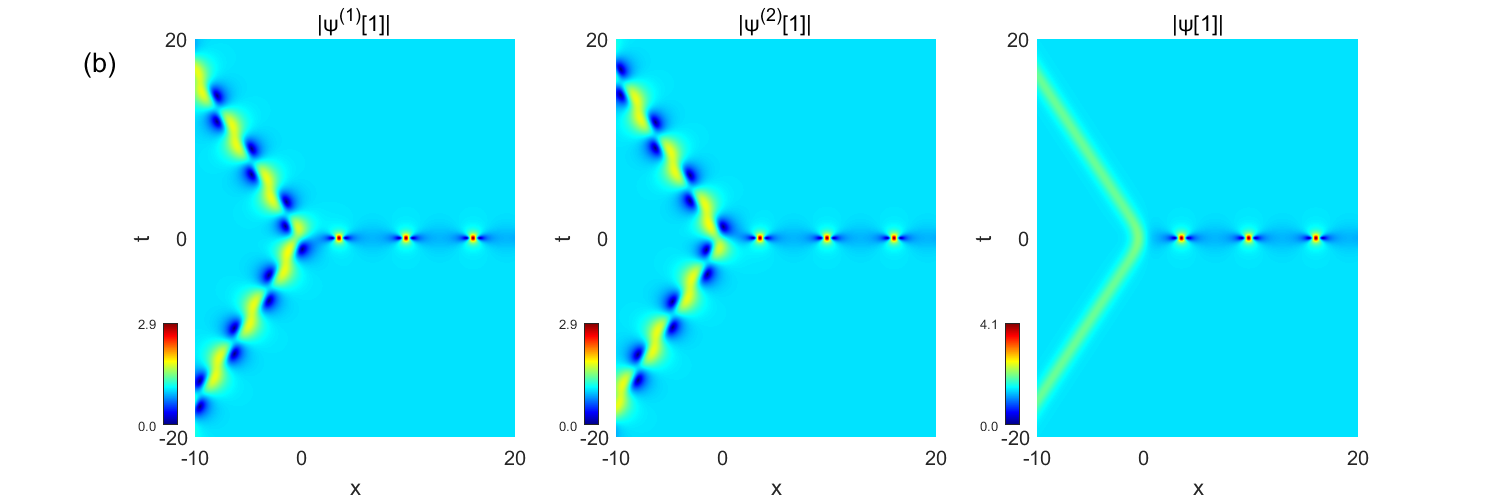}}
\caption{\footnotesize Amplitude profiles $|\psi^{(1)}[1]|,~|\psi^{(2)}[1]|$ and $|\psi[1]|$ of BS resonance modes corresponding to the eigenvalue $\chi_{a1}$ with (1,1,1). Panels (a) show the fission of BS with $\alpha=1$. Panels (b) show the reflection of BS with $\alpha=0$. Other parameters are $a=1, \varepsilon=0.01, \omega=1$.}
	\label{gozhen}
\end{figure}

When $\alpha=0$, two spectral roots satisfy $\chi_{br}=-\chi_{ar}$ and $\chi_{bi}=\chi_{ai}$. By calculation, spectral differences of the wave modes are given by $\chi_{b}-\chi_{a}=-\omega$ and $\Omega(\chi_{b})-\Omega(\chi_{a})=i[\omega\sqrt{8a^{2}-\omega^{2}}(\frac{1}{2}+\varepsilon(12a^{2}-\omega^{2}))]$. The former is real and the latter is purely imaginary. The resulting breather in the corresponding resonant modes is spatially periodic and temporally localized, thereby forming a standard AB pattern. For $\alpha\neq0$, both spectral differences $\chi_{b}-\chi_{a}=-\omega-i\alpha$ and $\Omega(\chi_{b})-\Omega(\chi_{a})=(i\omega-\alpha)\sqrt{8a^{2}-\omega^{2}}[\frac{1}{2}+\varepsilon(12a^{2}-(\omega^{2}+i\alpha))]$ are complex, giving rise to a moving TW with localization and periodic oscillations in the resonant modes. Figs.~\ref{gozhen} provide compelling confirmation of this interpretation.

We know that the composite patterns in Figs.~\ref{gozhen} involve eigenvalues $\chi_{a}$ and $\chi_{b}$, thereby producing the resonant modes. The two BS branches in Figs.~\ref{gozhen}~(a) correspond to Type-I and Type-II, respectively, whereas both BS branches in Figs.~\ref{gozhen}~(b) belong to Type-I. Importantly, the state transition condition for the TW branch ($\beta_{1}-\beta_{2}=0$) in Figs.~\ref{gozhen}~(a) differs from that given in Eqs.~(\ref{GB_ST2}). A recalculation yields
\begin{equation} \label{BS_ST2}
\begin{aligned}
\operatorname{Im}\!\frac{\Omega_{ab}}{\chi_{ab}}=0,
\end{aligned}
\end{equation}
where
\begin{equation}
\begin{aligned}
\Omega_{ab}=\Omega(\chi_{a})-\Omega(\chi_{b}),\quad
\chi_{ab}=\chi_{a}-\chi_{b}.
\end{aligned}
\end{equation}
Then the fourth-order term in Eq.~(\ref{BS_ST2}) is expressed by
\begin{equation}\label{BS_es}
\begin{aligned}
\varepsilon_{r}=-\frac{\lambda_{i}}{2[\lambda_{i}(12a^{2}-\omega^{2}+\alpha^{2})-2\omega\alpha\lambda_{r}]},
\end{aligned}
\end{equation}
with
\begin{equation}
\begin{aligned}
\lambda=\sqrt{(\omega+i\alpha)^{2}-8a^{2}}.
\end{aligned}
\end{equation}

\begin{figure}[H]
	\centering
{\includegraphics[width=400 bp,height=3 cm]{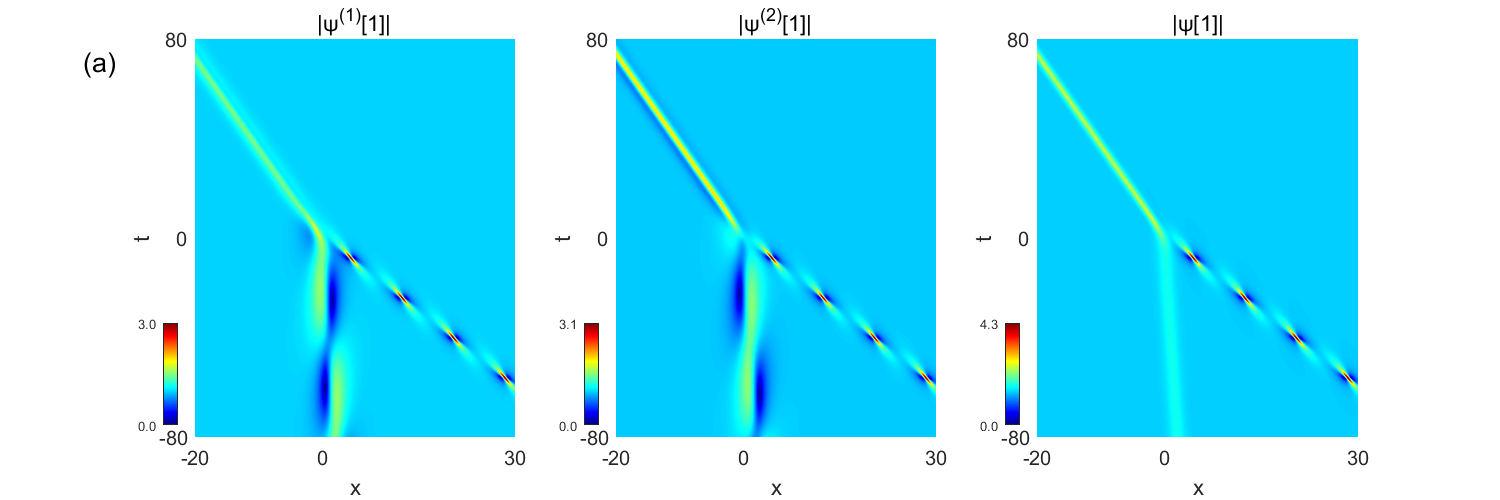}}
{\includegraphics[width=400 bp,height=3 cm]{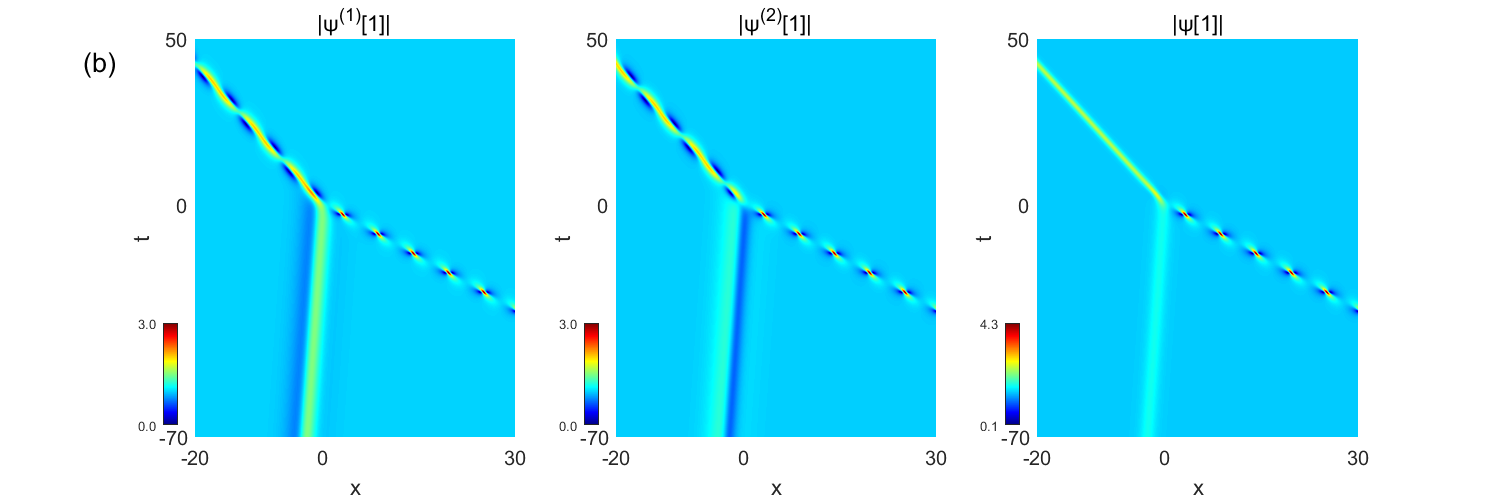}}
{\includegraphics[width=400 bp,height=3 cm]{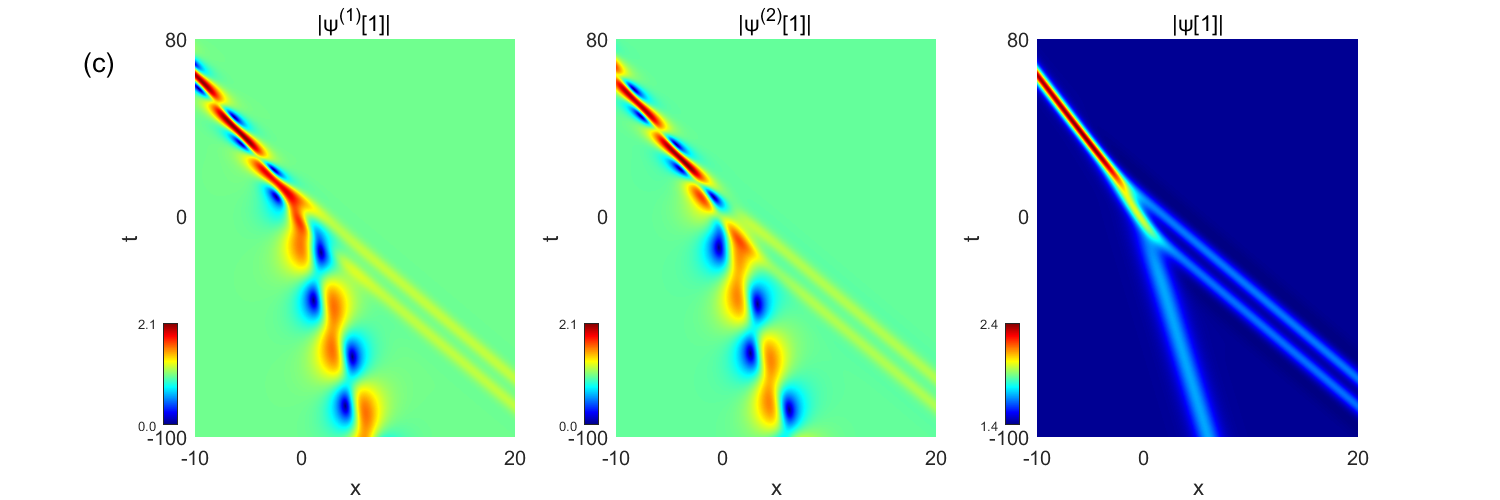}}
\caption{\footnotesize Amplitude profiles $|\psi^{(1)}[1]|,~|\psi^{(2)}[1]|$ and $|\psi[1]|$ of the transition solitons from each branch in BS resonance modes. Panels (a) show the kink solitons from Type-I BSs with $\varepsilon_{m}=-0.0484$. Panels (b) show the kink solitons from Type-II BSs with $\varepsilon_{m}=-0.0586$. Panels (c) show the multipeak solitons from the TW pattern with $\varepsilon_{r}=-0.0425$. Other parameters are the same as those in Figs.~\ref{gozhen}~(a).}
	\label{gozhen_ST1}
\end{figure}

Since three branches in the resonant modes share the same parameter $\varepsilon$, they cannot be transformed into solitons simultaneously with the same fourth-order transition parameter $\varepsilon_{m}$. Under condition (\ref{BS_ST3}), two BS branches can be transformed individually into the moving kink solitons in Figs.~\ref{gozhen_ST1}~(a) and (b). The transition between the TW branch and the multipeak soliton is described in Eq.~(\ref{BS_es}), as depicted in Figs.~\ref{gozhen_ST1}~(c). Moreover, the velocity of the multipeak soliton is analytically given by $V_{r}=-\frac{\Omega_{ab}}{\chi_{ab}}$. It is readily seen that the transformed solitons from the TW breathers also possess nonzero velocities.

\begin{figure}[H]
	\centering
{\includegraphics[width=400 bp,height=3 cm]{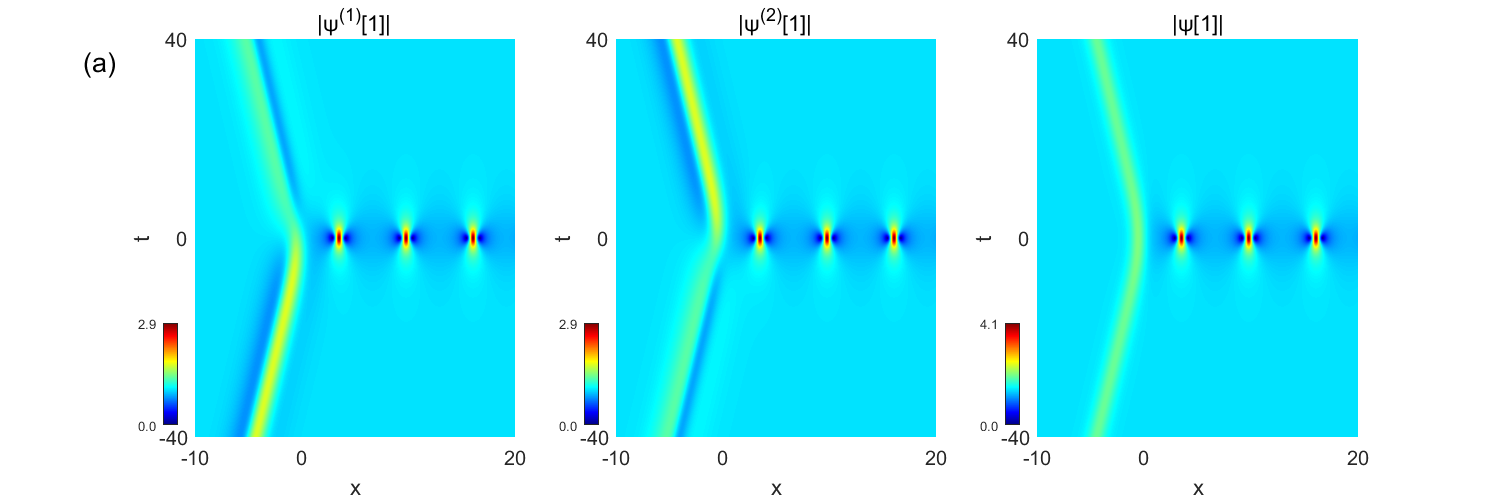}}
{\includegraphics[width=400 bp,height=3 cm]{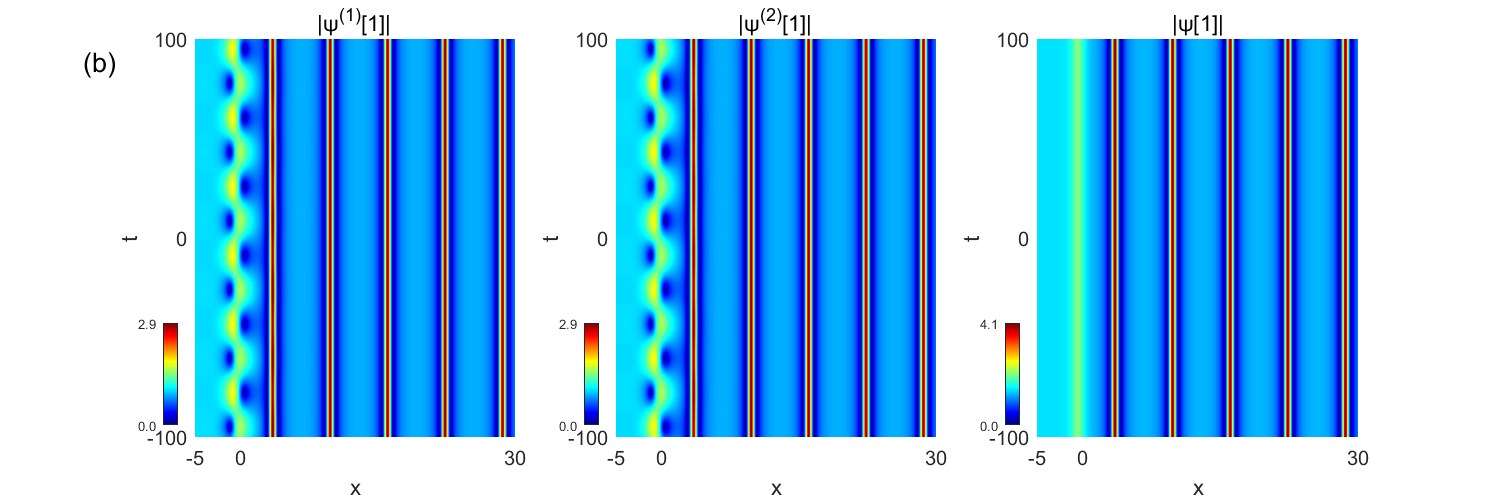}}
\caption{\footnotesize Amplitude profiles $|\psi^{(1)}[1]|,~|\psi^{(2)}[1]|$ and $|\psi[1]|$ of transition solitons from each branch in BS resonance modes. Panels (a) show the kink solitons from two BS branches with $\alpha_{1}=1, \varepsilon_{t}=-0.0556$. Panels (b) show periodic solitons from AB pattern with $\alpha_{1}=0, \varepsilon_{r}=-0.0455$. Other parameters are the same as those in Figs.~\ref{gozhen}~(b).}
	\label{gozhen_ST}
\end{figure}

When $\alpha=0$, all three spectral modes $\chi_{a},~\chi_{b}$ and $\chi_{c}$ still participate simultaneously. Since the two nonzero spectral roots $\chi_{a},~\chi_{b}$ have identical imaginary parts $\chi_{ai}=\chi_{bi}$ and opposite real parts $\chi_{ar}=\chi_{br}$, the resonant modes consist by two BS branches and a static AB pattern. The BS velocities satisfy $V_b=-V_a$, with $V_a=\omega\left[\frac{1}{2}+\varepsilon(12a^{2}-\omega^{2})\right]$, while the AB pattern remains static. The two BS branches in Figs.~\ref{gozhen}~(b) are both Type-I but propagate in opposite directions. Therefore, they share the same transition parameter $\varepsilon_{m}=\frac{1}{2(10a^{2}-\omega^{2})}$ and can be transformed into the moving kink solitons simultaneously in Figs.~\ref{gozhen_ST}~(a). Under the condition $\varepsilon_{r}=-\frac{1}{2(12a^{2}-\omega^{2})}$, the AB patterns can be transformed into periodic solitons with zero velocity. After the transition, velocities of the kink solitons change to $V_b=-V_a$, with $V_a=-\frac{\omega a^{2}}{10a^{2}-\omega^{2}}$. Both BSs overlap completely at the center $x=0$, as shown in Figs.~\ref{gozhen_ST}~(b).

\begin{figure}[H]
	\centering
{\includegraphics[width=400 bp,height=3 cm]{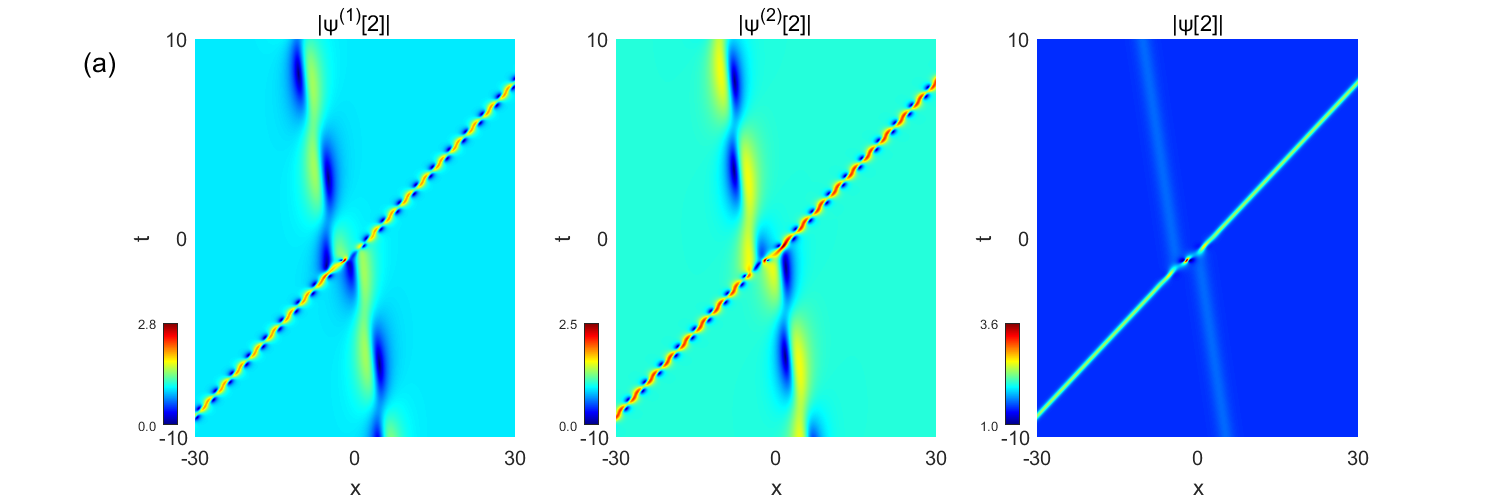}}
{\includegraphics[width=400 bp,height=3 cm]{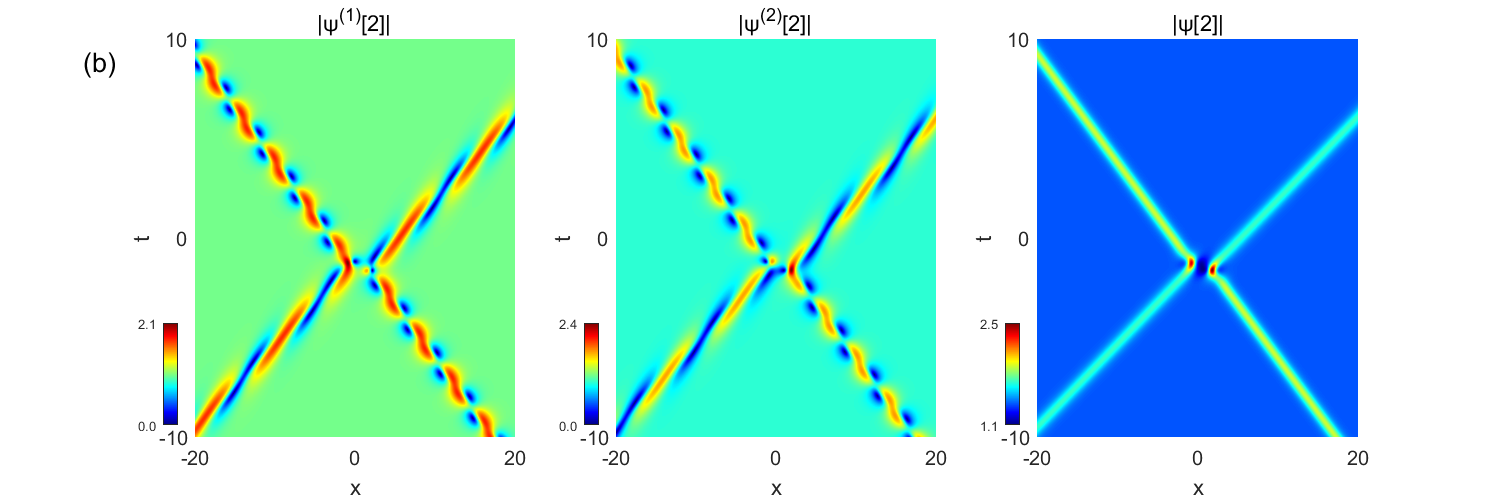}}
\caption{\footnotesize Amplitude profiles $|\psi^{(1)}[2]|,~|\psi^{(2)}[2]|$ and $|\psi[2]|$ of the second-order solutions formed by the nonlinear superposition of Type-I and Type-II BSs. The parameters are given in (a) $\alpha_{1}=1, \alpha_{2}=2, \omega_{1}=\omega_{2}=1$ and (b) $\alpha_{1}=\alpha_{2}=0, \omega_{1}=1, \omega_{2}=2$, respectively. Other parameters are $a=1$ and $\varepsilon=0.01$.}
	\label{MBS1}
\end{figure}

Next, we discuss the interaction between two types of BSs and their state transition dynamics. We know that the parameter $\alpha$ controls the localization of vector BSs, whereas $\omega$ determines their propagation velocity. When $\omega=0$, the interaction between two static BSs exhibits only limited wave patterns. Therefore, we focus on the interaction between two types BSs for $\omega\neq0$, considering the cases $\alpha=0$ and $\alpha\neq0$ separately. The structure resulting from the collision between two BSs depends on the velocity difference $|V_{g1}-V_{g2}|$ and envelope width $L_{n}=\frac{1}{\chi_{i}}$, respectively. Analytical examination of the second-order solutions (\ref{DT2}) shows that the phase difference of the cross term $\Psi[1]\Psi{\dagger}[1]$ is governed by $\Delta\mathcal{C}=(\chi_{a}-\chi_{b})x+(\Omega(\chi_{a})-\Omega(\chi_{b}))t$. The oscillation period of the short-lived structure is denoted by $\tau_{o}=\frac{2\pi}{Re(\Omega(\chi_{a})-\Omega(\chi_{b})+V_{c}(\chi_{a}-\chi_{b}))}$~($V_{c}\approx\frac{V_{g1}+V_{g2}}{2}$ represents the propagation velocity of the collision center). When the velocity difference $|V_{g1}-V_{g2}|$ between two BSs is small, their collision duration $T$ is close. When the parameters $\alpha_{1}, \alpha_{2},\omega_{1},\omega_{2}$ of two BSs do not satisfy condition $\alpha_{1}\cong \alpha_{2}$ or $\omega_{1}\cong \omega_{2}$, their trajectories form clear X-shaped patterns, with only a weak structure visible at the collision center in Figs.~\ref{MBS1}.

\begin{figure}[H]
	\centering
{\includegraphics[width=400 bp,height=3 cm]{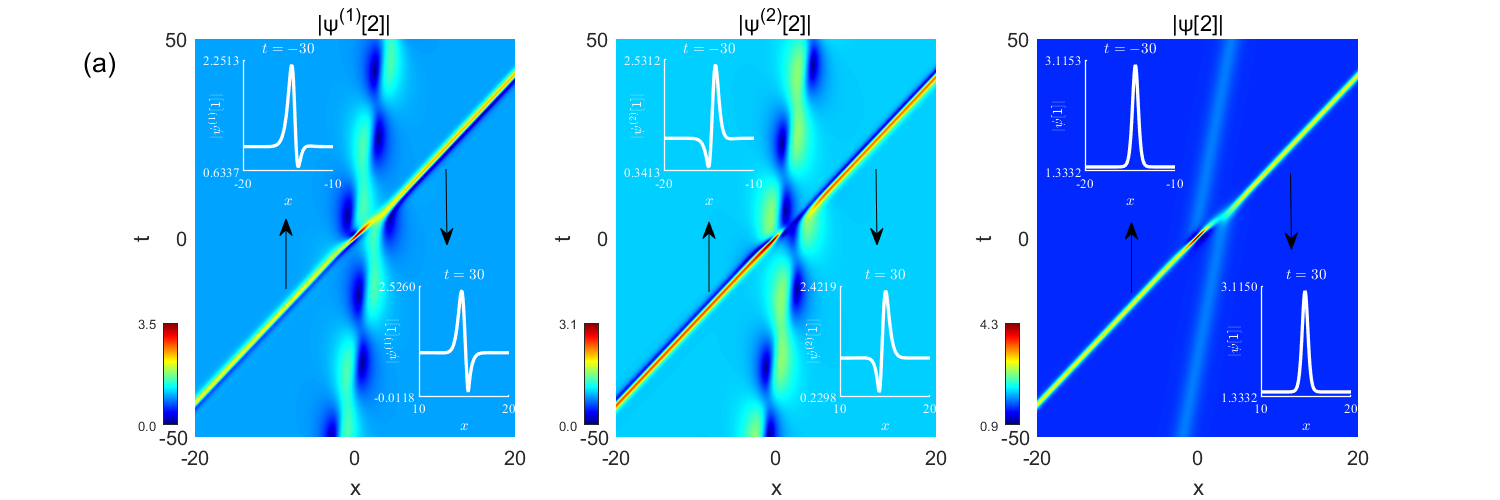}}
{\includegraphics[width=400 bp,height=3 cm]{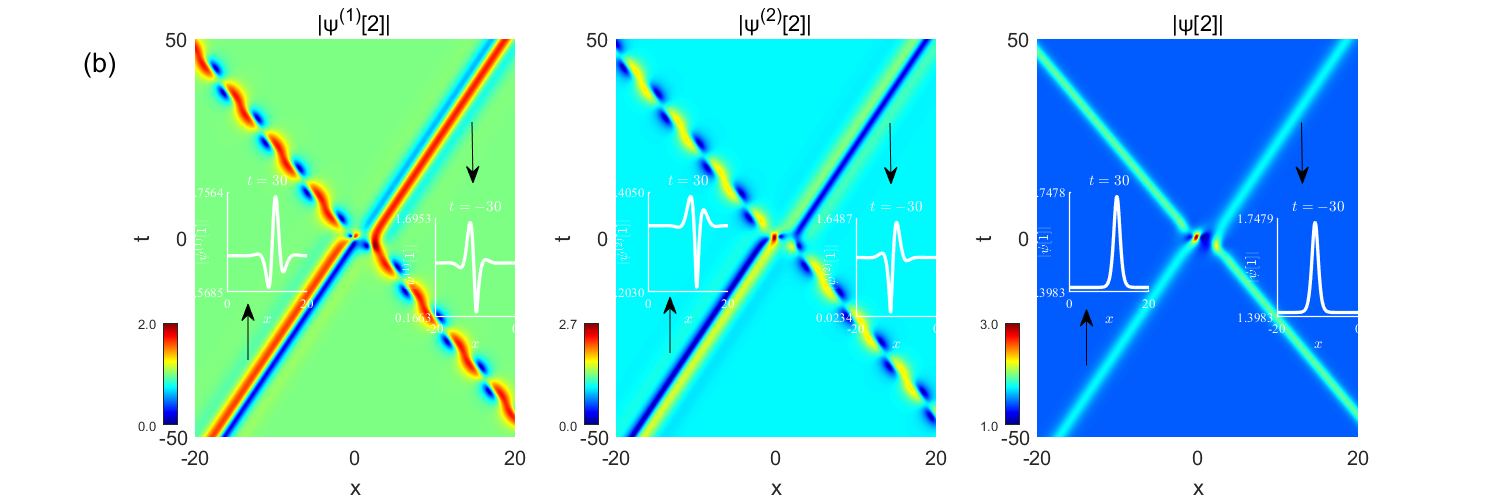}}
\caption{\footnotesize Amplitude profiles $|\psi^{(1)}[2]|,~|\psi^{(2)}[2]|$ and $|\psi[2]|$ of the second-order solutions formed by the nonlinear superposition between BS and the transformed soliton from BS. The transition parameters are given in (a) $\varepsilon=-0.0376$ and (b) $\varepsilon=-0.0833$, respectively. Other parameters are the same as those in Figs.~\ref{MBS1}. The two insets in each subplot respectively show the profiles at selected times in the positive and negative spatial regions.}
	\label{MBS1_ST}
\end{figure}

Because two vector BSs in Figs.~\ref{MBS1} belog to Type-I and Type-II, their corresponding transition parameters $\varepsilon_{m}$ are distinct. Therefore, they cannot simultaneously be transformed into the kink solitons. Figs.~\ref{MBS1_ST} illustrate the case in which only one type BSs are converted into the kink solitons. During the interaction, the remaining BSs undergo only positional shifts, while the transformed solitons exhibit pronounced deformations observed in the inset of Figs.~\ref{MBS1_ST}. These demonstrate the distinctive properties of the kink solitons transformed from the vector BSs.

\begin{figure}[H]
	\centering
{\includegraphics[width=400 bp,height=3 cm]{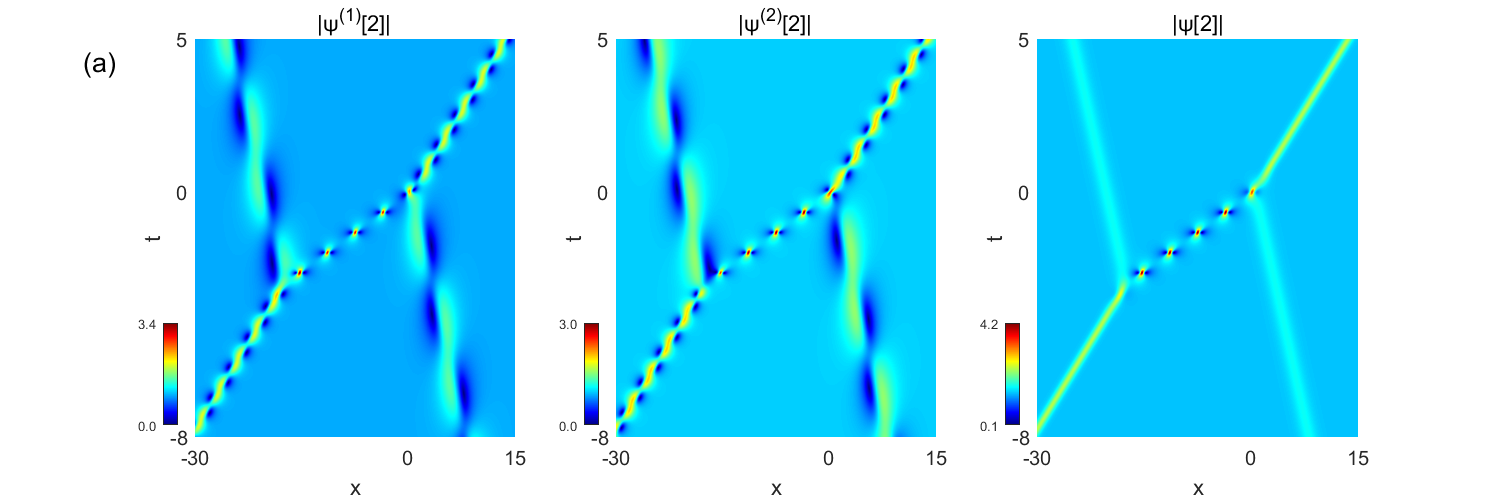}}
{\includegraphics[width=400 bp,height=3 cm]{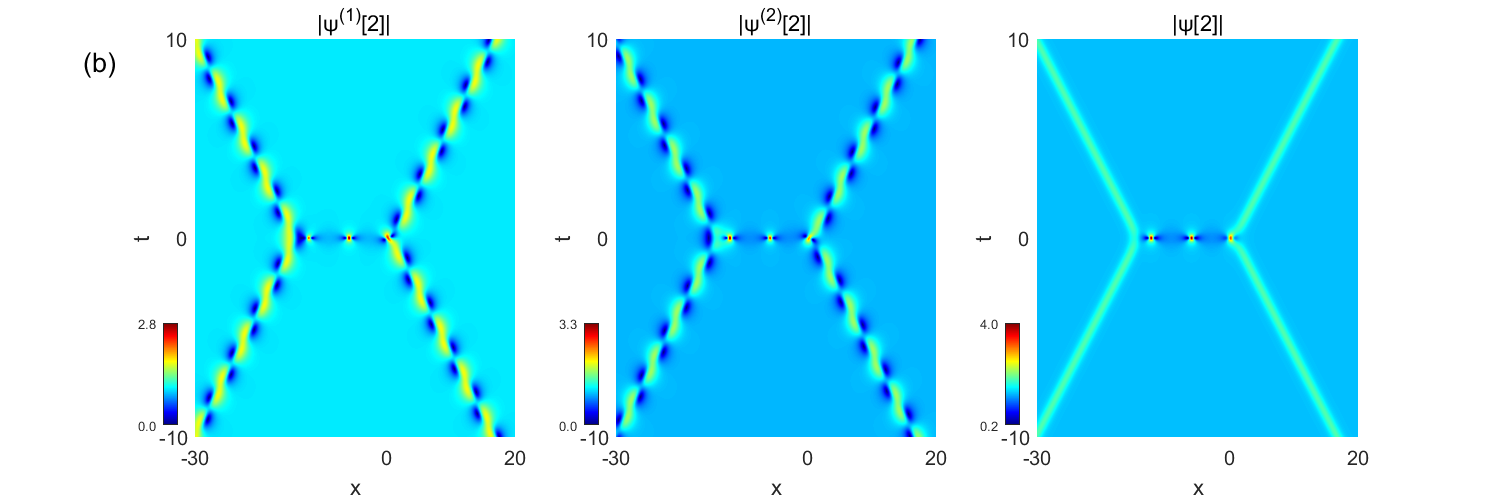}}
\caption{\footnotesize Amplitude profiles $|\psi^{(1)}[2]|,~|\psi^{(2)}[2]|$ and $|\psi[2]|$ of the short-lived patterns in the second-order solutions formed by the nonlinear superposition of Type-I and Type-II BSs. The parameters are (a) $\alpha_{1}=1, \alpha_{2}=1.00001, \omega_{1}=\omega_{2}=1$ and (b) $\alpha_{1}=\alpha_{2}=0, \omega_{1}=1, \omega_{2}=1.0000001$, respectively. Other parameters are $a=1$ and $\varepsilon=0.01$.}
	\label{MBS2}
\end{figure}

However, the collision becomes more complex when $(\alpha_{1},~\omega_{1})\cong(\alpha_{2},~\omega_{2})$. In this case, the corresponding spectral modes satisfy $\chi_{a}\approx\chi_{b}$ and $\Omega(\chi_{a})\approx\Omega(\chi_{b})$, causing two envelopes to remain overlapped for an extended period. This enhances the periodic oscillations and generates a short-lived structure at the collision center. Its characteristics are governed by the spectral differences $\Delta\chi=\chi_{b}-\chi_{a}$ and $\Delta\Omega=\Omega(\chi_{b})-\Omega(\chi_{a})$. When both $\alpha$ and $\omega$ are nonzero, $\Delta\chi$ and $\Delta\Omega$ are complex, yielding a moving TW-type pattern with nonzero group velocity $V_{c}$, as shown in Figs.~\ref{MBS2}~(a). When $\Delta\chi$ is real and $\Delta\Omega$ is purely imaginary, the central structure becomes an AB-type pattern that is spatially periodic and temporally localized, as illustrated in Figs.~\ref{MBS2}~(b).

\begin{figure}[H]
	\centering
{\includegraphics[width=400 bp,height=3 cm]{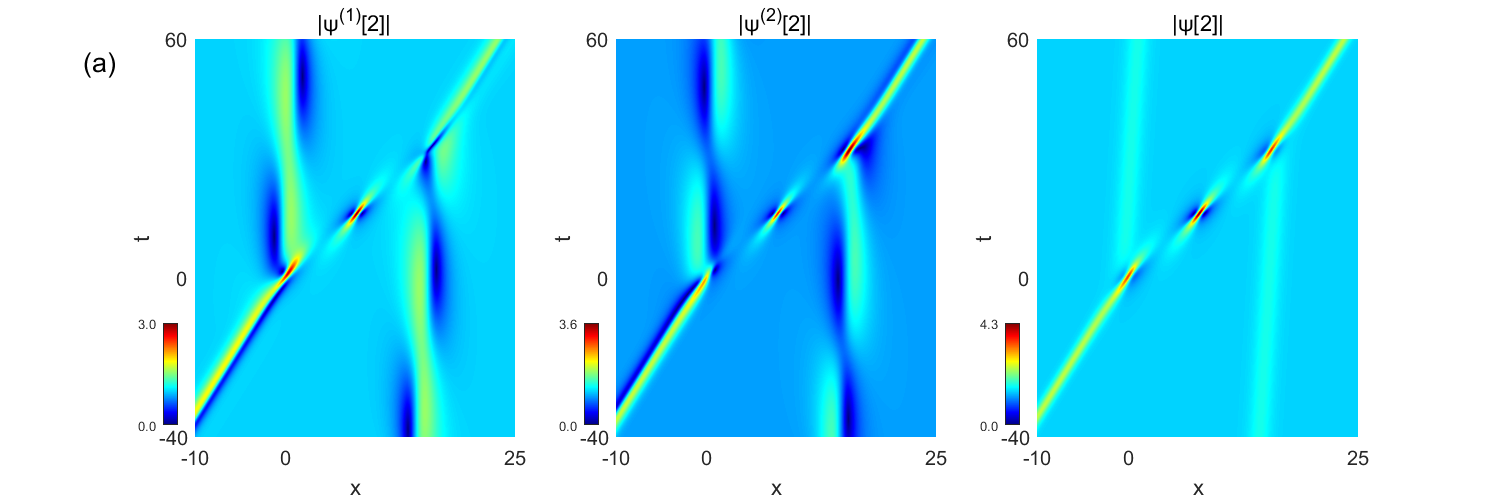}}
{\includegraphics[width=400 bp,height=3 cm]{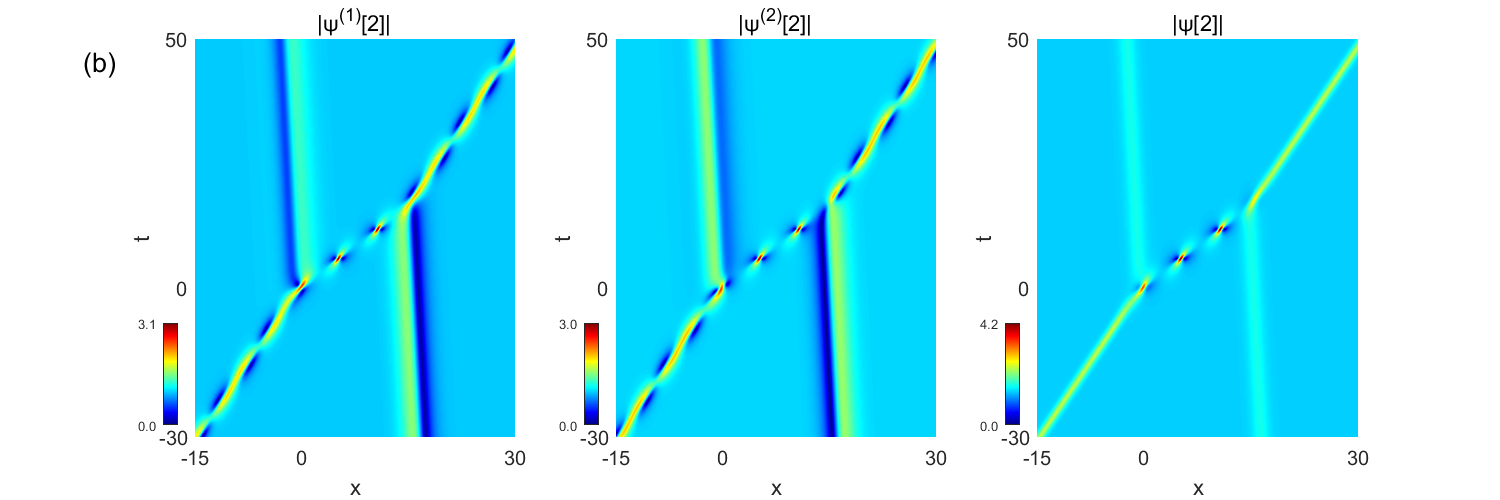}}
{\includegraphics[width=400 bp,height=3 cm]{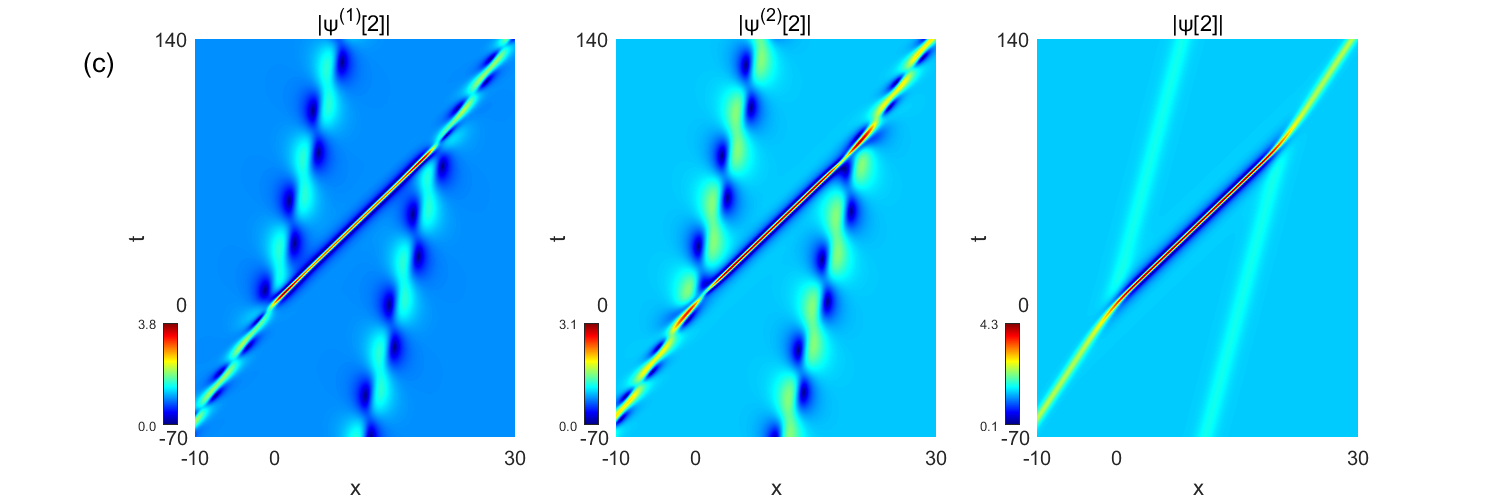}}
\caption{\footnotesize Amplitude profiles $|\psi^{(1)}[2]|,~|\psi^{(2)}[2]|$ and $|\psi[2]|$ of the short-lived TW-patterns in the second-order solution. Panels (a) show the interaction between Type-II BSs and the kink solitons from Type-I BSs with $\varepsilon_{m1}=-0.0483$. Panels (b) show the interaction between Type-I BSs and the kink solitons from Type-II BSs with $\varepsilon_{m2}=-0.0586$. Panels (c) show the multipeak solitons from short-lived TW-patterns in the interaction with $\varepsilon_{r}=-0.0425$. Other parameters are the same as those in Figs.~\ref{MBS2}~(a).}
	\label{MBS2_ST1}
\end{figure}

In Fig.~\ref{MBS2}, two BSs on either side propagate with equal speeds but in opposite directions $V_{a}=-V_{b}$. In comparison, the velocity of the short-lived structure is independent of two BSs and is given by $V_{B}=\frac{\operatorname{Im}\!\Delta\Omega}{\operatorname{Im}\!\Delta\chi}$. In the short-lived TW-patterns, two BSs on either side and the central breather satisfy the state transition conditions (\ref{BS_ST3}) and (\ref{BS_es}), respectively. However, two side BSs have distinct transition parameters, $\varepsilon_{m1}$ and $\varepsilon_{m2}$, and therefore cannot simultaneously transform into the kink solitons. As shown in Figs.~\ref{MBS2_ST1}, we explore their state transition case. Most notably, the central TW can also undergo the state transition, as illustrated in Figs.~\ref{MBS2_ST1}~(c).

\begin{figure}[H]
	\centering
{\includegraphics[width=400 bp,height=3 cm]{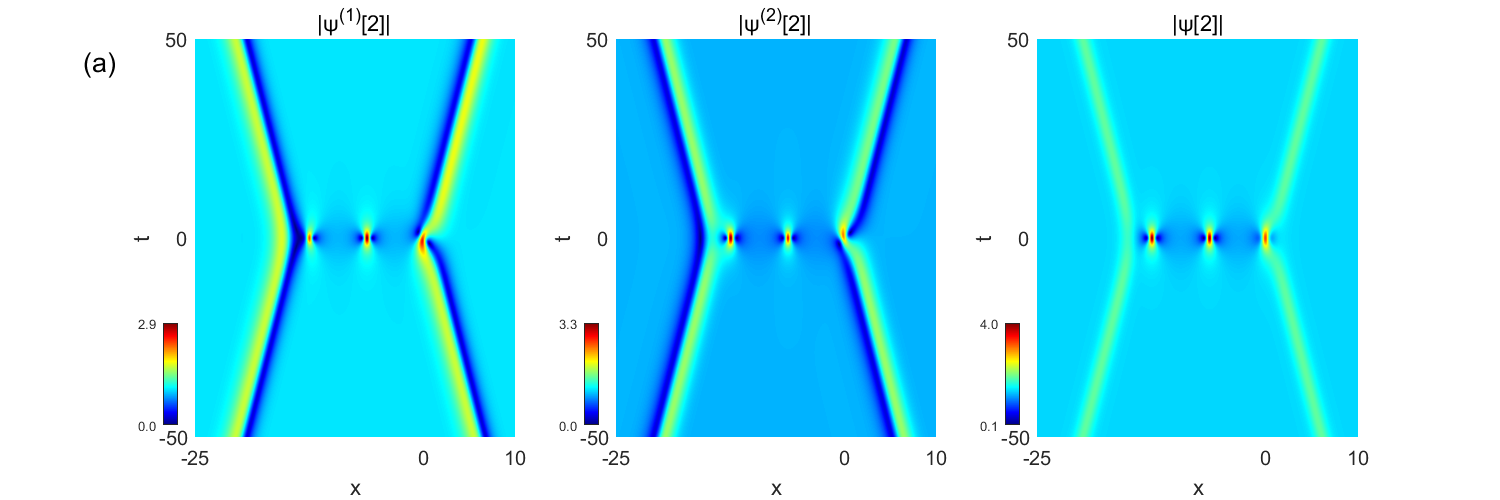}}
{\includegraphics[width=400 bp,height=3 cm]{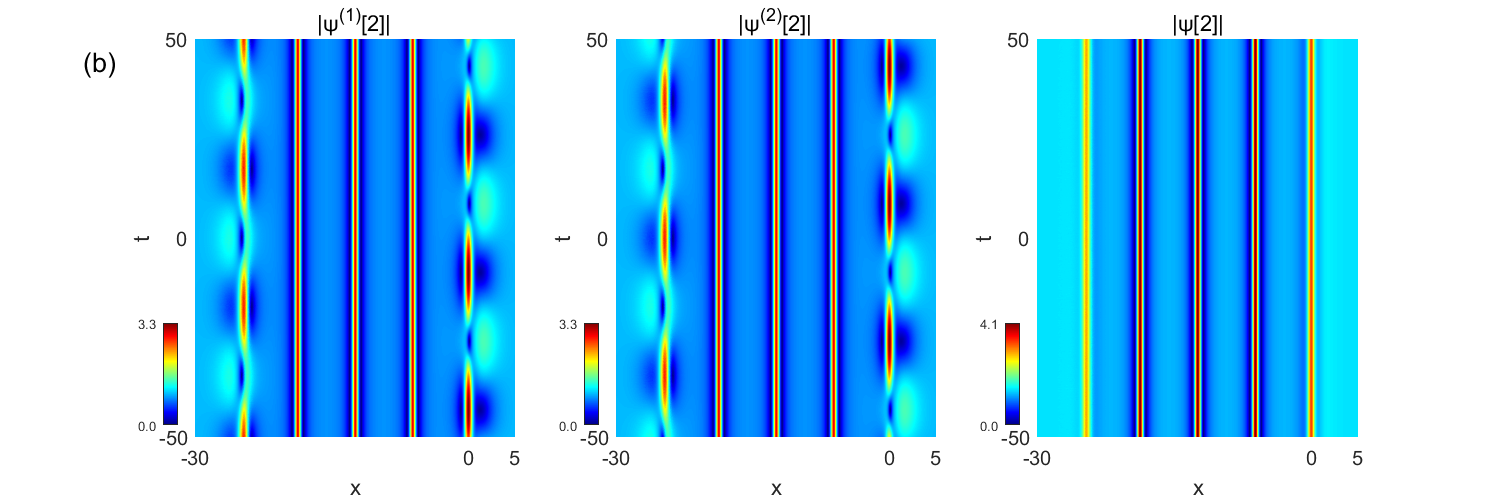}}
\caption{\footnotesize Amplitude profiles $|\psi^{(1)}[2]|,~|\psi^{(2)}[2]|$ and $|\psi[2]|$ of the short-lived AB-patterns in the second-order solution. Panels (a) show the interaction between two kink solitons from two Type-I BSs with $\varepsilon_{s}=-0.0556$ when $\alpha=1$. Panels (b) show the periodic solitons from the short-lived AB-patterns in the interaction with $\varepsilon_{r}=-0.0455$ when $\alpha=0$. Other parameters are the same as those in Figs.~\ref{MBS2}~(b).}
	\label{MBS2_ST2}
\end{figure}

For the short-lived AB-patterns in Figs.~\ref{MBS2}~(b), two side BSs have nearly identical parameters and exhibit the state transitions. As a result, they share the same transition condition $\varepsilon_s$ and can simultaneously evolve into the kink solitons, as illustrated in Figs.~\ref{MBS2_ST2}~(a). Meanwhile, Figs.~\ref{MBS2_ST2}~(b) present the conversion between the central AB and periodic solitons. These results provide a more complete characterization for the vector BSs.

\vspace{3mm}
\section{Exact analytic spectra} \label{eq:sec6}
\vspace{2mm}
Physical spectra~\cite{non-RW1,non-RW2,beating1} provide an important bridge between the theoretical description and experimental observation of the vector localized waves~\cite{15,16}. Here, we investigate the physical spectra of the vector localized waves under the fourth-order effect. The transition conditions are identifiable in the spectral domain for both the degenerate and non-degenerate cases. These reveal new dynamical features induced by the fourth-order effect.

\subsection{RWs spectra}
Since the TW solution~(\ref{eq:GB}) contains two independent parameters $\omega,~\alpha$, its spectrum cannot be directly represented in two dimensions using a single parameter. We therefore focus on the limiting case at $\omega=0, \alpha=0$, namely, the spectra of RW solutions~(\ref{eq:RW}). We rewrite~(\ref{eq:RW}) as
\begin{equation}\label{eq:RW1}
\begin{aligned}
\psi_{RW}^{(j)}(x,t)=\psi_{0}^{(j)}[1+\mathcal{J}_{j}(x,t)],
\end{aligned}
\end{equation}
where
\begin{equation}
\begin{aligned}
\mathcal{J}_{j}=\frac{2i(\chi_{r}+\beta_{j})(x+v_{r}t)-2i\chi_{i}v_{i}t-1}{((\chi_{r}+\beta_{j})^{2}+\chi_{i}^{2})
[(x+v_{r}t)^{2}+(v_{i}t)^{2}+\frac{1}{4\chi_{i}^{2}}]},
\end{aligned}
\end{equation}
$v=\chi(1+16\varepsilon a^{2}-4\varepsilon\chi^{2})$ is the velocity that governs the propagation direction of vector RWs. From \cite{non-RW1}, each of two wave components $\psi_{RW}^{(j)}(x,t)$ consists of an infinite number of spectral harmonics:
\begin{equation}\label{Fourier2}
\begin{aligned}
\psi^{(j)}_{x,t}=\int_{-\infty}^{\infty} \mathcal{F}^{(j)}_{\omega,t}e^{i\omega x}d\omega,
\end{aligned}
\end{equation}
where the Fourier components $\mathcal{F}^{(j)}_{\omega,t}$ are given by
\begin{equation}
\begin{aligned}
\mathcal{F}^{(j)}_{\omega,t}=\frac{1}{2\pi}\int_{-\infty}^{\infty}\psi^{(j)}_{x,t}e^{-i\omega x}dx.
\end{aligned}
\end{equation}
To simplify the calculations, Eq.~(\ref{Fourier2}) can be rewritten as
\begin{equation}\label{Fourier3}
\begin{aligned}
\psi^{(j)}_{x,t}=\int_{-\infty}^{\infty} \widetilde{\mathcal{F}}^{(j)}_{\omega,t}\psi_{0}^{(j)}e^{i\omega x}d\omega.
\end{aligned}
\end{equation}
Here, $\widetilde{\mathcal{F}}^{(j)}_{\omega,t}$ are the normalized spectra given by
\begin{equation}\label{F1}
\begin{aligned}
\widetilde{\mathcal{F}}^{(j)}_{\omega,t}=\frac{1}{2\pi}\int_{-\infty}^{\infty}\frac{\psi_{RW}^{(j)}(x,t)}{\psi_{0}^{(j)}}e^{-i\omega x}dx.
\end{aligned}
\end{equation}
Inserting the seed solution~(\ref{eq:seed solution}) into (\ref{Fourier3}), we obtain
\begin{equation}\label{Fourier5}
\begin{aligned}
\psi^{(j)}_{x,t}=\int_{-\infty}^{\infty} a_{j}\widetilde{\mathcal{F}}^{(j)}_{\omega,t}e^{ik_{j}t}e^{i(\omega+\beta_{j})x}d\omega.
\end{aligned}
\end{equation}
Comparing Eqs.~(\ref{Fourier3}) and (\ref{Fourier5}), we have
\begin{equation}\label{Fourier4}
\begin{aligned}
\mathcal{F}^{(j)}_{\omega,t}=a_{j}\widetilde{\mathcal{F}}^{(j)}_{\omega,t}e^{ik_{j}t}.
\end{aligned}
\end{equation}
Eq.~(\ref{Fourier4}) provides a simple transformation between the spectrum $\mathcal{F}^{(j)}_{\omega,t}$ and the normalized spectrum $\widetilde{\mathcal{F}}^{(j)}_{\omega,t}$. Now, we perform a change of variable to the complex one in the normalized spectrum (\ref{F1}): $x\rightarrow \mathcal{Z}$, so that
\begin{equation}\label{F2}
\begin{aligned}
\widetilde{\mathcal{F}}^{(j)}_{\omega,t}=\frac{1}{2\pi}\int_{-\infty}^{\infty}(1+\mathcal{J}_{j})e^{-i\omega \mathcal{Z}}d\mathcal{Z}.
\end{aligned}
\end{equation}
Eq.~(\ref{F2}) consists of the Dirac delta contribution from the background and the continuous spectrum generated by the localized perturbation of the RW, i.e., $\widetilde{\mathcal{F}}^{(j)}_{\omega,t}=\delta(\omega)+\mathcal{I}^{(j)}(\omega,t)$. Taking into account that `1' and the Dirac delta function are the Fourier transform pairs [i.e., $1\leftrightarrow \delta(\omega)$], the nontrivial part of the integral (\ref{F2}) is
\begin{equation}\label{Ij}
\begin{aligned}
\mathcal{I}^{(j)}=\frac{1}{2\pi}\int_{-\infty}^{\infty}\mathcal{J}_{j}e^{-i\omega \mathcal{Z}}d\mathcal{Z}.
\end{aligned}
\end{equation}
The roots of the binomial in the denominator of $\mathcal{J}_{j}$ are
\begin{equation}
\begin{aligned}
\mathcal{Z}_{1}=-\chi_{r}t+i\sqrt{(\chi_{i}t)^{2}+\frac{1}{4\chi_{i}^{2}}},\quad
\mathcal{Z}_{2}=-\chi_{r}t-i\sqrt{(\chi_{i}t)^{2}+\frac{1}{4\chi_{i}^{2}}}.
\end{aligned}
\end{equation}
The roots $\mathcal{Z}_{1}$ and $\mathcal{Z}_{2}$ are the first-order singularities for $\mathcal{J}_{j}$. Then, the residue $\mathcal{R}_{\mathcal{Z}}^{(j)}$ can be calculated using the formula
\begin{equation}\label{Rj}
\begin{aligned}
\mathcal{R}_{\mathcal{Z}_{1}}^{(j)}=\frac{i[2(\chi_{r}+\beta_{j})\sqrt{(\chi_{i}t)^{2}+\frac{1}{4\chi_{i}^{2}}}+1]-2\chi_{i}^{2}t}
{2i((\chi_{r}+\beta_{j})^{2}+\chi_{i}^{2})\sqrt{(\chi_{i}t)^{2}+\frac{1}{4\chi_{i}^{2}}}},\quad
\mathcal{R}_{\mathcal{Z}_{2}}^{(j)}=\frac{i[2(\chi_{r}+\beta_{j})\sqrt{(\chi_{i}t)^{2}+\frac{1}{4\chi_{i}^{2}}}-1]+2\chi_{i}^{2}t}
{2i((\chi_{r}+\beta_{j})^{2}+\chi_{i}^{2})\sqrt{(\chi_{i}t)^{2}+\frac{1}{4\chi_{i}^{2}}}}.
\end{aligned}
\end{equation}
Before using the residue theorem, let us pay more attention to the imaginary parts of $\mathcal{Z}_{1}$ and $\mathcal{Z}_{2}$. Namely, substituting $\mathcal{Z}_{1,2}$ into Eq.~(\ref{Rj}) and applying the residue theorem yields
\begin{equation}
\mathcal{I}^{(j)}
=\left\{
\begin{aligned}
\mathrm{i}\,\mathcal{R}_{\mathcal{Z}_{1}}^{\,j}
\mathrm{e}^{-\mathrm{i}\omega \mathcal{Z}_{1}},
& \operatorname{sgn}(\omega)
=-\operatorname{sgn}(\chi_i),\\[6pt]
\mathrm{i}\,\mathcal{R}_{\mathcal{Z}_{2}}^{\,j}
\mathrm{e}^{-\mathrm{i}\omega \mathcal{Z}_{2}},
& \operatorname{sgn}(\omega)
=\operatorname{sgn}(\chi_i).
\end{aligned}
\right.
\end{equation}
Collecting the results above, we have
\begin{equation}
\begin{aligned}
\widetilde{\mathcal{F}}^{(j)}=\mathcal{I}^{(j)}+\delta(\omega).
\end{aligned}
\end{equation}
The delta function $\delta(\omega)$ appears due to the infinite size of the background. Removing it does not affect the spectra. Using
the relation $\operatorname{sgn}(\omega)\operatorname{sgn}(\chi_i)=\operatorname{sgn}(\omega\chi_i)$, we obtain the spectra of the vector RWs:
\begin{equation}\label{IJJ}
\begin{aligned}
\mathcal{I}^{(j)}=\frac{e^{-\omega\frac{1+4(\chi_{i}v_{i}t)^{2}}{2\chi_i}}+i\omega v_{r}t}{(\chi_{r}+\beta_{j})^{2}+\chi_{i}^{2}}
[(\chi_{r}+\beta_{j})\operatorname{sgn}(\omega)-\frac{1+2i\chi_{i}v_{i}t}{4(\sqrt{1+4(\chi_{i}v_{i}t)^{2}}{\chi_i})}].
\end{aligned}
\end{equation}
After applying $\chi_i\operatorname{sgn}(\chi_i)=|\chi_i|$, the modulus of Eq.~(\ref{IJJ}) becomes
\begin{equation}\label{IJJ1}
\begin{aligned}
\left|\mathcal{I}_{j}(\omega,t)\right|
=
\begin{cases}
\left|\Phi_{0j}-\Phi_{sj}(t)\right|
\mathrm{e}^{\omega L(t)},
& \omega<0,\\[6pt]
\left|\Phi_{0j}+\Phi_{sj}(t)\right|
\mathrm{e}^{-\omega L(t)},
& \omega>0,
\end{cases}
\end{aligned}
\end{equation}
where
\begin{equation}
\begin{aligned}
\Phi_{0j}=-\frac{\chi_{r}+\beta_{j}}{\gamma_{j}}, \quad \Phi_{sj}(t)=\frac{\chi_{i}(1+2i\chi_{i}v_{i}t)}{2\gamma_{j}\sqrt{1+4(\chi_{i}v_{i}t)^{2}}{\chi_i}}, \quad
L(t)=\frac{\sqrt{1+4(\chi_{i}v_{i}t)^{2}}}{2\chi_i}.
\end{aligned}
\end{equation}

\begin{figure}[H]
	\centering
{\includegraphics[width=400 bp,height=3 cm]{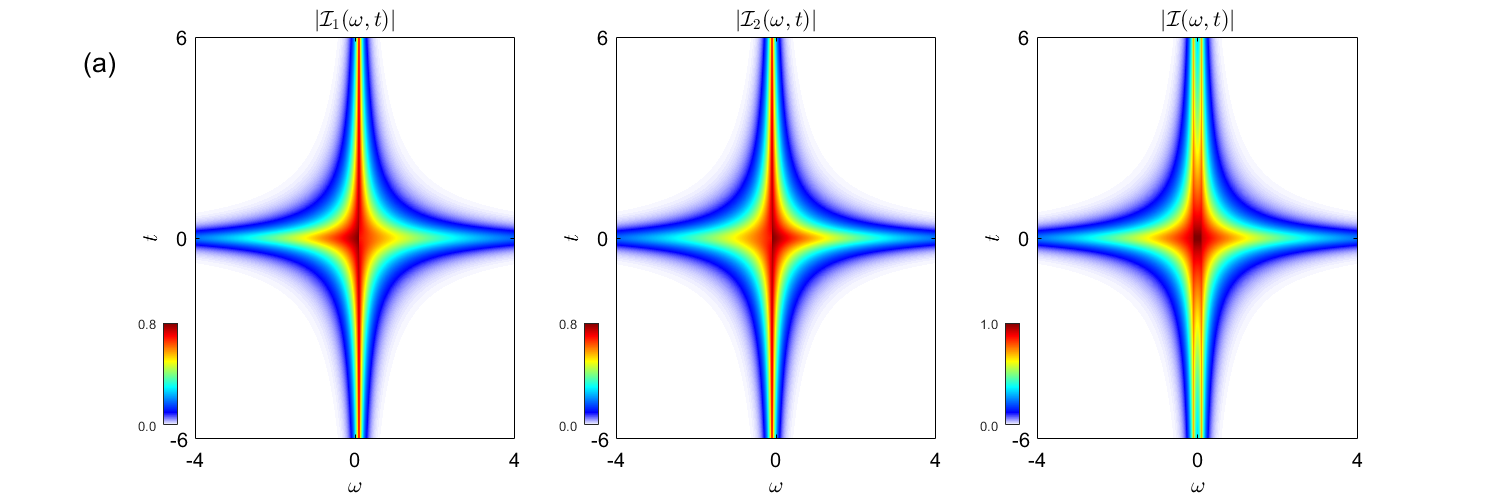}}
{\includegraphics[width=400 bp,height=3 cm]{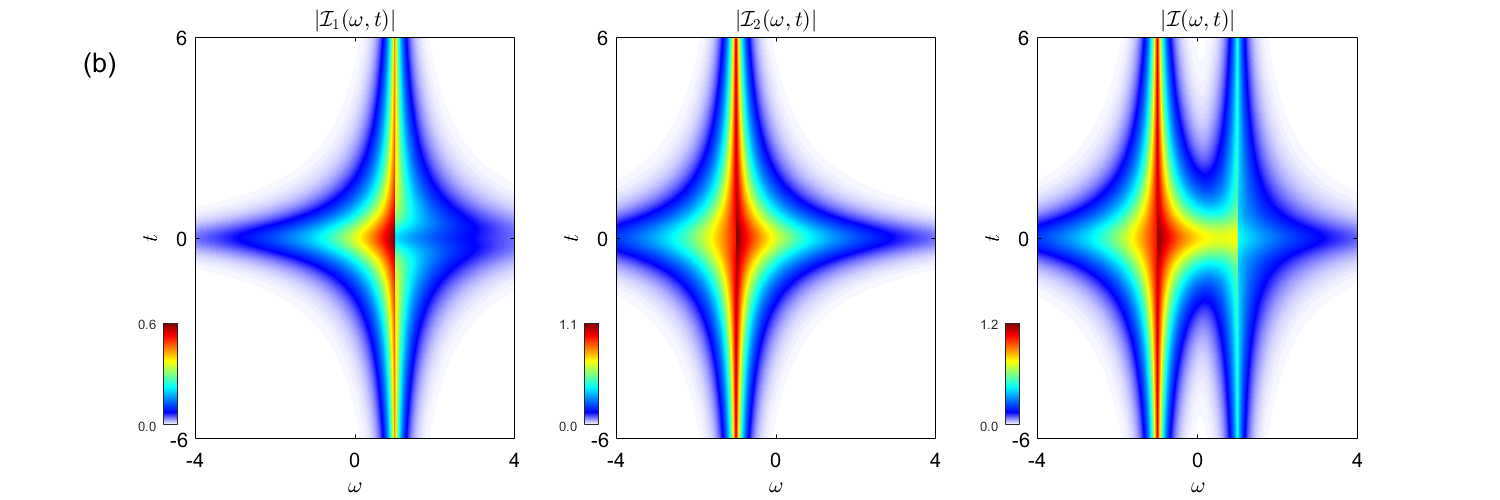}}
\caption{\footnotesize Evolution of different spectral components $|\mathcal{I}_{1}(\omega,t)|$, $|\mathcal{I}_{2}(\omega,t)|$ and the total spectral components $|\mathcal{I}(\omega,t)|=\sqrt{|\mathcal{I}_{1}(\omega,t)|^{2}+|\mathcal{I}_{2}(\omega,t)|^{2}}$ for (a) the degenerate ($\beta=0.1$) and (b) the non-degenerate ($\beta=1$) RWs given by Eqs.~(\ref{IJJ1}) on $\omega-t$ plane. These spectra correspond to vector RW solutions in Figs.~\ref{RW}.}
	\label{RW_PU}
\end{figure}

The spatially localized RWs in Figs.~\ref{RW} evolve in time and reach its maximum amplitude at a specific instant. Correspondingly, their spectra undergo expansion-contraction process. In the infinite period limit, the RW spectra exhibit a characteristic triangular profile with equal slopes on both sides and reach the maximum widths at the instant of strongest localizations~\cite{non-RW1}. Figs.~\ref{RW_PU} illustrate the spectral evolutions of the vector RWs. For each component, the spectra at negative and positive times are not individually symmetric under the frequency inversion $\omega\rightarrow-\omega$, although two spectral edges retain equal slopes. Instead, the degenerate RW exhibits a combined frequency inversion and component exchange symmetry, namely, $\left|\mathcal{I}_{1}(\omega,t)\right|=\left|\mathcal{I}_{2}(-\omega,t)\right|$. And their total spectra are symmetric about $(\omega=0)$. They reflect the similar dynamical behaviors of two wave modes, as shown in Figs.~\ref{RW_PU}~(a).

Conversely, this symmetry is broken for the non-degenerate RWs because two wave modes are associated with distinct eigenvalues and therefore exhibit different physical properties. The two components exhibit different spectral distributions, as illustrated in Figs.~\ref{RW_PU}~(b). The spectral symmetry therefore provides a simple criterion for distinguishing the degenerate and non-degenerate RW solutions. The spectral centers are determined by $\omega+\beta_j$ and shift accordingly with increasing $\beta_j$. A small $\beta_j$ produces only a slight spectral shift from $\omega=0$ in the degenerate regions, whereas a larger $\beta_j$ leads to a pronounced displacement in the non-degenerate regions. This provides another important distinction between the degenerate and non-degenerate RW spectra. However, this feature was obscured by the variable transformation $\omega\rightarrow\omega+\beta_j$ in previous studies~\cite{non-RW1,pan1}. Compared with the symmetric spectra of the degenerate RWs, those in the non-degenerate RW spectra show the symmetric-breaking property in both the individual components and the total spectrum.

\begin{figure}[H]
	\centering
{\includegraphics[width=400 bp,height=3 cm]{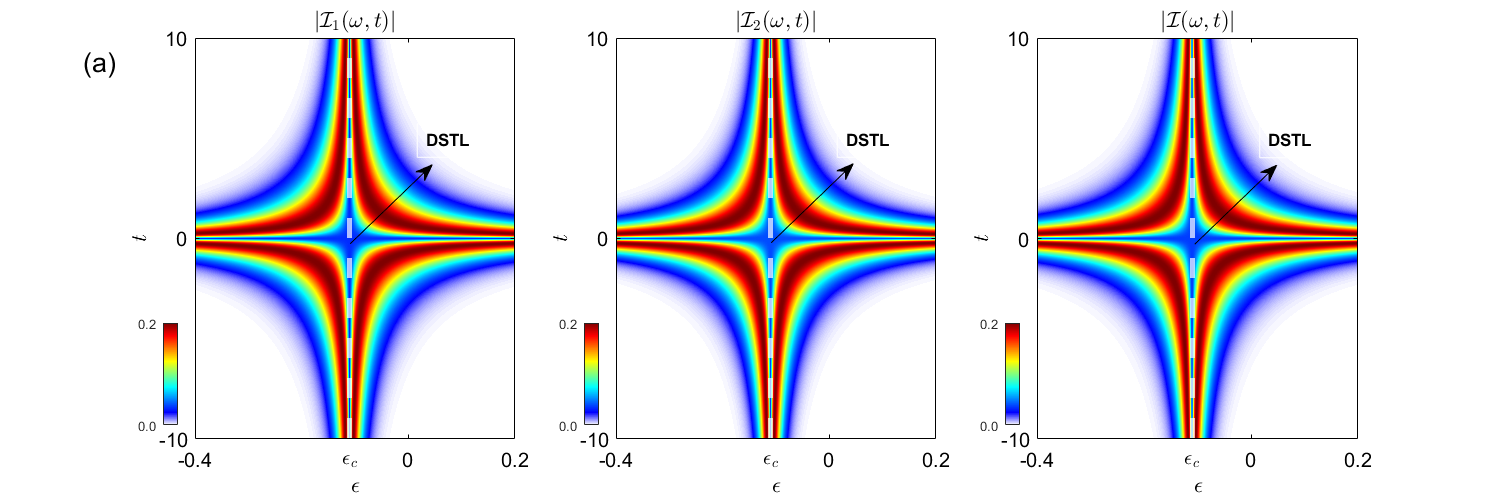}}
{\includegraphics[width=400 bp,height=3 cm]{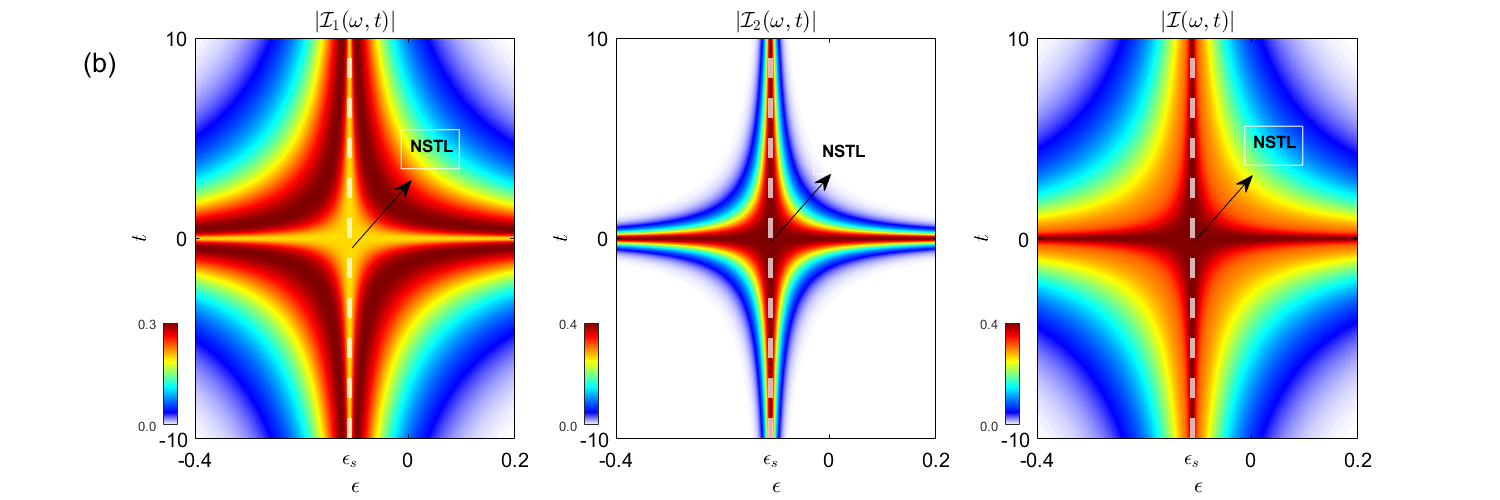}}
\caption{\footnotesize Evolution of different spectral components $|\mathcal{I}_{1}(\omega,t)|$, $|\mathcal{I}_{2}(\omega,t)|$ and $|\mathcal{I}(\omega,t)|$ of (a) the degenerate ($\beta=0.1$) and (b) the non-degenerate ($\beta=1$) RWs given by Eqs.~(\ref{IJJ}) on $\varepsilon-t$ plane. The white dashed lines $\varepsilon_c$ and $\varepsilon_s$ denote the static and moving solitons, while the short forms `DSTL' and `NSTL' represent the degenerate and non-degenerate state transition lines, respectively. Other parameters are $a=1, \omega=1.2$.}
	\label{RW_puso}
\end{figure}

Under the fourth-order effect, the physical spectra of vector RWs exhibit new dynamical behaviors. Figs.~\ref{RW_puso} show their spectral evolution on the $\varepsilon-t$ plane. As $t\rightarrow\pm\infty$, the RW perturbation vanishes and the nontrivial spectral contribution tends to zero. At $t=0$, the RW solutions reach the maximum amplitude, while the spectra attain the maximum width. More importantly, the state transition conditions are identified in both the degenerate and non-degenerate spectra domain. At $\varepsilon=\varepsilon_{c}$, the degenerate RWs convert into the static solitons, as marked by the white dashed lines in Figs.~\ref{RW_puso}~(a). For the non-degenerate case, the transition at $\varepsilon=\varepsilon_{s}$ yields the moving solitons, as illustrated in Figs.~\ref{RW_puso}~(b). This behavior fundamentally differs from that of the CH equations~\cite{pan1}, where the transition conditions can be identified only in the non-degenerate spectra domain. The spectral symmetry in $\varepsilon-t$ plane differs from that in Figs.~\ref{RW_PU}. The degenerate spectra are fully symmetric about $\varepsilon_{c}$. By comparison, the non-degenerate spectra show no corresponding symmetry about $\varepsilon_{s}$. This indicates that the fourth-order effect exhibits a pronounced symmetry under specific parameter conditions in spectral domain.

\subsection{BSs spectra}
We next consider the physical spectra of the vector BSs. They are given by the Fourier transform:
\begin{equation}\label{eq:beatingpu1}
\begin{aligned}
\mathcal{A}_{j}(\omega,t)=\frac{1}{2\pi}\int_{-\infty}^{\infty} \psi^{(j)}(x,t)e^{-i\omega x}dx.
\end{aligned}
\end{equation}
We rewrite the solutions (\ref{DT1}) with $\beta_{1}=\beta_{2}=\beta=0$ in the form
\begin{equation}
\begin{aligned}
\psi^{(j)}=\psi_{0}^{(j)}[1+(\lambda^{*}-\lambda)\psi_{k}^{(j)}(x,t)],
\end{aligned}
\end{equation}
where the new function $\psi_{k}^{(j)}(x,t)$ is given by
\begin{equation}
\begin{aligned}
\psi_{k}^{(j)}(x,t)=\frac{\pm B_{1}(t)e^{-i\chi_{kr}x}+B_{2}(t)e^{-\chi_{ki}x}}{D_{1}(t)e^{\chi_{ki}x}+D_{2}(t)e^{-\chi_{ki}x}},\quad k=a,b,
\end{aligned}
\end{equation}
with
\begin{equation}
\begin{aligned}
&B_{1}(t)=\frac{1}{a}e^{-i\Gamma_{kr}t},\quad
B_{2}(t)=\frac{1}{\chi_{kr}}e^{-\Gamma_{ki}t},\\
&D_{1}(t)=2e^{\Gamma_{ki}t},\quad
D_{2}(t)=(1+\frac{2a^{2}}{|\chi_{k}|^{2}})e^{-\Gamma_{ki}t},\\
&\Gamma_{k}=-\varepsilon\chi_{k}^{4}+(\frac{1}{2}+8\varepsilon a^{2})\chi_{k}^{2}.
\end{aligned}
\end{equation}
Here, $\Gamma_{kr}$ governs the temporal oscillation of the beating pattern, while $\Gamma_{ki}$ controls the motion of the envelope in the $x-t$ plane. The envelope velocity and period of the BS spectra are given by $V_{g}=-\frac{\Gamma_{ki}}{\chi_{ki}}$ and $T_{sp}=\frac{2\pi}{|\Gamma_{kr}|}$, respectively. The integral (\ref{eq:beatingpu1}) has its trivial part which is the Dirac delta function $\delta(\omega)$ due to the presence of the infinite background $\psi_{0}^{(j)}$. It will be omitted below. The nontrivial part of the integral (\ref{eq:beatingpu1}) is
\begin{equation}\label{eq:beatingpu2}
\begin{aligned}
I_{j}=\frac{1}{2\pi}\int_{-\infty}^{\infty} \psi_{k}^{(j)}(t,x)e^{-i\omega x}dx.
\end{aligned}
\end{equation}
The integral (\ref{eq:beatingpu2}) can be calculated analytically using the residue theorem. Namely, $I=2\pi iR$, where $R$ is the residue
of the corresponding singularity of $\psi_{k}^{(j)}(x,t)$ in $x$. The complex function $\psi_{k}^{(j)}(t,x)$ has two singularities $x_{1}$ and $x_{2}$ which are given by
\begin{equation}
\begin{aligned}
x_{1}=\frac{1}{\chi_{ki}}\log[-i\frac{\sqrt{D_{2}}}{\sqrt{D_{1}}}],\quad
x_{2}=\frac{1}{\chi_{ki}}\log[i\frac{\sqrt{D_{2}}}{\sqrt{D_{1}}}], \quad k=a,b.
\end{aligned}
\end{equation}
The explicit expressions for the corresponding residues at $x=x_{1}$ and $x=x_{2}$, i.e., $R_{x_{1}}$ and $R_{x_{2}}$, are given by
\begin{equation}
\begin{aligned}
&R_{x_{1}}^{(j)}=\frac{(B_{2}\sqrt{D_{1}}e^{-i\chi_{kr}x_{1}}\mp iB_{1}\sqrt{D_{2}})e^{-i(\chi_{kr}+\omega)x_{1}}}{2\sqrt{D_{1}}D_{2}\chi_{ki}},\\
&R_{x_{2}}^{(j)}=\frac{(B_{2}\sqrt{D_{1}}e^{-i\chi_{kr}x_{2}}\pm iB_{1}\sqrt{D_{2}})e^{-i(\chi_{kr}+\omega)x_{2}}}{2\sqrt{D_{1}}D_{2}\chi_{ki}}.
\end{aligned}
\end{equation}
One of the singularities, $x_{1}$ is located on the upper complex half-plane while the other one, $x_{2}$ is located on the lower one. Then, $I=2\pi iR_{x_{1}}$ for $\omega<0$, while $I=2\pi iR_{x_{2}}$ for $\omega>0$. Thus, the exact analytic expressions for BS spectra
can be written as
\begin{equation}\label{BSPU1}
\begin{aligned}
\mathcal{A}_{j}(\omega,t)&=i\psi_{0}^{(j)}[\lambda^{*}-\lambda]R_{x_{1}}^{(j)},\quad \omega<0,\\
\mathcal{A}_{j}(\omega,t)&=i\psi_{0}^{(j)}[\lambda^{*}-\lambda]R_{x_{2}}^{(j)}, \quad \omega>0.
\end{aligned}
\end{equation}

\begin{figure}[H]
	\centering
{\includegraphics[width=400 bp,height=3 cm]{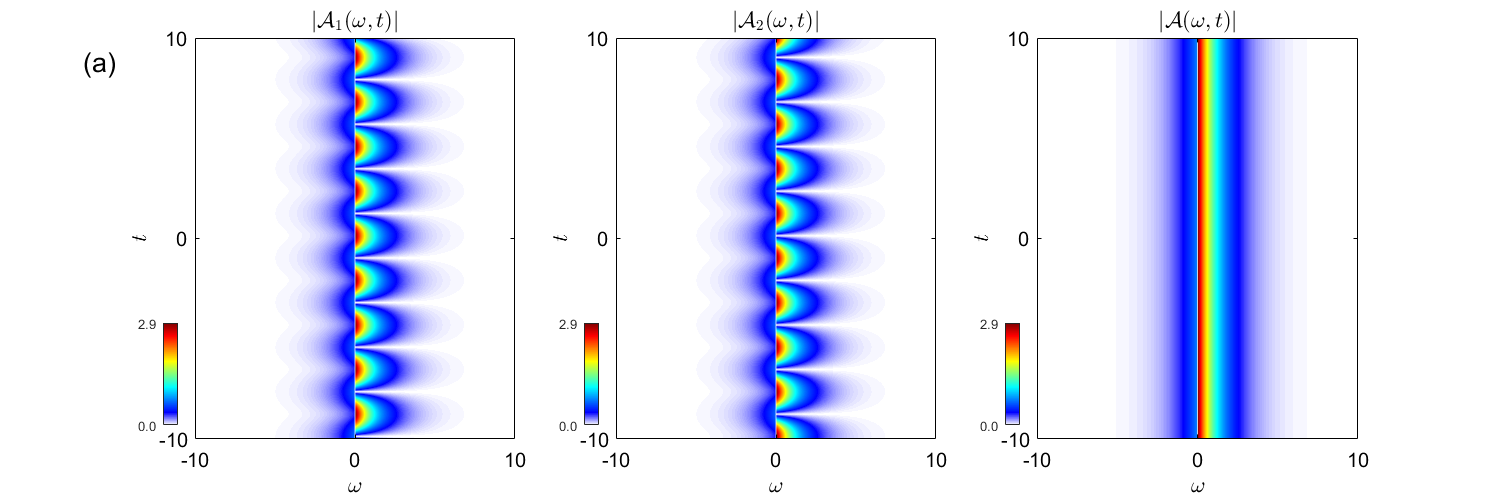}}
{\includegraphics[width=400 bp,height=3 cm]{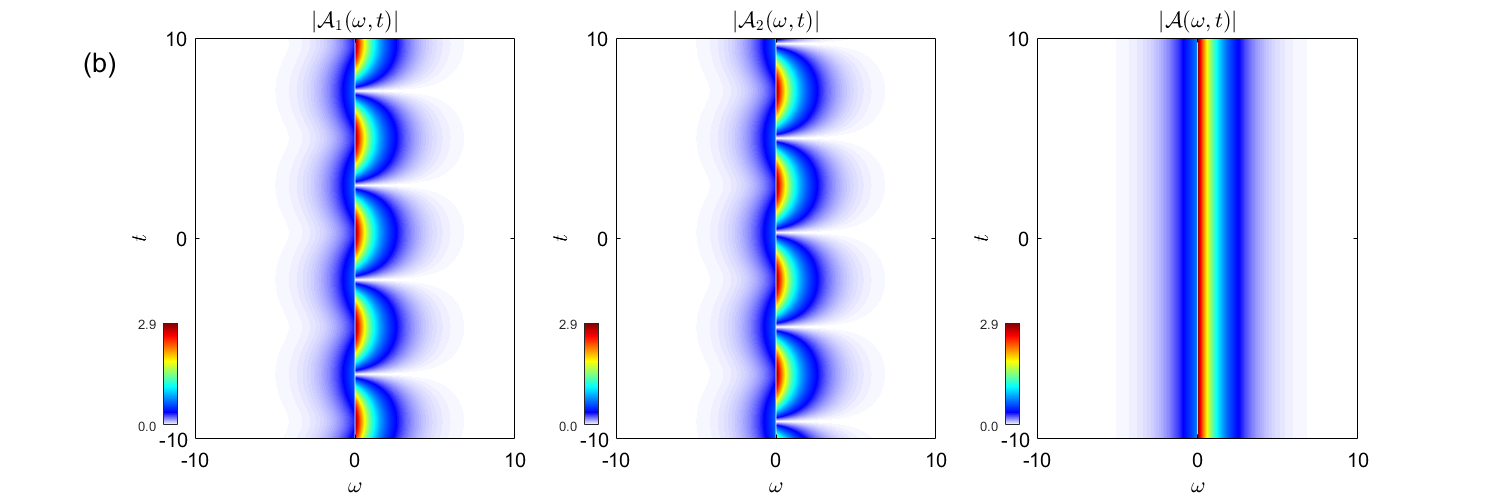}}
\caption{\footnotesize Evolution of different spectral components $|\mathcal{A}_{1}(\omega,t)|$, $|\mathcal{A}_{2}(\omega,t)|$ and the total spectral components $|\mathcal{A}(\omega,t)|=\sqrt{|\mathcal{A}_{1}(\omega,t)|^{2}+|\mathcal{A}_{2}(\omega,t)|^{2}}$ of (a) Type-I and (b) Type-II moving BSs with $\omega=1, \alpha=1$ which are given by Eq.~(\ref{BSPU1}) on $\omega-t$ plane. Other parameters are $a=1, \varepsilon=0.1$.}
	\label{BSPU_m}
\end{figure}

\begin{figure}[H]
	\centering
{\includegraphics[width=400 bp,height=3 cm]{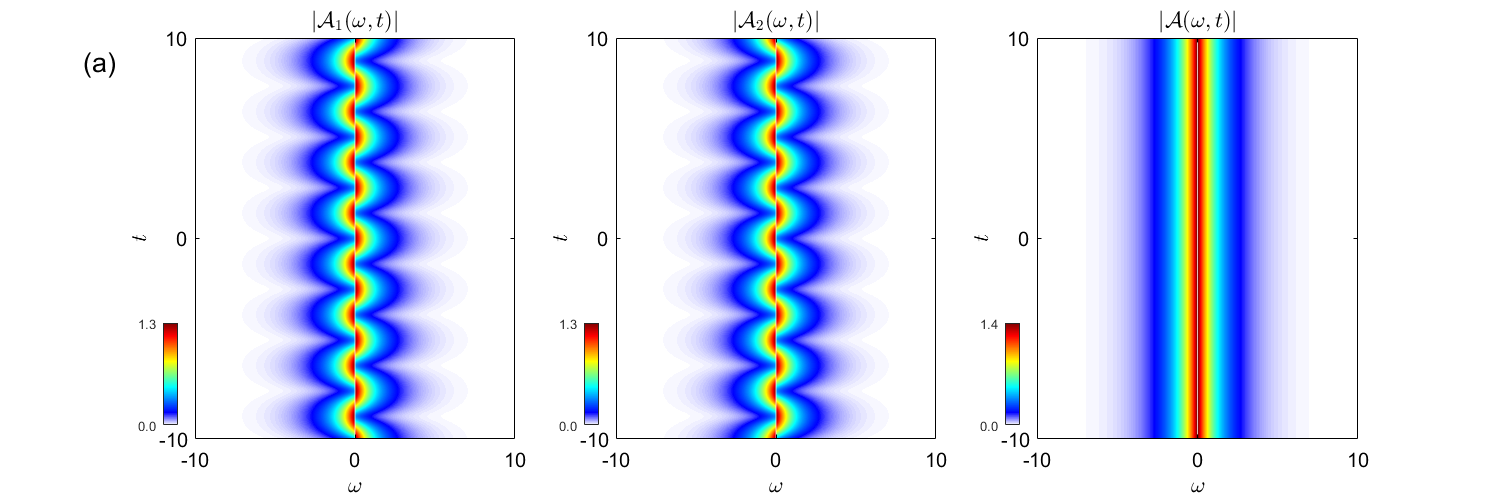}}
{\includegraphics[width=400 bp,height=3 cm]{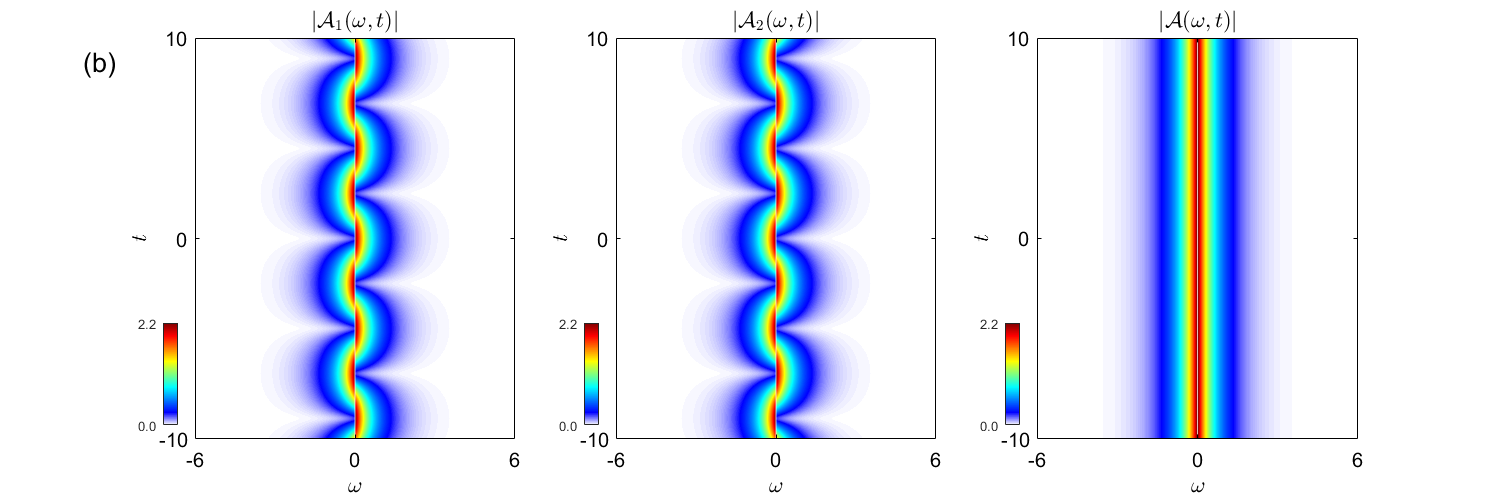}}
\caption{\footnotesize Evolution of different spectral components $|\mathcal{A}_{1}(\omega,t)|$, $|\mathcal{A}_{2}(\omega,t)|$ and $|\mathcal{A}(\omega,t)|$ of (a) Type-I and (b) Type-II static BSs with $\omega=0, \alpha=1$, which are given by the Eq.~(\ref{BSPU1}) on $\omega-t$ plane. Other parameters are $a=1, \varepsilon=0.1$.}
	\label{BSPU_t}
\end{figure}

Figs.~\ref{BSPU_m} show the spectral evolutions of the moving Type-I and Type-II BSs associated with the profiles in Figs.~\ref{beating1}. The static counterparts are depicted in Figs.~\ref{BSPU_t}, corresponding to Figs.~\ref{beating3}. For both types BSs, the spectra of the individual components exhibit periodic oscillations in $t$, whereas the total spectrum is non-periodic. Moreover, the total spectra provide clear distinctions between the moving and static BSs. The moving BSs are asymmetric under $\omega\rightarrow-\omega$ but those of the static BSs are symmetric. These spectral characteristics differ markedly from those of the vector breathers reported previously~\cite{15,16}.

\begin{figure}[H]
	\centering
{\includegraphics[width=200 bp,height=3 cm]{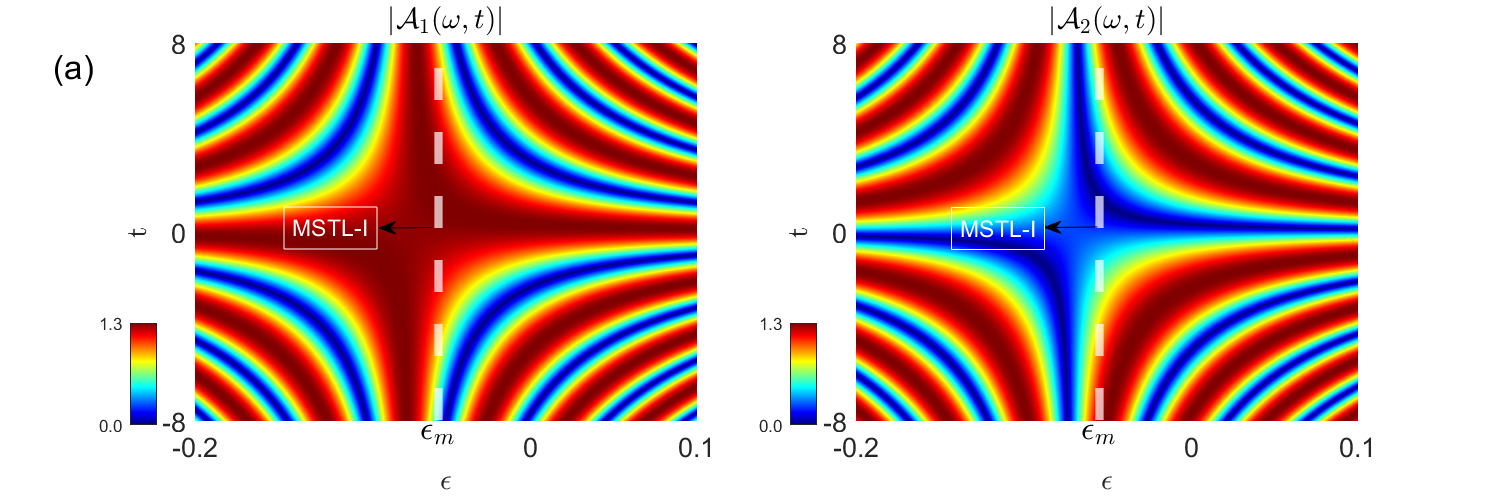}}
{\includegraphics[width=200 bp,height=3 cm]{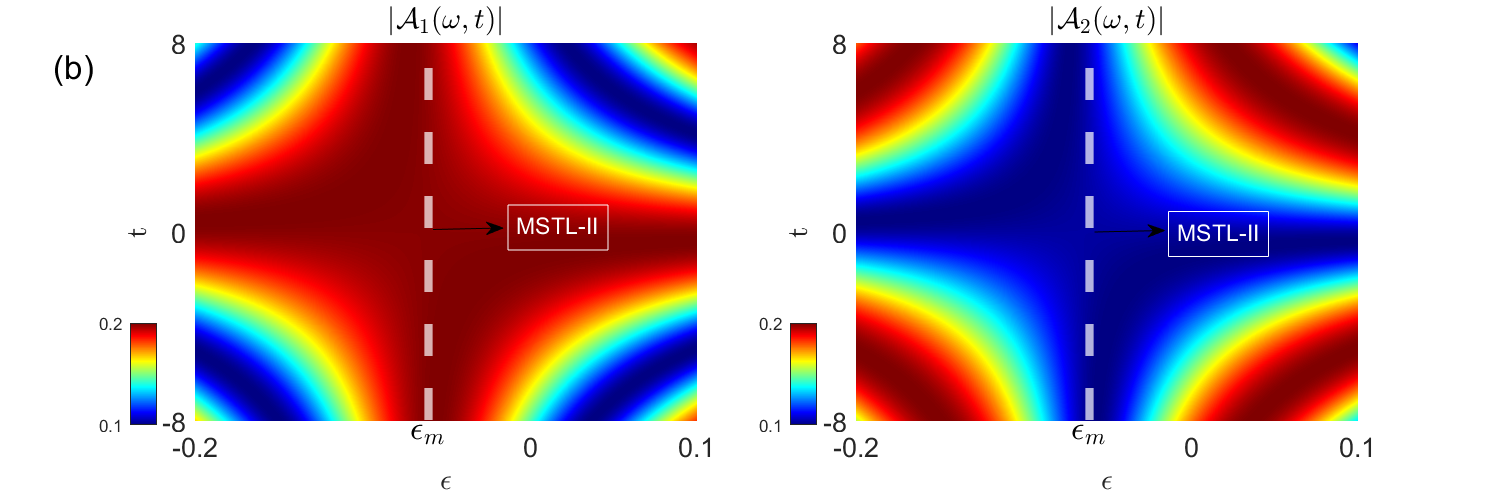}}
{\includegraphics[width=200 bp,height=3 cm]{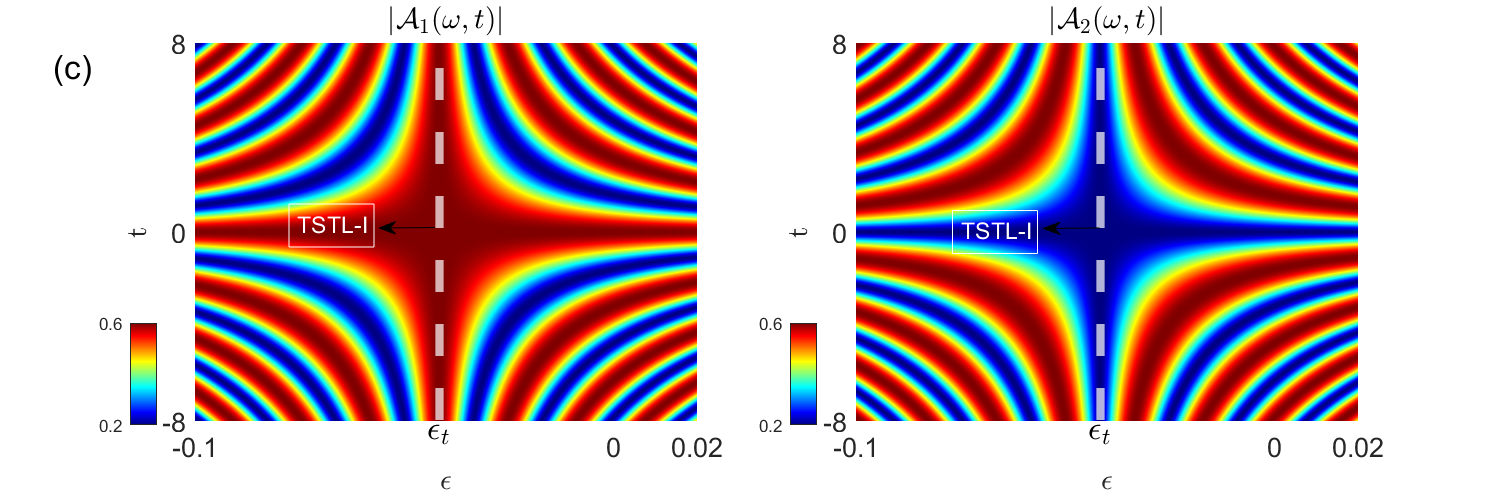}}
{\includegraphics[width=200 bp,height=3 cm]{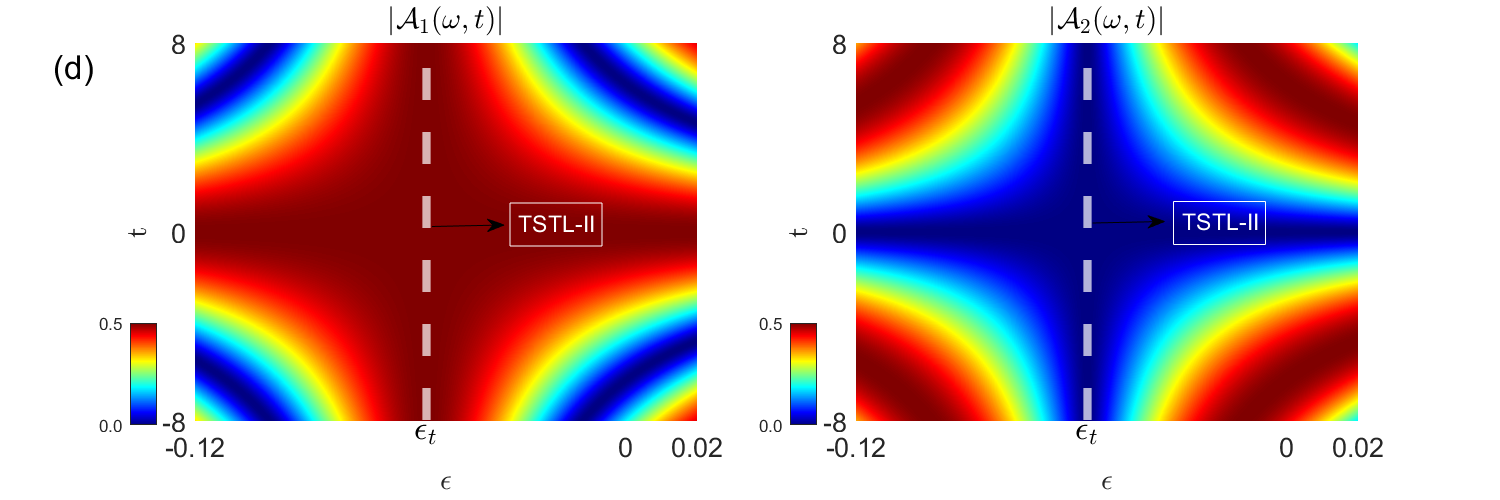}}
\caption{\footnotesize Evolution of different spectral components $|\mathcal{A}_{1}(\omega,t)|$ and $|\mathcal{A}_{2}(\omega,t)|$ on $\varepsilon-t$ plane. The white dashed lines $\varepsilon_m$ and $\varepsilon_t$ denote the moving and static kink solitons, while the short forms `MSTL' and `TSTL' represent the moving and static state transition lines (-I and II are type of BSs), respectively. Other parameters are the same as those in Figs.~\ref{BSPU_m} and \ref{BSPU_t}.}
	\label{BSPU_ST}
\end{figure}

Vector BSs can be transformed into the kink solitons, and the corresponding transition conditions allow to be identified in the spectral domain on $\varepsilon-t$ plane. For the moving BSs, the transition conditions $\varepsilon=\varepsilon_{m}$ are marked by the white dashed line labeled `MSTL'. While for static BS, the conditions $\varepsilon=\varepsilon_{t}$ are indicated by the line labeled `TSTL'. Remarkably, the total BS components already exhibit soliton-like structures, so no corresponding transition conditions appear in the spectral domain. They are therefore omitted for brevity. As observed in Figs.~\ref{BSPU_ST}, the static BS spectra are symmetric about the transition line $\varepsilon_{t}$, whereas the moving BS spectra are asymmetric respect to $\varepsilon_{m}$. Compared with the symmetric spectra of the static BSs, those in the moving BS spectra show the symmetric-breaking property in $\varepsilon-t$ plane. This result reveals a pronounced spectral symmetry induced by the fourth-order effect under specific parameter conditions.

\vspace{3mm}
\section{Numerical simulations} \label{eq:sec7}
\begin{figure}[H]
	\centering
{\includegraphics[width=400 bp,height=3 cm]{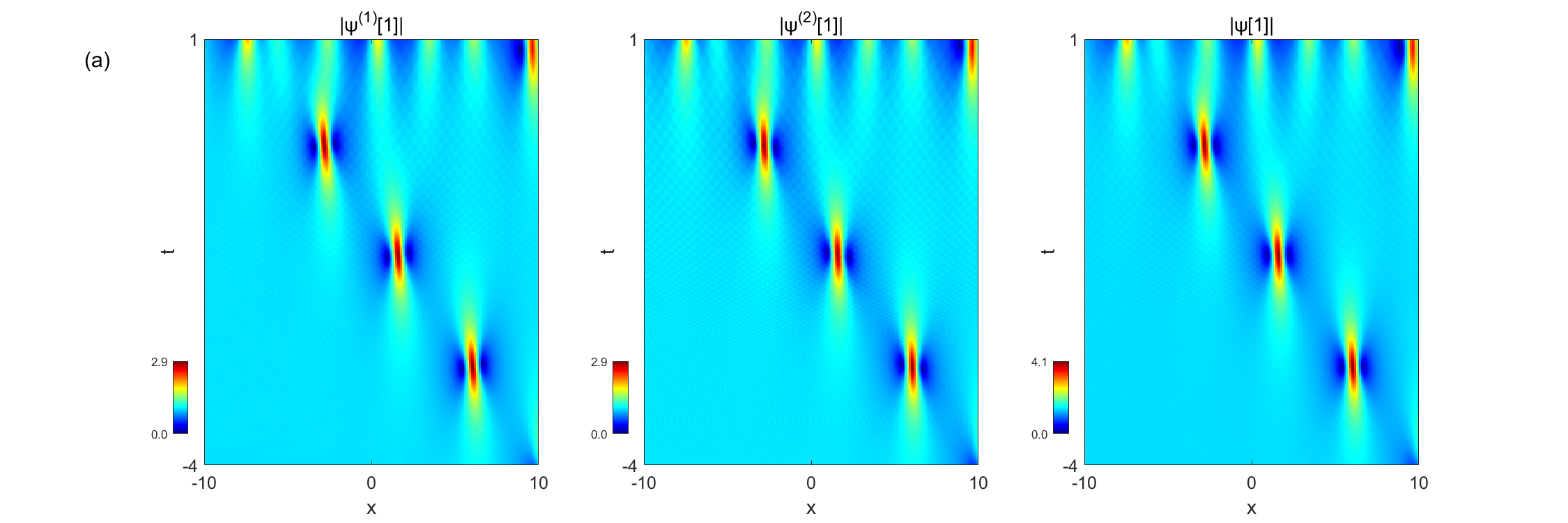}}
{\includegraphics[width=400 bp,height=3 cm]{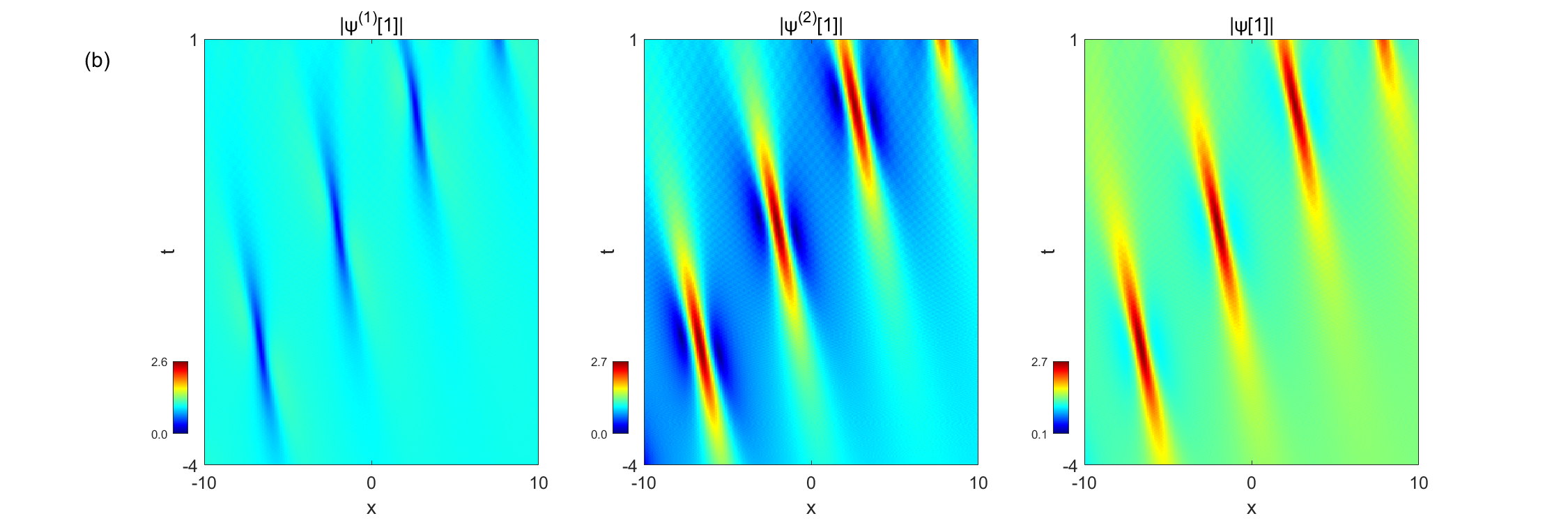}}
\caption{\footnotesize Numerical simulations of the (a) degenerate ($\beta=0.1$) and (b) non-degenerate ($\beta=1$) TWs developed from the exact solutions (\ref{eq:GB}) with $\varepsilon=0.01$. Other parameters are chosen in Figs.~\ref{GB}}
  \label{smgb}
\end{figure}

\begin{figure}[H]
	\centering
{\includegraphics[width=400 bp,height=3 cm]{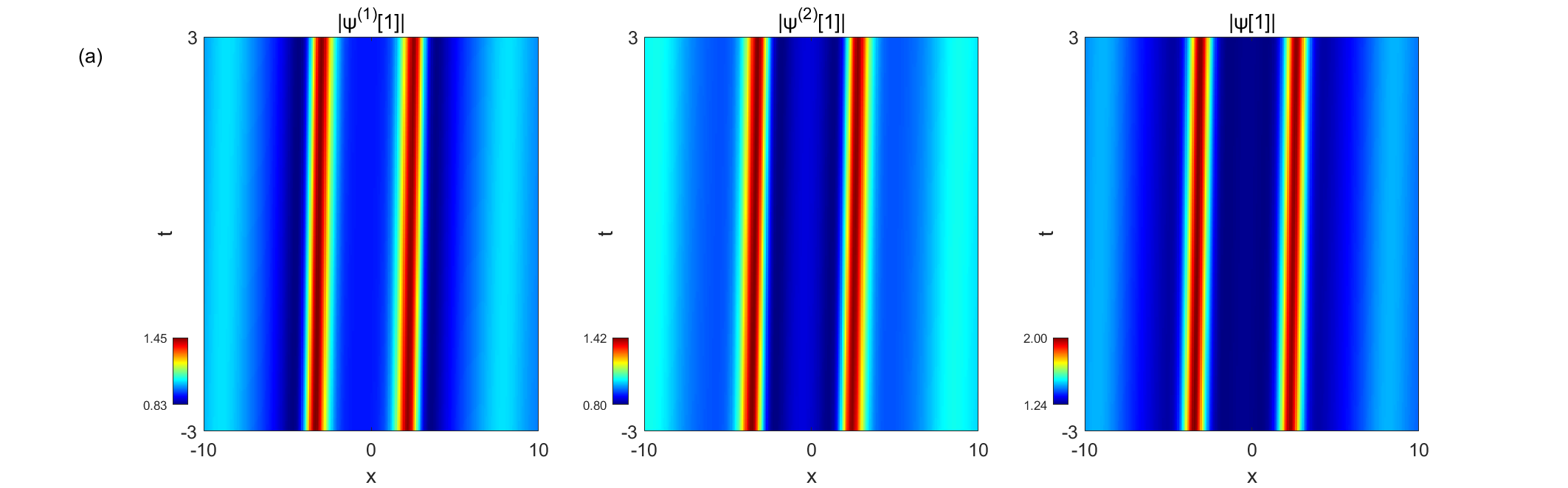}}
{\includegraphics[width=400 bp,height=3 cm]{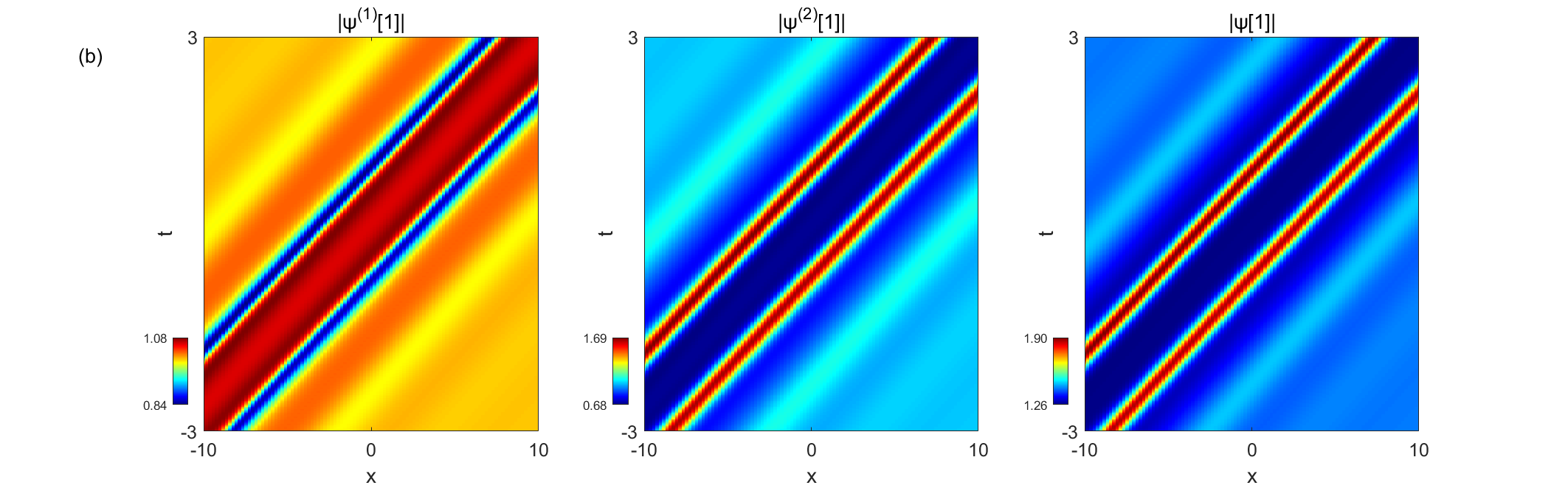}}
\caption{\footnotesize Numerical simulations of the multipeak solitons from the (a) degenerate ($\beta=0.1$) and (b) non-degenerate ($\beta=1$) TW breathers developed by Eq.~(\ref{eq:GB}) with $\varepsilon_{tr}$. Other parameters are chosen in Figs.~\ref{abtsGB}.}
  \label{smgb1}
\end{figure}

\vspace{2mm}
In this section, we use the initial conditions to excite both the vetcor degenerate and non-degenerate localized wave solutions, including TWs, RWs and BSs. An important consideration from an experimental perspective is what kinds of initial conditions can excite these vector localized waves. Clearly, our exact solutions (\ref{eq:GB}), (\ref{eq:RW}) and (\ref{BS}) provide ideal initial conditions at any given $t$. Here $t\in R$. Therefore, our numerical results confirm the excitation of both the localized waves and their transformed solitons under the fourth-order effect. Below, we present numerical results that confirm the analytical predictions. By using Eq.~(\ref{eq:GB}), the degenerate and non-degenerate TWs and their transformed solitons are excited in Figs.~\ref{smgb} and \ref{smgb1}.

\begin{figure}[H]
	\centering
{\includegraphics[width=400 bp,height=3 cm]{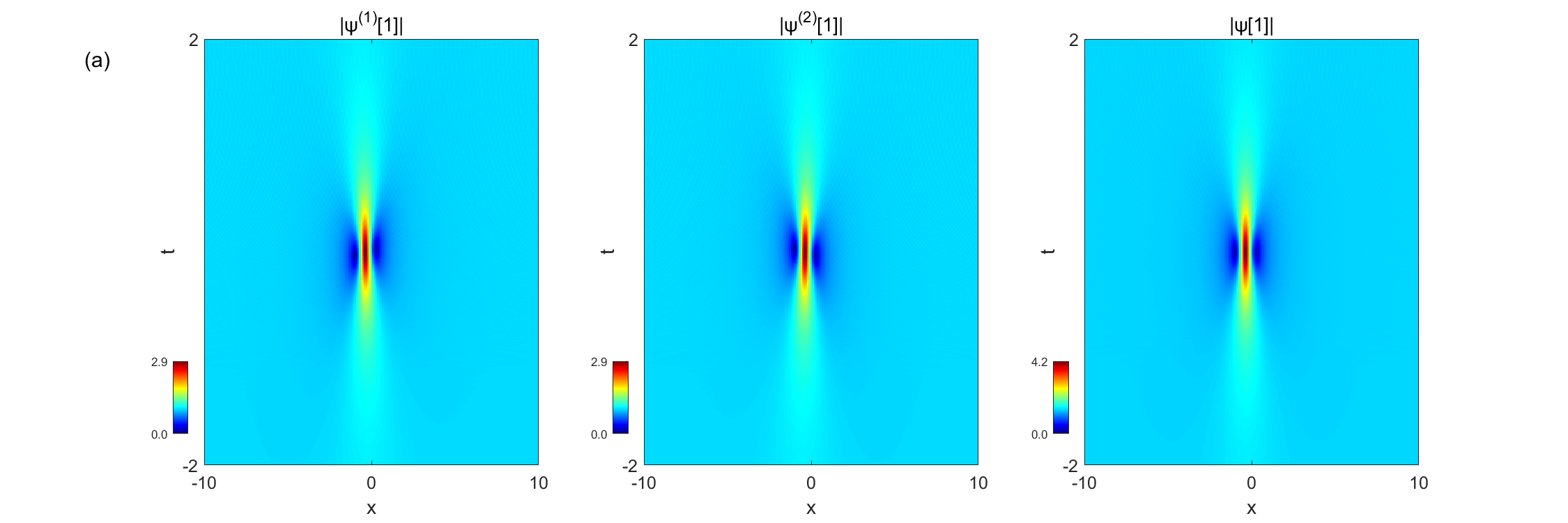}}
{\includegraphics[width=400 bp,height=3 cm]{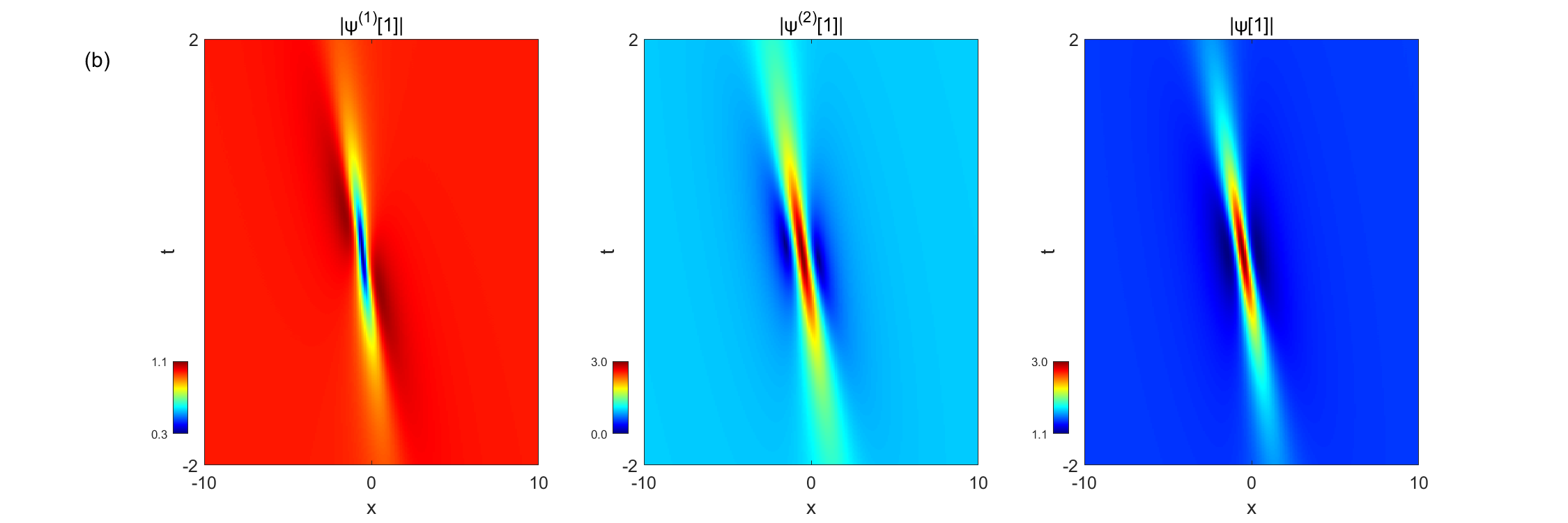}}
\caption{\footnotesize Numerical simulations of the (a) degenerate ($\beta=0.1$) and (b) non-degenerate ($\beta=1$) RWs developed from the exact solutions (\ref{eq:RW}) with $\varepsilon=0.01$. Other parameters are chosen in Figs.~\ref{RW}.}
  \label{smrw}
\end{figure}

\begin{figure}[H]
	\centering
{\includegraphics[width=400 bp,height=3 cm]{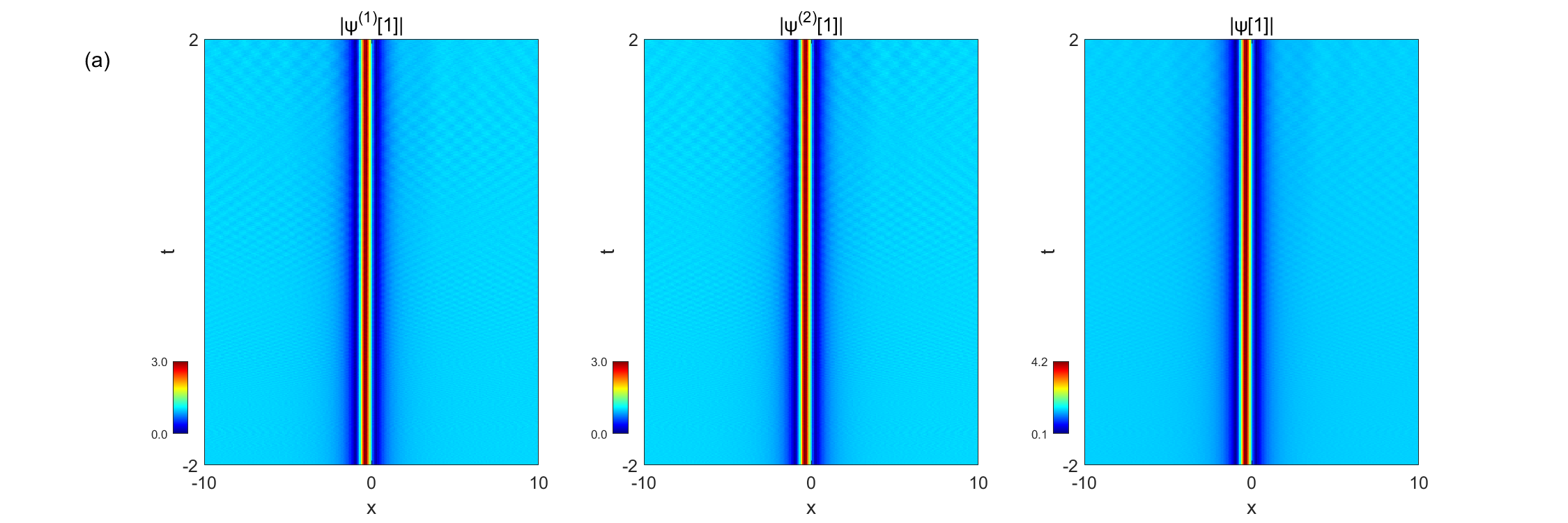}}
{\includegraphics[width=400 bp,height=3 cm]{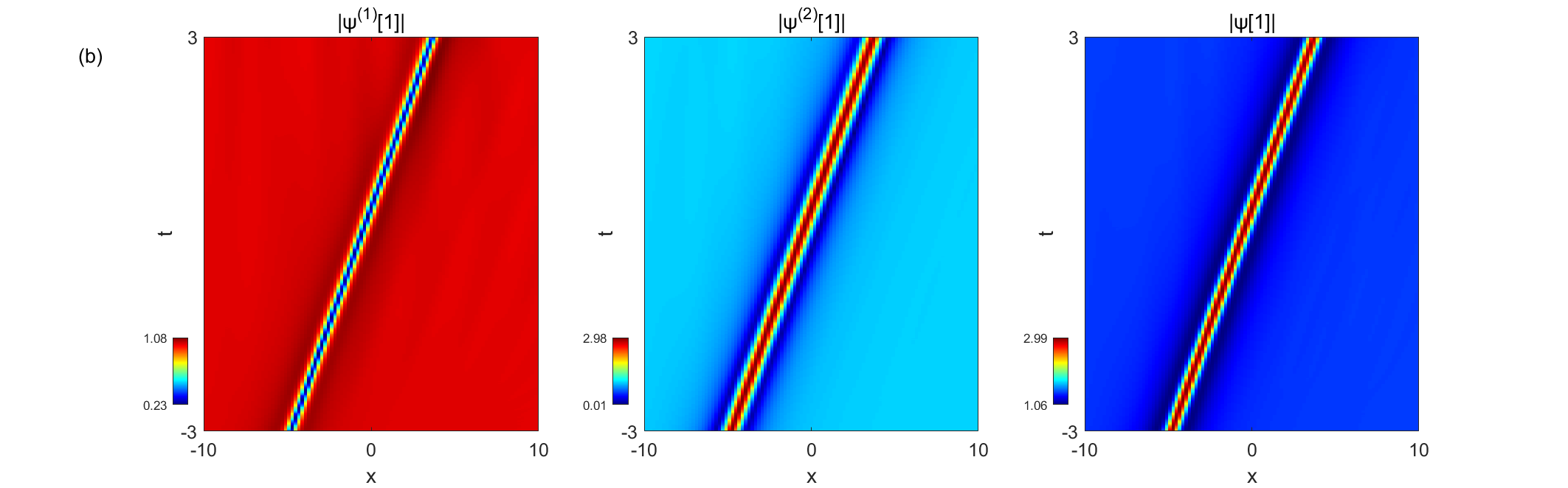}}
\caption{\footnotesize Numerical simulations of the rational solitons from the (a) degenerate ($\beta=0.1$) and (b) non-degenerate ($\beta=1$) RWs developed by Eq.~(\ref{eq:RW}) with $\varepsilon_{s}$. Other parameters are chosen in Figs.~\ref{abts1}.}
  \label{smrw1}
\end{figure}

\begin{figure}[H]
	\centering
{\includegraphics[width=400 bp,height=3 cm]{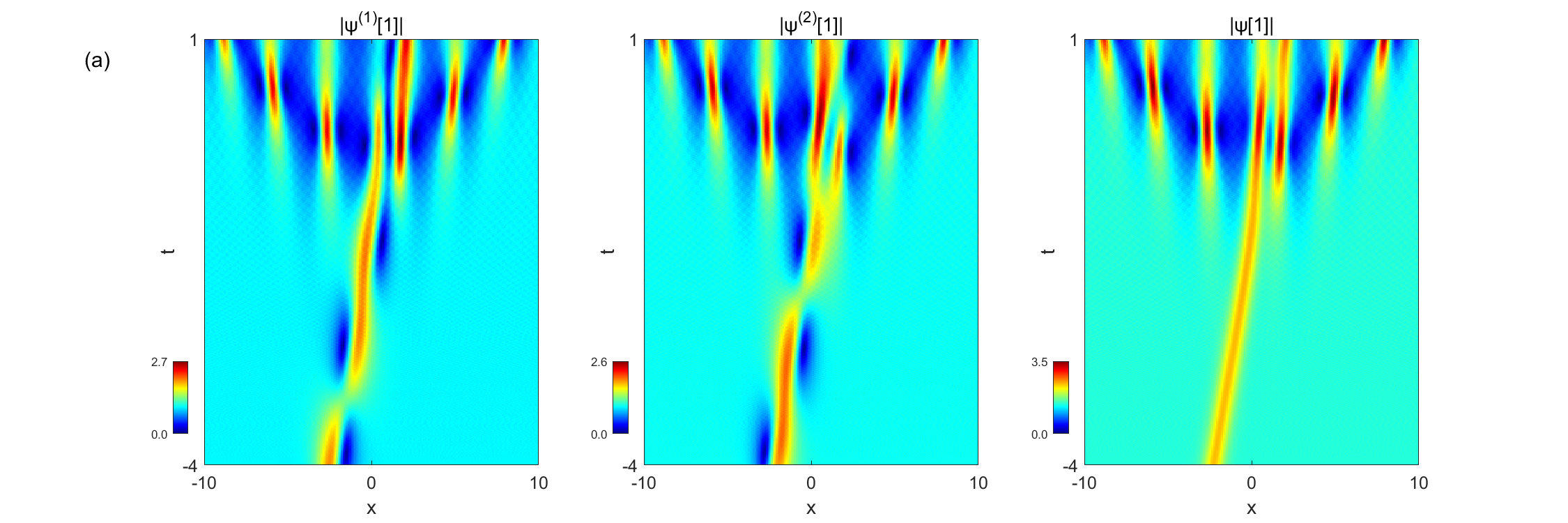}}
{\includegraphics[width=400 bp,height=3 cm]{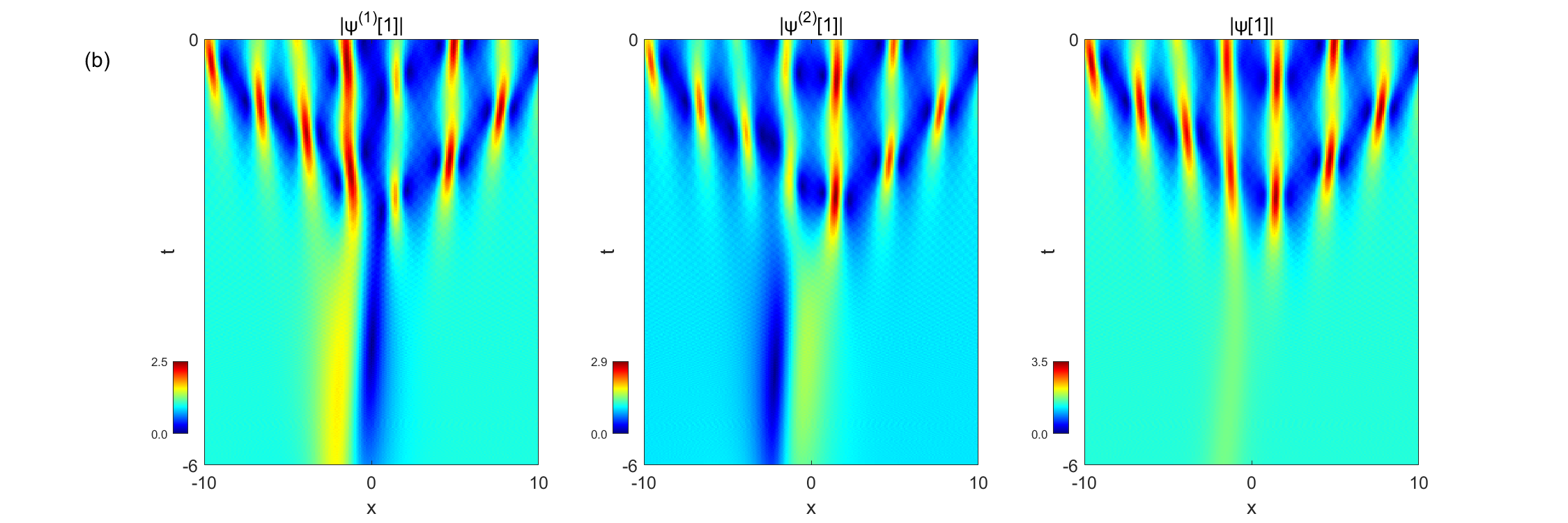}}
\caption{\footnotesize Numerical simulations of (a) Type-I and (b) Type-II moving BSs developed from the exact solutions (\ref{BS}) with $\varepsilon=0.01$. Other parameters are chosen in Figs.~\ref{beating1}.}
  \label{smbs}
\end{figure}

\begin{figure}[H]
	\centering
{\includegraphics[width=400 bp,height=3 cm]{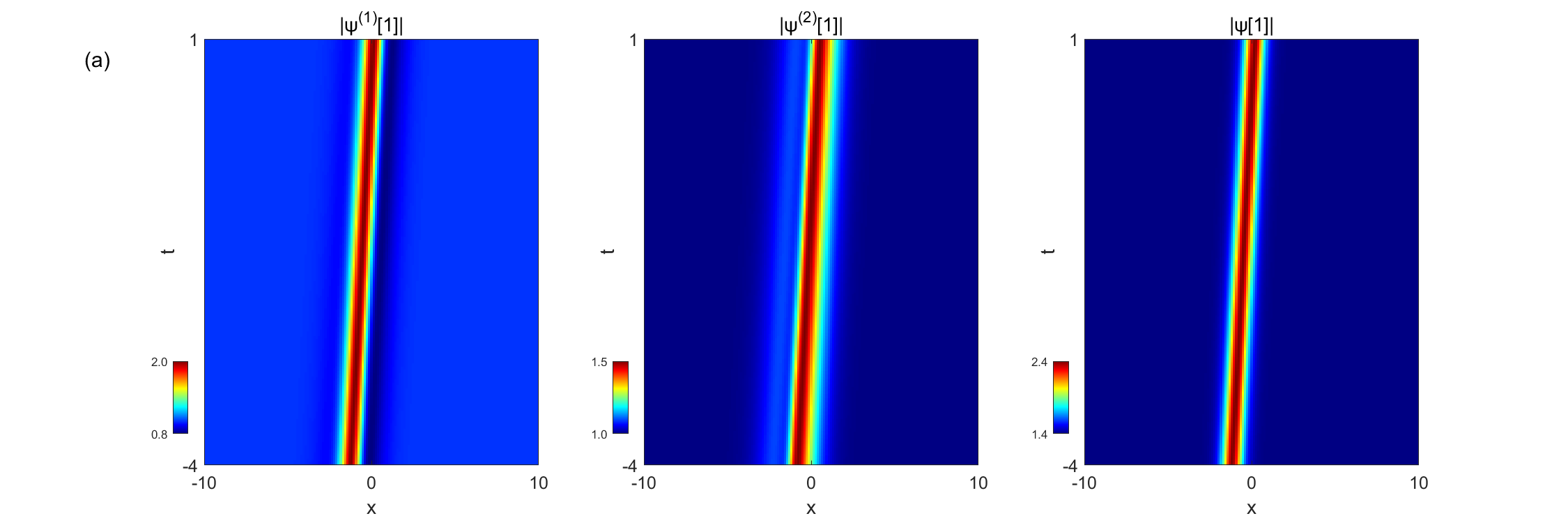}}
{\includegraphics[width=400 bp,height=3 cm]{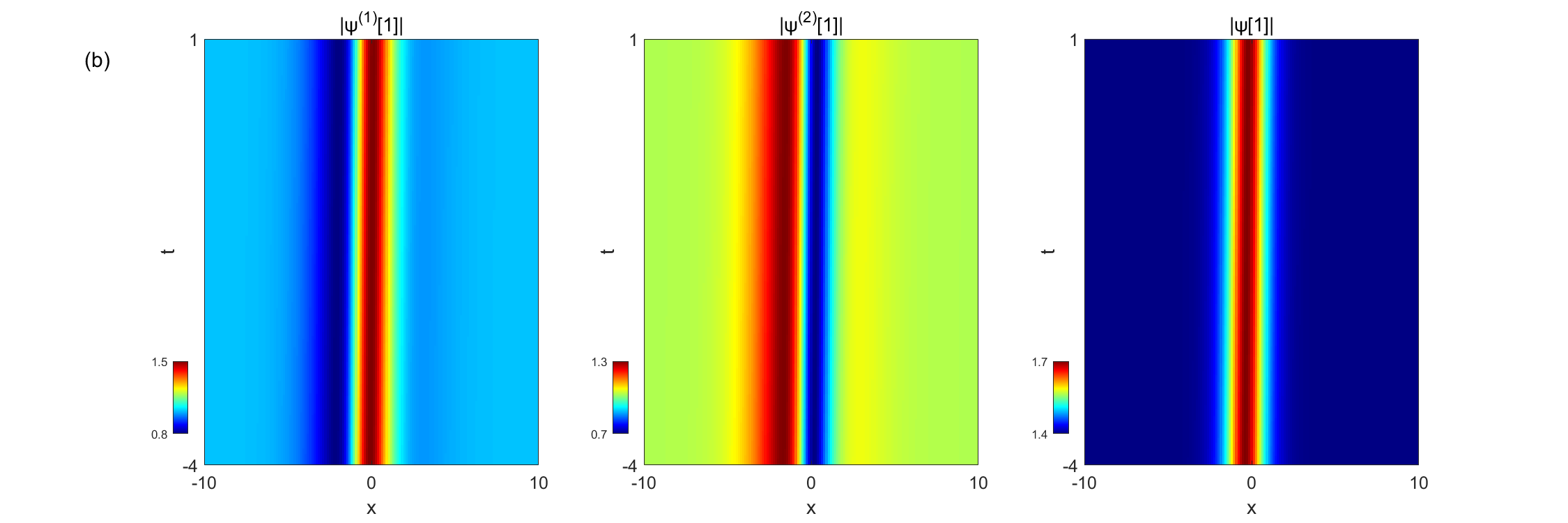}}
\caption{\footnotesize Numerical simulations of the moving kink solitons from (a) Type-I and (b) Type-II BSs developed by Eq.~(\ref{BS}) with $\varepsilon_{m}$. Other parameters are chosen in Figs.~\ref{beating3}.}
  \label{smbs1}
\end{figure}

Next, by means of the exact solutions (\ref{eq:RW}) as initial conditions, we report numerical simulation results for the degenerate and non-degenerate RWs and their transformed solitons, as depicted in Figs.~\ref{smrw} and \ref{smrw1}. In parallel, we present numerical simulations of the two types of the moving BSs and their corresponding transformed solitons, using the initial conditions given by Eqs.~(\ref{BS}), as shown in Figs.~\ref{smbs} and \ref{smbs1}.

\vspace{3mm}
\section{Conclusions} \label{eq:sec8}
\vspace{2mm}
We investigated the degenerate and non-degenerate localized wave solutions of the CLPD equations, which described the dynamics of the Heisenberg ferromagnetic spin chain. Using DT method, we constructed vector TWs, RWs, BSs and resonant modes and analyzed their state transitions induced by the fourth-order effect. Both the degenerate and non-degenerate TWs and RWs can be transformed into solitons. The solitons converted from the degenerate TWs have velocities approaching zero, whereas those from the degenerate RWs are strictly stationary, in contrast to the generally moving solitons obtained in the non-degenerate cases. Such degenerate transformed solutions are absent in the CH equation, indicating a new transition phenomenon induced by the fourth-order effect. Individual branches in the resonant modes also exhibit state transitions, although three branches can not be transformed simultaneously because they do not share the same fourth-order transition parameter. In particular, the TW branch is transformed into the multipeak solitons, while the RW branch is converted into the rational solitons. Most notably, both the moving and static BSs can be transformed into stable solitons. Such transformed states are cannot be observed in the Manakov system. We further analyzed the physical spectra of these localized waves. The spectra of RWs and BSs were obtained explicitly, allowing their state transition conditions to be identified directly. Finally, by using the exact solutions as initial conditions, numerical simulations confirm our analytical results. Although the Manakov, CH and CLPD equations share the same spatial spectral problem in their Lax pairs, only the CLPD equations reveal the state transitions in the degenerate regions. The underlying mechanism responsible for this phenomenon will be investigated in our future work.

\vspace{3mm}
\	\section*{CRediT authorship contribution statement}
	
	\textbf{Rong Han}: Methodology, analysis, simulations, calculations, visualization, code, writing. \textbf{Muchen Dong}: Methodology, analysis, writing. \textbf{Lei Wang}: Idea, conceptualization, methodology, analysis, writing, supervision, project administration.
	\section*{Declaration of competing interest}
	The authors declare that they have no known competing financial interests or personal relationships that could have appeared to influence the work reported in this paper.
	\section*{Data availability statements}
	Data availability is not applicable to this article as no new data were created or analyzed in this study.
	\section*{Acknowledgment}
	We express our sincere thanks to all the members of our discussion group for their valuable comments. The authors acknowledge financial support from the National Natural Science Foundation of China (No.~12375002).

\bibliographystyle{unsrtnat}
\bibliography{bibliography}


\end{document}